\documentclass[10pt,onecolumn]{IEEEtran}
\usepackage{amsthm}
\usepackage{amssymb}
\usepackage{graphicx}
\usepackage{multirow}
\usepackage[cmex10]{amsmath}
\usepackage{subfig}
\usepackage{longtable}
\usepackage{supertabular}

\newcommand{\abs}[1]{\vert #1 \vert }

\newcommand{\defn}{\triangleq}

\newcommand{\msf}[1]{\mathsf{#1}}

\newtheorem{theorem}{Theorem}
\newtheorem{lemma}{Lemma}
\newtheorem{cor}{Corollary}
\newtheorem{define}{Definition}

\newtheorem{proposition}{Proposition}
\allowdisplaybreaks
\begin{document}
\title{Detecting Arbitrary Attacks Using Continuous Secured Side Information in Wireless Networks}
\author{Ruohan Cao
  \thanks{R.~Cao is with the
     the School of Information and Communication Engineering,
    Beijing University of Posts and Telecommunications (BUPT), Beijing
    100876, China (e-mail: caoruohan@bupt.edu.cn). }}
\maketitle

\begin{abstract}
This paper focuses on Byzantine attack detection for Gaussian two-hop one-way relay network, where an amplify-and-forward relay may conduct Byzantine attacks by forwarding altered symbols to the destination.
For facilitating attack detection, we utilize the openness of wireless medium to make the destination observe some secured signals that are not attacked. 
Then, a detection scheme is developed for the destination by using its secured observations to statistically check other observations from the relay. 
On the other hand, notice the Gaussian channel is continuous, which allows the possible Byzantine attacks 
to be conducted within continuous alphabet(s). The existing work on discrete channel is not applicable for investigating the performance of the proposed scheme.
The main contribution of this paper is to prove that if and only if the wireless relay network satisfies a non-manipulable channel condition, the proposed detection scheme achieves asymptotic errorless performance against arbitrary attacks that allow the stochastic distributions of altered symbols to vary arbitrarily and depend on each other. No pre-shared secret or secret transmission is needed for the detection. Furthermore, we also prove that the relay network is non-manipulable as long as all channel coefficients are non-zero, which is not essential restrict for many practical systems. 
\end{abstract}

\section{Introduction}
Relay nodes are widely employed in modern communication networks to
  enhance coverage and connectivity of the networks. This dependence
  on the relaying infrastructure may increase the risk on security as
  malicious relays may forward false information in order to deceive the
  intended participants into accepting counterfeit information. These attacks, referred to as Byzantine
    attacks, impose significant ramifications on the design of
  network protocols \cite{Buttyan2006Security}\cite{bloch2011physical}. With the presence of Byzantine attacks, the attack detection technique, which 
determines whether Byzantine attacks are conducted or not, is one of the key steps 
supporting secure communication. 

 The work on attack detection starts above physical-layer, where
 each link is treated as a unit-capacity bit-pipe, while specific physical-layer characteristics are 
shielded. 
Based on this setting, 
cryptography keys are often used to make attacks detectable \cite{papadimitratos2006secure}, \cite{hu2005ariadne},
while requiring
    the cryptographic keys, to which the
    relays are not privy, to be shared between the source and
    destination before the communication takes place.
 Without using cryptography keys,
information theoretic detection schemes are proposed for multicast system or Caterpillar Network \cite{ho2008byzantine, kosut2009nonlinear}.  
These schemes are able to achieve errorless performance in probability, yet assuming that at least one relay or
link is absolutely trustworthy.


Besides these schemes treating channels as noiseless bit-pipes, there are
also many other attack detection schemes designed according to specific characteristics of physical-layer channels
for varying application scenarios. These schemes are mainly enabled by utilizing tracing symbols, or 
self-information provided by network topology structure.
In tracing-based schemes \cite{mao2007tracing}-\cite{nonherentsCL}, source node inserts tracing symbols into a sequence of information bits, and then sends them together to
the destination.  
Relying on the \emph{priori} knowledge of tracing symbols,
the destination could detect attacks by comparing its observed tracing symbols and the ground truth of tracing symbols.
%
%
This tracing-based method is applicable
with perfect CSI \cite{mao2007tracing, Tradeoff} or no need of CSI \cite{noncoherent, nonherentsCL} for varying network scenarios.
The tracing-symbol based schemes commonly assume that the value and insertion location of the tracing symbols are known only at the source
and the destination, which indeed requires a additional  tracing-symbol distribution
mechanism implemented between communication parties. Besides that,  these schemes assume that each malicious relay garbles
its received symbols according to independent and identically
distributed (i.i.d.) stochastic distributions. This model of
i.i.d. attacks may not always be valid in practice, although it makes
analysis simple. The Byzantine attack detection methods presented in
\cite{mao2007tracing}-\cite{nonherentsCL} may no longer be provably unbreakable for non-i.i.d. attacks.

Notice that all the above-mentioned schemes detect attacks by inserting redundancy, the overhead cost thus increases.
In contrast, the schemes, utilizing side information (SI) provided by network topology structure, do not need to insert any 
redundancy or just insert negligible redundancy \cite{KimTWC}-\cite{CaoTIFS}. The detectability of parts of these schemes
are beyond i.i.d. attacks \cite{he2013strong}-\cite{CaoTIFS}.
Generally, upon the network topology structure, most of these schemes are able to gain secured SI, that it is absolutely not attacked, and it is statistically
correlated to the observations from the relay when the relay is non-malicious. 
Then, the SI can be used to check the signals observed from the relay.  
For instance, in \cite{KimTWC} and \cite{KimCL}, a direct link between source and destination
allows the destination to observe signals from the source directly. These observations
are secured SI according to its safeness and statistic dependence on other observations.
Then, attacks are detectable by comparing the observations from the relay with the secured SI.
Similarly, in \cite{OFDM} and \cite{ARQ}, source node detects attacks using its transmitted signals as secured SI. 
The detection performance of  \cite{KimTWC}-\cite{ ARQ} are impacted by channel fading and signal-to-noise ratio (SNR).
Some denoising measures, such as perfect correction codes
(ECCs), are often required. However, due to channel fading, the achieved performance still cannot approach to errorless asymptotically. 
Especially, the performance of \cite{KimTWC} and \cite{KimCL} highly depends
on the the quality of direct link, such that it may not work well in the network where direct link does not
exist or suffers deep fading.

On the contrary, \cite{he2013strong}-\cite{CaoTIFS} could detect attacks with arbitrary small error probability. In particular, 
\cite{he2013strong}-\cite{GravesISIT13} consider two-way relaying (TWR) protocol for the typical three-node network, where
communication parties could use its transmitted signals as secured SI to detect arbitrary attacks. 
To elaborate a little further, in \cite{he2013strong}, 
communication parties are required 
to simultaneously transmit signal to relay with the same power constraint,
then each node's own lattice-coded transmitted
symbols are employed to simultaneously support secret transmission and
construct an algebraic manipulation detection (AMD) code to detect
arbitrary Byzantine attacks in TWR networks with Gaussian channels.
It is difficult to
extend this scheme to non-Gaussian channels and to the destination with restricted power. 
In our previous work
\cite{GravesINFOCOM12}-\cite{GravesISIT13}, focusing on two-way relay system, we show
that for discrete memoryless channels (DMCs), it is possible to detect arbitrary Byzantine attacks
dispensing any AMD
code or cryptographic keys. The basic idea is 
that each node utilizes its own
transmitted symbols as secured SI for
statistically checking against the other node's symbols forwarded by the
relay. This scheme is difficult to
extend beyond DMC channels. 
We extend the method proposed in \cite{GravesINFOCOM12}-\cite{GravesISIT13} to the DMC one-way relay system composing of two potential malicious relays. Since 
all observations of the destination are prone to be attacked \cite{CaoTIFS}, this network setup only provides
unsecured SI for the destination. This work indicates that due to the lack of
secured SI, we cannot properly protect communication parties against arbitrary attacks.

In this paper, we consider attack detection problem for the 
wireless typical three-node network. One-way relaying protocol is performed in the network. 
 The goal of this paper is to make the destination probabilistically detect arbitrary attacks, despite of i.i.d or non-i.i.d attacks, without using any AMD code or secret transmission. 
To that end, we facilitate attack detection by utilizing the openness of wireless medium to make the destination obtain secured SI, i.e., the signals directly heared from the source, etc. 
On the other hand, due to  properties of wireless medium again, all observed signals are continuous, 
the possible attacks are also continuous in the sense that they are likely to be conducted within continuous alphabet(s). Our previous work focusing on utilizing secured SI in DMCs \cite{GravesINFOCOM12}-\cite{CaoTIFS}
are not applicable to investigate the detectability of continuous attacks. This paper proves the detectability of continuous attacks is equivalent to the non-manipulability of wireless channels, also proves the non-manipulability of general wireless channels.
The main contributions of this paper are summarized as follows:
\begin{enumerate}
\item  We prove that under a non-manipulable condition of wireless channel, all
  Byzantine attacks, despite of i.i.d or non-i.i.d, are asymptotically
  detectable by simply using secured SI to statistically check the observations
  from the relay. The proposed scheme does not use any AMD code or secret transmission, while achieving 
  asymptotically errorless performance. This result is summarized in Theorem 1 of Section III.
%
  
%
%

\item We prove that the non-manipulable channel condition is satisfied as long as the secured SI is not vacancy.
It is not essential restrict for the general wireless relay networks with direct channels.
It further indicates the system does not need much power burden to generate the secured SI. 
This result is summarized in Proposition 1 of Section III.

\end{enumerate}
The rest of this paper is organized as follows. In Section II, we 
discuss the system model and formalize the problem to be addressed.
We detail our main contribution as stated above in Section III.  Numerical Examples are
presented in Section IV. 
The conclusions are drawn in Section V. In Appendices A, we detail the proof of Proposition 1.
In Appendices B and C, we detail the necessity and sufficiency proofs of 
Theorem 1, respectively.

\section{System Model}

\subsection{Notation}

Let $A$ be an 2-dimensional $M\!\times\! N$ matrix. For $i=1,2,\ldots,M$ and $j=1,2,\ldots,N$,
$[A]_{i,j}$ denotes the $(i,j)$th entry of $A$. 
Let $B$ be an 3-dimensional $M\!\times\! N\!\times\! K$ matrix. For $i=1,2,\ldots,M$, $j=1,2,\ldots,N$,
and $k=1,2,\ldots,K$, $[B]_{i,j,k}$ denotes the $(i,j,k)$th entry of $B$. 
Whenever there is no
ambiguity, we will employ the notation with no brackets for
simplicity.
The identity and zero
matrices of any dimension are denoted by the generic symbols $I$ and
$0$, respectively.


For continuous random variables, we use capital letters
and lower-case letters to denote the random variables and variables of corresponding probability density functions (PDFs), respectively.
For instance, suppose $U$ and $V$ are random variables defined over $\left(-\infty,\:+\infty\right)$.
$f_{U}\left(u\right)$ and $f_{V}\left(v\right)$ denote PDFs of $U$ and $V$, respectively.
$f_{V\left|U\right.}\left(v\left|u\right.\right)$ denotes the conditional PDF 
of $V$ given $U$.  $F_{V\left|U\right.}\left(v\left|u\right.\right)$ denotes the conditional cumulative distribution function (CDF) 
of $V$ given $U$. We employ $U^n$ to denote a sequence of $n$ continuous random variables defined over
$\left(-\infty,\:+\infty\right)$, and $U_i$ to denote the $i$th random variable in
$U^n$. $f_{V^n\left|U^n\right.}\left(v^n\left|u^n\right.\right)$ denotes the conditional PDF 
of $V^n$ given $U^n$.

For the discrete random variables, we use upper-case script letters and
serif-font letters to denote the corresponding discrete alphabets and
elements in the alphabets, respectively.  For instance, suppose that
we denote a finite alphabet by
$\mathcal{X} = \{ \mathsf{x}_1, \mathsf{x}_2, \ldots,
\mathsf{x}_{|\mathcal{X}|} \}$,
where $\abs{\mathcal{X}}$ is the cardinality of $\mathcal{X}$. Then
$X$ is a generic random yariable oyer $\mathcal{X}$, and $\mathsf{x}$
is a generic element in $\mathcal{X}$.
For a pair of random variables $X$ and $Y$, we use $P_{X}(\mathsf{x})$
and $P_{X|Y}( \mathsf{x}| \mathsf{y})$ to denote the marginal
distribution of $X$ and the conditional distribution of $X$ given $Y$,
respectively. 
We also employ $X^n$ to denote a sequence of $n$ random variables defined over
$\mathcal{X}$, and $X_i$ to denote the $i$th random variable in
$X^n$. The counting function $N(\mathsf{x}|X^n)$ records the number of
occurrences of the element $\mathsf{x}$ in the sequence $X^n$. The
indicator function $1_i(\mathsf{x}|X^n)$ tells whether $X_i$ is
$\mathsf{x}$. We may also use $1_i(\mathsf{x})$ instead of
$1_i(\mathsf{x}|X^n)$ for simplicity, whenever the meaning is clear
from the context. We may
trivially extend the aforementioned notations to a tuple of symbols
drawn from the corresponding alphabets.  

\begin{figure}
\begin{centering}
\includegraphics[scale=0.5]{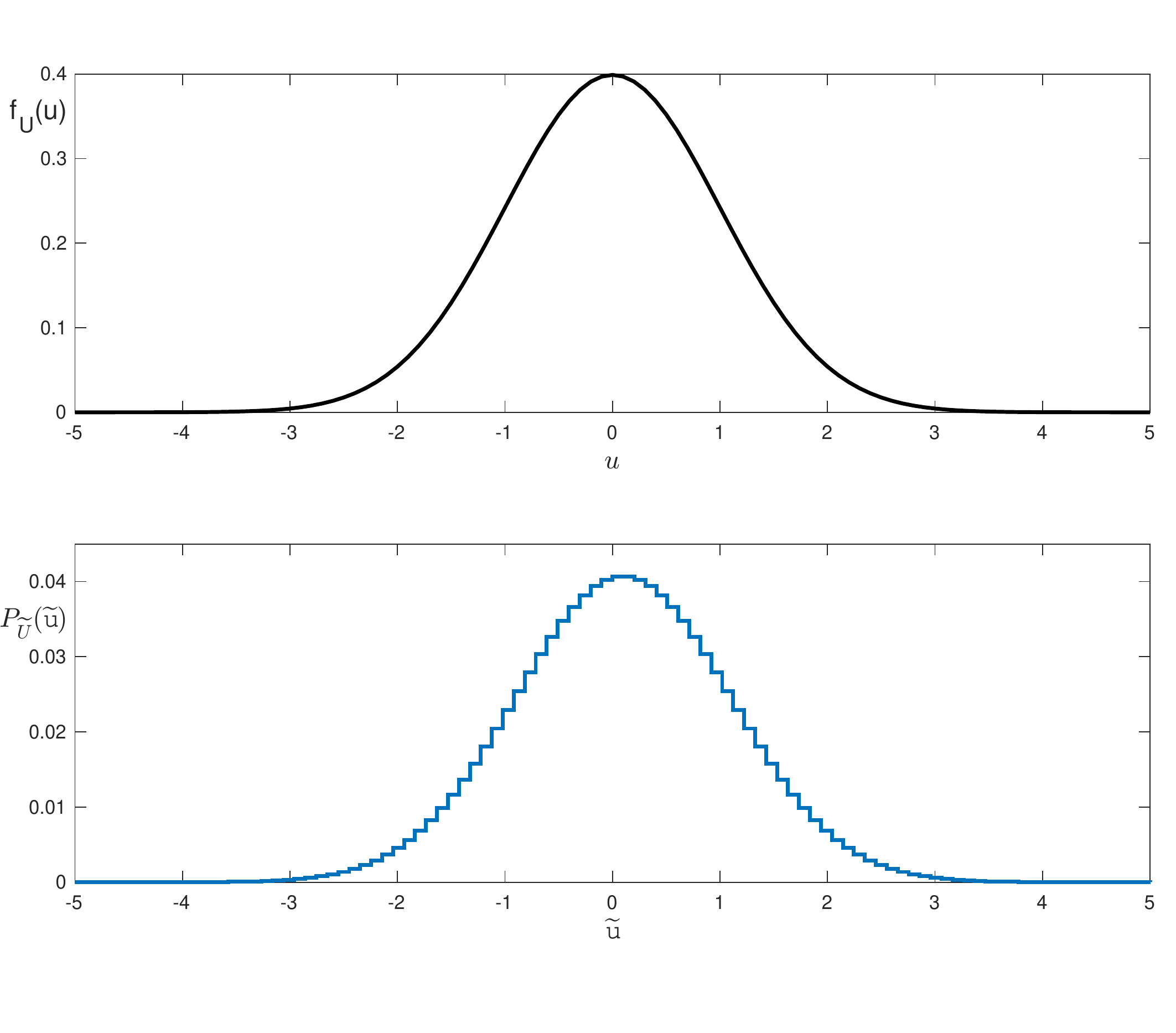}
\par\end{centering}

\caption{$f_{U}\left(u\right)$ and $P_{\widetilde{U}}(\mathsf{\widetilde{u}})$, where $(\alpha_{1}, \beta_{1}, n')=(-5, 5, 100)$.}

\end{figure}

For the continuous random variables, and the random variables 
whose continuity or discreteness are not definitely known, 
there are corresponding discrete 
random variables generated by quantifying the random variables.
Let use the quantization of continuous random variable $U$ as an example as follows.
Suppose one $n'$-length sequence $\mathsf{\widetilde{u}}_{1},\mathsf{\widetilde{u}}_{2},\ldots,\mathsf{\widetilde{u}}_{n'}$. 
Then, for $j=1,\ldots,n'$, $\mathcal{B}\left(\mathsf{\widetilde{u}}_{j}\right)$ denotes the domain consisting of $\mathsf{\widetilde{u}}_{j}$. They 
satisfy the constraints as follows.
\begin{align}
\label{q_u1}
&\alpha_{1}=\mathsf{\widetilde{u}}_{1}<\mathsf{\widetilde{u}}_{2}<\mathsf{\widetilde{u}}_{3}\ldots<\mathsf{\widetilde{u}}_{n'-1}=\beta_{1},\:\beta_{1}<\mathsf{\widetilde{u}}_{n'},\\
\label{q_u2}
&\widetilde{\mathsf{u}}_{j}-\widetilde{\mathsf{u}}_{j-1}=\frac{\beta_{1}-\alpha_{1}}{n'-2},j=2,3,\ldots,n'-1,\\
\label{q_u3}
&\mathcal{B}\left(\widetilde{\mathsf{u}}_{j}\right)=\begin{cases}
\begin{array}{cc}
\left(\widetilde{\mathsf{u}}_{j-1},\,\widetilde{\mathsf{u}}_{j}\right], & j=2,3,\ldots,n'-1\\
\left(-\infty,\,\alpha_{1}\right], & j=1\\
\left(\beta_{1},\,+\infty\right), & j=n'
\end{array},\end{cases}
\end{align}where $\alpha_{1}$ and $\beta_{1}$ are assumed to depend on $n'$, $\triangle_{u}=\frac{\beta_{1}-\alpha_{1}}{n'-2}$, and $\lim_{n'\rightarrow\infty}\triangle_{u}=0$.
Based on the definition of $\mathcal{B}\left(\widetilde{\mathsf{u}}_{j}\right)$, the continuous variable $U$
can be quantized to discrete $\widetilde{U}$. In particular, if $U\in\mathcal{B}\left(\widetilde{\mathsf{u}}_{j}\right)$, then $\widetilde{U}=\widetilde{\mathsf{u}}_{j}$.
In other words, $\widetilde{U}\triangleq\sum_{j=1}^{n'}1\left(U\in\mathcal{B}\left(\widetilde{\mathsf{u}}_{j}\right)\right)\widetilde{\mathsf{u}}_{j}$.
Correspondingly, $U^n$ is quantized to $\widetilde{U}^n$, where $\widetilde{U}_i\triangleq\sum_{j=1}^{n'}1\left(U_i\in\mathcal{B}\left(\widetilde{\mathsf{u}}_{j}\right)\right)\widetilde{\mathsf{u}}_{j}$. To illustrate motivation of the proposed quantization, let us take an example where $U$ follows standard Gaussian distribution, and it is quantized to $\widetilde{U}$ by setting $(\alpha_{1}, \beta_{1}, n')=(-5, 5, 100)$. 
Fig. 1 presents the PDF and probability mass function (PMF) of $U$ and $\widetilde{U}$, i.e., $f_{U}\left(u\right)$ and $P_{\widetilde{U}}(\mathsf{\widetilde{u}})$, respectively. It is observed that the curve of $P_{\widetilde{U}}(\mathsf{\widetilde{u}})$ is similar to the curye of $f_{U}\left(u\right)$. $P_{\widetilde{U}}(\mathsf{\widetilde{u}})$ could be used for fitting $f_{U}\left(u\right)$. This observation motivates us to make 
the random variables be simulated by discrete variables, whose discreteness facilitate our analysis work given later.

In this paper, $\widetilde{U}$ is referred as quantized version of $U$. $\left(\alpha_{1},\beta_{1},n'\right)$  is referred as quantization parameter. 
We also term above-mentioned quantization of $U$ as that $U$ is quantified according to $\left(\alpha_{1},\beta_{1},n'\right)$. These notations
could be applied to other random variables and their corresponding quantized versions. 
Obviously, the quantized variables are discrete, hence, their notations follow the format of discrete variables given above. More details are listed in Table I.  
{\begin{table*}[!t]
\centering
\caption{Notation Table}
\scriptsize
\begin{tabular}{|c|c|}
\hline
  $S$&
 	\text{Source symbol, $S\in\left\{ -1,1\right\}$ }\\
\hline	
$U$&
 	\text{Symbol received by the relay, $U\in\left(-\infty,\:+\infty\right)$}\\
\hline
 $V$&
 	\text{Symbol forwarded by the relay, $V\in\left(-\infty,\:+\infty\right)$}\\
\hline
$Y$&
 	\text{The destination's received symbol from the relay, $Y\in\left(-\infty,\:+\infty\right)$}\\
\hline	
$X$&	
	\text{The destination's received symbol from the source, $X\in\left(-\infty,\:+\infty\right)$}\\
\hline	
$\widetilde{U}$&
      \text{Quantized version of $U$ according to $\left(\alpha_{1},\beta_{1},n'\right)$, please refer to (\ref{q_u1})-(\ref{q_u3}) for detailed definition.}\\	
\hline      
\multirow{4}{*}{$\widetilde{V}$}& Quantized version of $V$ according to {\small{$\left(\alpha_{2},\beta_{2},n_v\right)$, $\widetilde{V}\triangleq\sum_{j=1}^{n'}1\left(V\in\mathcal{B}\left(\widetilde{\mathsf{v}}_{j}\right)\right)\widetilde{\mathsf{v}}_{j}$}}, where \\
&{\small{$\alpha_{2}=\mathsf{\widetilde{v}}_{1}<\mathsf{\widetilde{v}}_{2}<\mathsf{\widetilde{v}}_{3}\ldots<\mathsf{\widetilde{v}}_{n_v-1}=\beta_{2},\:\beta_{2}<\mathsf{\widetilde{v}}_{n_v}$}}\\
&{\small{$\widetilde{\mathsf{v}}_{j}-\widetilde{\mathsf{v}}_{j-1}=\frac{\beta_{2}-\alpha_{2}}{n_v-2},j=2,3,\ldots,n_v-1$}},\\
&{\small{$\mathcal{B}\left(\widetilde{\mathsf{v}}_{j}\right)=\begin{cases}
\begin{array}{cc}
\left(\widetilde{\mathsf{v}}_{j-1},\,\widetilde{\mathsf{v}}_{j}\right], & j=2,3,\ldots,n_v-1\\
\left(-\infty,\,\alpha_{2}\right], & j=1\\
\left(\beta_{2},\,+\infty\right), & j=n_y
\end{array}\end{cases}$}}\\
\hline
\multirow{4}{*}{$\widetilde{Y}$}& Quantized version of $Y$ according to {\small{$\left(\alpha_{y},\beta_{y},n_y\right)$, $\widetilde{Y}\triangleq\sum_{j=1}^{n'}1\left(Y\in\mathcal{B}\left(\widetilde{\mathsf{y}}_{j}\right)\right)\widetilde{\mathsf{y}}_{j}$}}, where \\
&{\small{$\alpha_{3}=\mathsf{\widetilde{y}}_{1}<\mathsf{\widetilde{y}}_{2}<\mathsf{\widetilde{y}}_{3}\ldots<\mathsf{\widetilde{y}}_{n_y-1}=\beta_{y},\:\beta_{3}<\mathsf{\widetilde{y}}_{n_y}$}}\\
&{\small{$\widetilde{\mathsf{y}}_{j}-\widetilde{\mathsf{y}}_{j-1}=\frac{\beta_{3}-\alpha_{3}}{n_y-2},j=2,3,\ldots,n_y-1$}},\\
&{\small{$\mathcal{B}\left(\widetilde{\mathsf{y}}_{j}\right)=\begin{cases}
\begin{array}{cc}
\left(\widetilde{\mathsf{y}}_{j-1},\,\widetilde{\mathsf{y}}_{j}\right], & j=2,3,\ldots,n_y-1\\
\left(-\infty,\,\alpha_{3}\right], & j=1\\
\left(\beta_{3},\,+\infty\right), & j=n_y
\end{array}\end{cases}$}}\\
\hline
\multirow{4}{*}{$\widetilde{X}$}& Quantized version of $X$ according to $\left(\alpha_{4},\beta_{4},n_x\right)$, $\widetilde{Y}\triangleq\sum_{j=1}^{n'}1\left(X\in\mathcal{B}\left(\widetilde{\mathsf{x}}_{j}\right)\right)\widetilde{\mathsf{x}}_{j}$, where \\
&$\alpha_{4}=\mathsf{\widetilde{x}}_{1}<\mathsf{\widetilde{x}}_{2}<\mathsf{\widetilde{x}}_{3}\ldots<\mathsf{\widetilde{x}}_{n_x-1}=\beta_{x},\:\beta_{4}<\mathsf{\widetilde{x}}_{n_x}$\\
&$\widetilde{\mathsf{x}}_{j}-\widetilde{\mathsf{x}}_{j-1}=\frac{\beta_{4}-\alpha_{4}}{n_x-2},j=2,3,\ldots,n_x-1$,\\
&$\mathcal{B}\left(\widetilde{\mathsf{x}}_{j}\right)=\begin{cases}
\begin{array}{cc}
\left(\widetilde{\mathsf{x}}_{j-1},\,\widetilde{\mathsf{y}}_{j}\right], & j=2,3,\ldots,n_x-1\\
\left(-\infty,\,\alpha_{4}\right], & j=1\\
\left(\beta_{4},\,+\infty\right), & j=n_x
\end{array}\end{cases}$\\
\hline
$\mathcal{S}$, $\mathcal{\widetilde{U}}$, $\mathcal{\widetilde{V}}$, $\mathcal{\widetilde{Y}}$, $\mathcal{\widetilde{X}}$&
\text{Alphabets of $S$, $\widetilde{U}$, $\widetilde{V}$, $\widetilde{Y}$, and $\widetilde{X}$, respectively }\\
\hline
$\mathsf{s}_{i}$, $\mathsf{\widetilde{u}}_{i}$, $\mathsf{\widetilde{v}}_{i}$, $\mathsf{\widetilde{y}}_{i}$, $\mathsf{\widetilde{x}}_{i}$&
\text{The $i$-th elements in $\mathcal{S}$, $\mathcal{\widetilde{U}}$, $\mathcal{\widetilde{V}}$, $\mathcal{\widetilde{Y}}$, and $\mathcal{\widetilde{X}}$, respectively }\\
\hline
$\mathsf{s}$, $\mathsf{\widetilde{u}}$, $\mathsf{\widetilde{v}}$, $\mathsf{\widetilde{y}}$, $\mathsf{\widetilde{x}}$&
\text{Generic elements in $\mathcal{S}$, $\mathcal{\widetilde{U}}$, $\mathcal{\widetilde{V}}$, $\mathcal{\widetilde{Y}}$, and $\mathcal{\widetilde{X}}$, respectively }\\
\hline
$P_{\widetilde{U}}(\mathsf{\widetilde{u}})$&
\text{PMF of $\widetilde{U}$}\\
\hline
$1_{i}\left(\widetilde{V}_{i}=\widetilde{\mathsf{v}}_{k}\right)$& \text{Indicator of whether $\widetilde{V}_{i}=\widetilde{\mathsf{v}}_{k}$ is true or not}\\
\hline
$1_{i}\left(\widetilde{U}_{i}=\widetilde{\mathsf{u}}_{j}\right)$& \text{Indicator of whether $\widetilde{U}_{i}=\widetilde{\mathsf{v}}_{j}$ is true or not}\\
\hline
{\small{$N\left(\widetilde{\mathsf{u}}_{j}\left|\widetilde{U}^{n}\right.\right)$}}
&{\small{$\sum_{i}^{n}1_{i}\left(\widetilde{\mathsf{u}}_{j}\left|\widetilde{U}^{n}\right.\right)$}}\\ \hline
$P_{\widetilde{U}\left|\widetilde{X}\right.}\left(\widetilde{\mathsf{u}}_{j}\left|\widetilde{\msf{x}}\right.\right)$& \text{Conditional probability of $\left\{ \widetilde{U}=\widetilde{\mathsf{u}}_{j}\right\} $ given $\left\{ \widetilde{X}=\widetilde{\msf{x}}\right\} $}\\ \hline
$\triangle_{u}$, $\triangle_{v}$ &
$\triangle_{u}=\frac{\beta_{1}-\alpha_{1}}{n'-2}$, $\triangle_{v}=\frac{\beta_{2}-\alpha_{2}}{n_v-2}$\\
\hline
$\Phi\left(\cdot\right)$ & \text{pulse response function}\\

\hline
\end{tabular}
\end{table*}}

\subsection{Channel Model}
\begin{figure}
\begin{centering}
\includegraphics[scale=0.5]{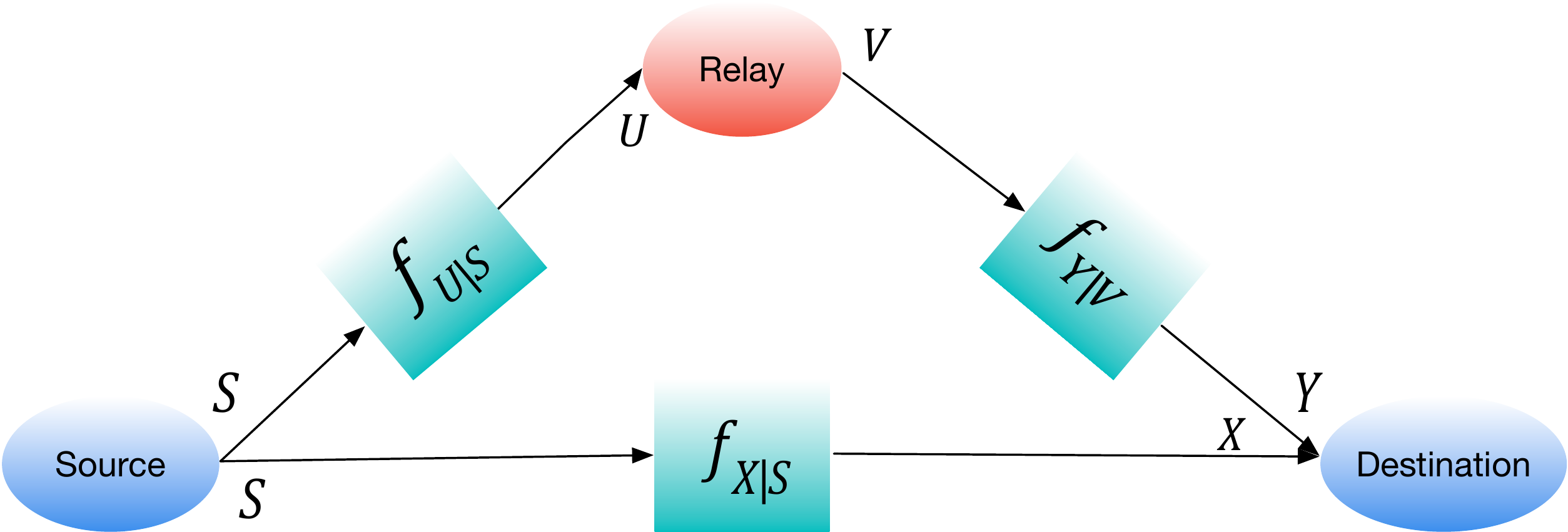}
\par\end{centering}

\caption{System model.}

\end{figure}

Let us focus on the two-hop one-way relay example, where one source and one destination exchange information
via a relay system. The communication takes place in two stages. Each stage includes $n$ instants. In the first $n$ instants,
the source sends $n$-length sequence $S^{n}$ to the relay node, $S$ is a discrete random variable. Correspondingly, the relay observes $n$-length sequence $U^n$.
Without loss of generalization, we assume that $S$ is equiprobability binary symbol generated from alphabet $(+1, -1)$
The channel from the source to the relay is specified by $U =h_1S+N_r$, where $U$ is the received signal of the relay in each instant, $h_1$ is the constant channel coefficient,
and $N_r$ is AWGN existed in the channel. $N_r$ is random variables following standard Gaussian distribution. Then, the pdf of $U$ conditioned on $S=\mathsf{s}$ is given as 
\begin{equation}\label{R1_channel}
f_{U\left|S\right.}\left(u\left|\mathsf{s}\right.\right)=\frac{1}{\sqrt{2\pi}}\exp\left(-\frac{\left(u-h_{1}\mathsf{s}\right)^{2}}{2}\right).
\end{equation} 

Meanwhile, due to the open nature of wireless scenario, the destination
could hear sequence $X^n$ from the direct channel.
The direct channel is specified by $X =h_3S+N_d$, where $h_3$ is the constant coefficient of the direct channel, and $N_d$ is standard AWGN existed in the direct channel.
\begin{equation}\label{D_channel}
f_{X\left|S\right.}\left(x\left|\mathsf{s}\right.\right)=\frac{1}{\sqrt{2\pi}}\exp\left(-\frac{\left(x-h_{3}\mathsf{s}\right)^{2}}{2}\right).
\end{equation}
Based on (\ref{R1_channel}) and (\ref{D_channel}), the PDF of $U$ conditioned on $X$ is given by 
\begin{equation}
f_{U\left|X\right.}\left(u\left|{x}\right.\right)=\sum_{i=1}^{2}P_{S\left|X\right.}\left(\mathsf{s}_{i}\left|{x}\right.\right)f_{U\left|S\right.}\left(u\left|\mathsf{s}_{i}\right.\right),
\end{equation}where $f_{U\left|S\right.}\left(u\left|\mathsf{s}_{i}\right.\right)$ is given by (\ref{R1_channel}), and $P_{S\left|X\right.}\left(\mathsf{s}_{i}\left|{x}\right.\right)=\frac{\exp\left(-\frac{\left({x}-h_{3}\mathsf{s}_{i}\right)^{2}}{2}\right)}{\sum_{j=1}^{2}\exp\left(-\frac{\left({x}-h_{3}\mathsf{s}_{j}\right)^{2}}{2}\right)}$. We assume the
channels from source to relay and to destination are stationary and memoryless. As a result, $U^n$ and $X^n$ are i.i.d. sequences, we thus have $f_{U^{n}\left|X^{n}\right.}\left(u^{n}\left|{x^{n}}\right.\right)=\prod_{i=1}^{n}f_{U\left|X\right.}\left(u_{i}\left|{x_{i}}\right.\right)$, where $u_{i}$ and $x_i$ are the $i$-th symbol of generic sequences $u^n$ and $x^n$, respectively.

Secondly, in the instants $n+1, n+2, \ldots, 2n$, the relay forwards sequence $V^n$ to the destination. 
$V$ is the variable denoting the forwarded signal of the relay. Depending on the relay's attack act,
the alphabet of $V$ may be discrete or continuous. Furthermore, we allow the relay to conduct arbitrary attacks as long as $X^n$, $U^n$, and $V^n$ satisfy a markov constraint, i.e., $f_{V^{n}\left|U^{n},X^{n}\right.}\left(v^{n}\left|{u^{n},x^{n}}\right.\right)=f_{V^{n}\left|U^{n}\right.}\left(v^{n}\left|{u^{n}}\right.\right)$. It indicates the relay conducts attacks only upon its own observation.

The channel from the relay to the destination is specified by $Y =h_2V+N'_d$
where $h_2$ is the constant channel coefficient, $N'_d$ is standard AWGN existed in the channel. The pdf of $Y$ conditioned on $V$ is given as $f_{Y\left|V\right.}\left(y\left|{v}\right.\right)=\frac{1}{\sqrt{2\pi}}\exp\left(-\frac{\left(y-h_{2}{v}\right)^{2}}{2}\right)$. Correspondingly, 
the cumulative distribution function (cdf) of $Y$ conditioned on $V$ is given as 
\begin{equation}
F_{Y\left|V\right.}\left(t\left|{v}\right.\right)=\int_{-\infty}^{t}\frac{1}{\sqrt{2\pi}}\exp\left(-\frac{\left(y-h_{2}{v}\right)^{2}}{2}\right)dy.
\end{equation}
We assume the destination knows $f_{U\left|X\right.}\left(u\left|{x}\right.\right)$ and $F_{Y\left|V\right.}\left(t\left|{v}\right.\right)$ for facilitating attack detection.
$\left(f_{U\left|X\right.}\left(u\left|{x}\right.\right),\; F_{Y\left|V\right.}\left(t\left|{v}\right.\right)\right)$ is referred as to observation channel. In the instants $n+1, n+2, \ldots, 2n$, the destination receives sequence $Y^n$ from the relay. The channel from the relay to the destination is also stationary and memoryless,  thus pdf of $Y_i$ given $V^n$ is given by $f_{Y_{i}\left|V^{n}\right.}\left(y\left|{v^{n}}\right.\right)=f_{Y\left|V\right.}\left(y\left|{v_{i}}\right.\right)$, where $v_{i}$ is the $i$-th symbol of sequence $v^n$, $Y_i$ denotes the $i$-th variable in random sequence $Y^n$.

\subsection{Attack Model}

\begin{figure}
\begin{centering}
\includegraphics[scale=0.5]{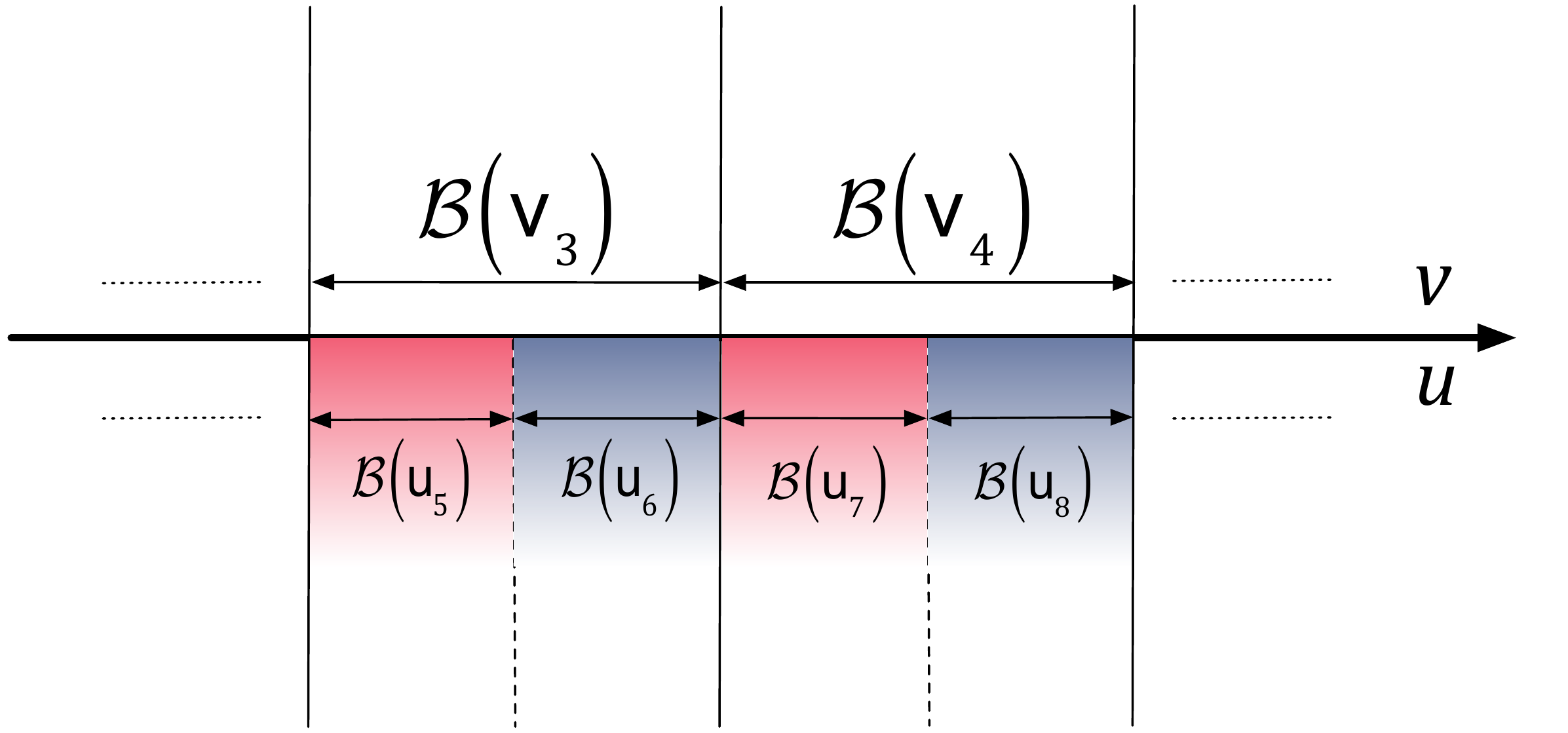}
\par\end{centering}

\caption{An instance of quantization setup employed by Section II. C, where each $\mathcal{B}\left(\widetilde{\mathsf{u}}_{j}\right)$
($j=1,\ldots, n'$) is included by one of $\{\mathcal{B}\left(\widetilde{\mathsf{v}}_{1}\right), \mathcal{B}\left(\widetilde{\mathsf{v}}_{2}\right),
\ldots, \mathcal{B}\left(\widetilde{\mathsf{v}}_{n_v}\right)\}$ in whole.}

\end{figure}

The potential malicious relay receives sequence $U^n$, and then forwards sequence $V^n$.
The relay may perform arbitrary attacks, including i.i.d. and non-i.i.d. attacks, while the parameter of attack is unknown to the destination. Thus, $V^n$ may be
a non-i.i.d. sequence, and distribution of $U^n$ and $V^n$, i.e., $f_{V^{n},U^{n}}\left(v^{n},u^{n}\right)$, is also unknown to the destination. 
This property motivates us to develop a non-parameter method to model the attacks.

For modeling the attack, we first notice the $U$ is continuous random variable. The alphabet of $V$ is not
definite, it is possible to be any arbitrary value in $\left(-\infty,\:+\infty\right)$. We quantize $U$ and $V$ according to quantized parameters $\left(\alpha_{1},\beta_{1},n'\right)$
and  $\left(\alpha_{2}, \beta_{2},n_v\right)$ where $\alpha_{2}$, $\beta_{2}$,  where $n_v$ depend on $n'$, $n_v$ approaches to infinity as $n'\rightarrow\infty$,
and then get quantized variables $\widetilde{U}$ and $\widetilde{V}$, respectively. Furthermore, to describe the maliciousness of relay, $n_v$ and $n'$ are properly chosen such that for $j=1,2,\ldots,n'$ and $k=1,2,\ldots,n_v$, $\mathcal{B}\left(\widetilde{\mathsf{u}}_{j}\right)$ and $\mathcal{B}\left(\widetilde{\mathsf{v}}_{k}\right)$ have either $\mathcal{B}\left(\widetilde{\mathsf{u}}_{j}\right)\subseteq\mathcal{B}\left(\widetilde{\mathsf{v}}_{k}\right)$ or $\mathcal{B}\left(\widetilde{\mathsf{u}}_{j}\right)\bigcap\mathcal{B}\left(\widetilde{\mathsf{v}}_{k}\right)=\varnothing$. We illustrate an instance of this quantization setup in Fig. 3.

Based on the quantization, the function $\triangle F_{\widetilde{V}^{n}\left|\widetilde{U}^{n}\right.}^{(n')}\left(v\left|u\right.\right)$ for sequence pair ($U^n$, $V^n$) is defined as
\begin{equation}\label{de_F}
\triangle F_{\widetilde{V}^{n}\left|\widetilde{U}^{n}\right.}^{(n')}\left(v\left|u\right.\right)=\begin{cases}
\begin{array}{cc}
\frac{\sum_{i=1}^{n}1_{i}\left(\widetilde{V}_{i}=\widetilde{\mathsf{v}}_{k}\right)1_{i}\left(\widetilde{U}_{i}=\widetilde{\mathsf{u}}_{j}\right)}{N\left(\widetilde{\mathsf{u}}_{j}\left|\widetilde{U}^{n}\right.\right)}, & N\left(\widetilde{\mathsf{u}}_{j}\left|\widetilde{U}^{n}\right.\right)\neq0,\, u\in\mathcal{B}\left(\widetilde{\mathsf{u}}_{j}\right),\, v\in\mathcal{B}\left(\widetilde{\mathsf{v}}_{k}\right)\\
0, & otherwise
\end{array}\end{cases},
\end{equation}
where $\frac{\sum_{i=1}^{n}1_{i}\left(\widetilde{V}_{i}=\widetilde{\mathsf{v}}_{k}\right)1_{i}\left(\widetilde{U}_{i}=\widetilde{\mathsf{u}}_{j}\right)}{N\left(\widetilde{\mathsf{u}}_{j}\left|\widetilde{U}^{n}\right.\right)}$ is the empirical transition probability from 
$\{\widetilde{U}=\widetilde{\mathsf{u}}_{j}\}$ to $\{\widetilde{V}=\widetilde{\mathsf{v}}_{k}\}$. Consider the case that the probability is strictly non-zero while $\mathcal{B}\left(\widetilde{\mathsf{u}}_{j}\right)\bigcap\mathcal{B}\left(\widetilde{\mathsf{v}}_{k}\right)=\varnothing$, then it indicates the relay partly modifies its received symbol (i.e., $U$) that belongs to $\mathcal{B}\left(\widetilde{\mathsf{u}}_{j}\right)$ to another disparate domain $\mathcal{B}\left(\widetilde{\mathsf{v}}_{k}\right)$.
To be more precise, if the relay is absolutely reliable, we must always have {\small{$\sum_{j=1}^{{n_{v}}}\sum_{i=1}^{n'}\left|\triangle F_{\widetilde{V}^{n}\left|\widetilde{U}^{n}\right.}^{\left(n'\right)}\left(\widetilde{\mathsf{v}}_{j}\left|\widetilde{\mathsf{u}}_{i}\right.\right)-\left[W_{0}^{(n')}\right]_{i,j}\right|=0$}}, where
  \begin{equation}
 \left[W_{0}^{\left(n'\right)}\right]_{i,j}=\begin{cases}
\begin{array}{cc}
1, & \mathcal{B}\left(\widetilde{\mathsf{u}}_{i}\right)\subseteq\mathcal{B}\left(\widetilde{\mathsf{v}}_{j}\right)\\
0, & otherwise
\end{array}\end{cases}
 \end{equation}
From this intuitively understanding, the malicious relay is defined
as follows. 
\begin{define} \label{def:maliciousness}
\textbf{(Malicious Relay)} The relay is said to be non-malicious if {\small{$\sum_{j=1}^{{n_{v}}}\sum_{i=1}^{n'}\left|\triangle F_{\widetilde{V}^{n}\left|\widetilde{U}^{n}\right.}^{\left(n'\right)}\left(\widetilde{\mathsf{v}}_{j}\left|\widetilde{\mathsf{u}}_{i}\right.\right)-\left[W_{0}^{(n')}\right]_{i,j}\right|
 \rightarrow 0$}} in probability as $n$ and $n'$ approach to infinity. Otherwise,
  the relay is considered malicious.\end{define}

Note that Definition~\ref{def:maliciousness} mainly tolerates two kinds of manipulation.
In the one manipulating scenario, the modification always makes $U\neq V$ as well as $\mathcal{B}\left(\widetilde{U}\right)\subseteq\mathcal{B}\left(\widetilde{V}\right)$. In other words, when the relay receives a symbol of $\mathcal{B}\left(\widetilde{\mathsf{u}}_{i}\right)$, 
and according to the quantization setup, there exists $j\in\{1,2,\ldots, n_v\}$ satisfying $\mathcal{B}\left(\widetilde{\mathsf{u}}_{i}\right)\subseteq\mathcal{B}\left(\widetilde{\mathsf{v}}_{j}\right)$. Then, the relay modifies the symbol to arbitrary another symbol that belongs to $\mathcal{B}\left(\widetilde{\mathsf{v}}_{j}\right)$.
It yields to 
  \begin{equation}
\triangle F_{\widetilde{V}^{n}\left|\widetilde{U}^{n}\right.}^{\left(n'\right)}\left(\widetilde{\mathsf{v}}_{j}\left|\widetilde{\mathsf{u}}_{i}\right.\right)=\begin{cases}
\begin{array}{cc}
1, & \mathcal{B}\left(\widetilde{\mathsf{u}}_{i}\right)\subseteq\mathcal{B}\left(\widetilde{\mathsf{v}}_{j}\right)\\
0, & otherwise
\end{array}\end{cases}
 \end{equation}for $i=1,\ldots,n'$ and $j=1,\ldots,n_v$. 
According to Definition~\ref{def:maliciousness}, such modification is considered
to be non-malicious. 
The modification is conducted within $\mathcal{B}\left(\widetilde{\mathsf{v}}\right)$. As $n$ and $n_v$ approach to infinity,
we have $\left|U-V\right|\leq\frac{\beta_{v}-\alpha_{v}}{n_{v}-2}$ in probability. Hence, for sufficient large $n$ and $n_v$,
the modification error is controllable and negligible.
 
In another manipulating scenario, the relay only modifies negligible fraction of symbols, such that $\triangle F_{\widetilde{V}^{n}\left|\widetilde{U}^{n}\right.}^{\left(n'\right)}\left(\widetilde{\mathsf{v}}_{j}\left|\widetilde{\mathsf{u}}_{i}\right.\right)$ is close to $\left[W_{0}^{(n')}\right]_{i,j}$ for $i=1,\ldots,n'$ and $j=1,\ldots,n_v$. 
 This relaxation
has essentially no effect on the information rate from the source to
the destination across the relay. We allow these two kinds of manipulation for mathematical convenience.

Recall from Section II.B that in the instants $n+1, n+2, \ldots, 2n$, the relay transmits $V^n$ to the destination. Conrrespondingly, the destination 
observes $Y^n$ from the relay. On the other hand, in the instants $1, 2, \ldots, n$, the destination also observes $X^n$ from the source.
The destination is free to use $Y^n$ and $X^n$ to detect the existence of malicious relay.
Since $X^n$ is delivered via the direct channel, it is guaranteed not to be attacked and is statistically
correlated to $Y^n$ when the relay is non-malicious. Thus, $X^n$ could work as secured SI to check whether $Y^n$ is attacked by the relay.
In particular, the destination uses $X^n$ and $Y^n$ for getting functions
\begin{equation}
F_{Y^{n}\left|\widetilde{X}^{n}\right.}^{n}\left(t\left|\widetilde{\msf{x}}\right.\right)=\begin{cases}
\begin{array}{cc}
\frac{\sum_{i=1}^{n}1_{i}\left(Y_{i}<t\right)1_{i}\left(\widetilde{X}_{i}=\widetilde{\msf{x}}\right)}{N\left(\widetilde{\msf{x}}\left|\widetilde{X}^{n}\right.\right)}, &  N\left(\widetilde{\msf{x}}\left|\widetilde{X}^{n}\right.\right)\neq0,\\
0, & otherwise
\end{array}\end{cases}.
\end{equation}Then, in lemma \ref{Alem2}, we will show that as $n\rightarrow\infty$, $n'\rightarrow\infty$, the convergence 
\begin{equation}\label{CI}
F_{Y^{n}\left|\widetilde{X}^{n}\right.}^{n}\left(t\left|\widetilde{\msf{x}}\right.\right)\rightarrow\sum_{k=1}^{n_{v}}\sum_{j=1}^{n'}P_{\widetilde{U}\left|\widetilde{X}\right.}\left(\widetilde{\mathsf{u}}_{j}\left|\widetilde{\msf{x}}\right.\right)\triangle F_{\widetilde{V}^{n}\left|\widetilde{U}^{n}\right.}^{(n')}\left(\widetilde{\mathsf{v}}_{k}\left|\widetilde{\mathsf{u}}_{j}\right.\right)F_{Y\left|V\right.}\left(t\left|\widetilde{\mathsf{v}}_{k}\right.\right)
\end{equation}is established in probability for arbitrary value of $t$ and arbitrary distribution of $\left\{ \widetilde{V}^{n},\widetilde{U}^{n}\right\} $. The convergence characterized by (\ref{CI}) allows the destination to  
determine whether {\small{$\sum_{j=1}^{{n_{v}}}\sum_{i=1}^{n'}\left|\triangle F_{\widetilde{V}^{n}\left|\widetilde{U}^{n}\right.}^{\left(n'\right)}\left(\widetilde{\mathsf{v}}_{j}\left|\widetilde{\mathsf{u}}_{i}\right.\right)-\left[W_{0}^{(n')}\right]_{i,j}\right|$}} is far away from $0$ or not by using its observation of $F_{Y^{n}\left|\widetilde{X}^{n}\right.}^{n}\left(t\left|\widetilde{\msf{x}}\right.\right)$. In other words, the attack detection could be implemented based on physical-layer observation, i.e., $X^n$ and $Y^n$. 
We will propose the detection method only using $X^n$ and $Y^n$, and prove that as $n\rightarrow\infty$, the error probability of the proposed detection method approaches to $0$.

\section{Main Results}
Notice that the convergence (\ref{CI}) could be rewrote as
\begin{equation}
F_{Y^{n}\left|\widetilde{X}^{n}\right.}^{n}\left(t\left|\widetilde{\msf{x}}\right.\right)\rightarrow\sum_{k=1}^{n_{v}}\sum_{j=1}^{n'}\frac{P_{\widetilde{U}\left|\widetilde{X}\right.}\left(\widetilde{\mathsf{u}}_{j}\left|\widetilde{\msf{x}}\right.\right)}{\triangle_{u}}\frac{\triangle F_{\widetilde{V}^{n}\left|\widetilde{U}^{n}\right.}^{(n')}\left(\widetilde{\mathsf{v}}_{k}\left|\widetilde{\mathsf{u}}_{j}\right.\right)}{\triangle_{v}}F_{Y\left|V\right.}\left(t\left|\widetilde{\mathsf{v}}_{k}\right.\right)\triangle_{v}\triangle_{u}
\end{equation}where we set that $n_v$ approaches to infinity as $n'\rightarrow\infty$, hence $\lim_{n'\rightarrow\infty}\triangle_{u}=0$, and $\lim_{n'\rightarrow\infty}\triangle_{v}=0$. On the other hand, we will also prove that the convergence (\ref{CI}) is established for arbitrary sufficient large $n'$. Then, according to the definition of 
 integral, for sufficient large $n'$, we could get that 
 \begin{equation}\label{r1}
F_{Y^{n}\left|\widetilde{X}^{n}\right.}^{n}\left(t\left|\widetilde{\msf{x}}\right.\right)\rightarrow\int_{\alpha_{1}}^{\beta_{1}}\int_{\alpha_{2}}^{\beta_{2}}f_{U\left|X\right.}\left(u\left|x\right.\right)f_{\widetilde{V}^{n}\left|\widetilde{U}^{n}\right.}^{(n')}\left(v\left|u\right.\right)F_{Y\left|V\right.}\left(y\left|v\right.\right)dudv \end{equation}where $x\in\mathcal{B}\left(\widetilde{\msf{x}}\right)$, $f_{\widetilde{V}^{n}\left|\widetilde{U}^{n}\right.}^{(n')}\left(v\left|u\right.\right)=\frac{\triangle F_{\widetilde{V}^{n}\left|\widetilde{U}^{n}\right.}^{(n')}\left(v\left|u\right.\right)}{\triangle_{v}}$. Furthermore, if $f_{\widetilde{V}^{n}\left|\widetilde{U}^{n}\right.}^{(n')}\left(v\left|u\right.\right)$
converges to $\lim_{n'\rightarrow\infty}f_{\widetilde{V}^{n}\left|\widetilde{U}^{n}\right.}^{(n')}\left(v\left|u\right.\right)$ as $n'\rightarrow\infty$, we could also obtain that 
 \begin{equation}\label{r2}
 F_{Y^{n}\left|\widetilde{X}^{n}\right.}^{n}\left(t\left|\widetilde{\msf{x}}\right.\right)\rightarrow\int_{-\infty}^{+\infty}\int_{-\infty}^{+\infty}f_{U\left|X\right.}\left(u\left|x\right.\right)\lim_{n'\rightarrow\infty}f_{\widetilde{V}^{n}\left|\widetilde{U}^{n}\right.}^{(n')}\left(v\left|u\right.\right)F_{Y\left|V\right.}\left(y\left|v\right.\right)dudv.
 \end{equation}Either (\ref{r1}) or (\ref{r2}) indicates the observation channel $\left(f_{U\left|X\right.}\left(u\left|{x}\right.\right),\; F_{Y\left|V\right.}\left(t\left|{v}\right.\right)\right)$ plays a key role to the attack detection using observation of $F_{Y^{n}\left|\widetilde{X}^{n}\right.}^{n}\left(t\left|\widetilde{\msf{x}}\right.\right)$. 
 Intuitively, if observation channel $\left(f_{U\left|X\right.}\left(u\left|{x}\right.\right),\; F_{Y\left|V\right.}\left(t\left|{v}\right.\right)\right)$ makes $\lim_{n'\rightarrow\infty}f_{\widetilde{V}^{n}\left|\widetilde{U}^{n}\right.}^{(n')}\left(v\left|u\right.\right)=\Phi\left(v-u\right)$ is the single pdf solution to $$\int_{-\infty}^{+\infty}\int_{-\infty}^{+\infty}f_{U\left|X\right.}\left(u\left|x\right.\right)\lim_{n'\rightarrow\infty}f_{\widetilde{V}^{n}\left|\widetilde{U}^{n}\right.}^{(n')}\left(v\left|u\right.\right)F_{Y\left|V\right.}\left(y\left|v\right.\right)dudv=\int_{-\infty}^{+\infty}f_{U\left|X\right.}\left(u\left|x\right.\right)F_{Y\left|V\right.}\left(y\left|u\right.\right)du,$$then the destination 
 is able to determine whether $\lim_{n'\rightarrow\infty}f_{\widetilde{V}^{n}\left|\widetilde{U}^{n}\right.}^{(n')}\left(v\left|u\right.\right)$ is far away from $\Phi\left(v-u\right)$ or not by comparing  $\int_{-\infty}^{+\infty}f_{U\left|X\right.}\left(u\left|x\right.\right)F_{Y\left|V\right.}\left(y\left|u\right.\right)du$ with $F_{Y^{n}\left|\widetilde{X}^{n}\right.}^{n}\left(t\left|\widetilde{\msf{x}}\right.\right)$. Obviously, the distance between $\lim_{n'\rightarrow\infty}f_{\widetilde{V}^{n}\left|\widetilde{U}^{n}\right.}^{(n')}\left(v\left|u\right.\right)$ and $\Phi\left(v-u\right)$ indicates maliciousness of the relay. 
As a beneficial result, the attacks 
 are detectable upon $F_{Y^{n}\left|\widetilde{X}^{n}\right.}^{n}\left(t\left|\widetilde{\msf{x}}\right.\right)$. This intuition leads to the following dichotomy on all AWGN observation channels.

  \begin{define}
  \textbf{Non-manipulable AWGN Relay Channel} The observation channel $\left(f_{U\left|X\right.}\left(u\left|{x}\right.\right),\; F_{Y\left|V\right.}\left(y\left|{v}\right.\right)\right)$ is non-manipulable, if there exists function $\Psi\left(v\left|u\right.\right)$ that satisfies the following three conditions
  \begin{enumerate}
  \item $\Psi\left(v\left|u\right.\right)$ is a conditional pdf.
\item $\ensuremath{\Psi\left(v\left|u\right.\right)}\neq\Phi\left(v-u\right)$.
 \item {\small{$\int_{-\infty}^{+\infty}\int_{-\infty}^{+\infty}f_{U\left|{X}\right.}\left(u\left|x\right.\right)\Psi\left(v\left|u\right.\right)F_{Y\left|V\right.}\left(y\left|v\right.\right)dudv=\int_{-\infty}^{+\infty}f_{U\left|{X}\right.}\left(u\left|x\right.\right)F_{Y\left|V\right.}\left(y\left|u\right.\right)du$}}.
  \end{enumerate}
\end{define}Otherwise, the observation channel $\left(f_{U\left|X\right.}\left(u\left|{x}\right.\right),\; F_{Y\left|V\right.}\left(y\left|{v}\right.\right)\right)$ is non-manipulable.
The following theorem will show the non-manipulability of relay network is equivalent
to the detectability of maliciousness.
\begin{theorem}\textbf{(Maliciousness detectability )} \label{thm:main2}
 The observation channel
 $\left(f_{U\left|X\right.}\left(u\left|{x}\right.\right),\; F_{Y\left|V\right.}\left(y\left|{v}\right.\right)\right)$ is
  non-manipulable is a necessary and sufficient condition for the existence of a
  sequence of decision statistics $\left\{ D^{n} \right\}$
  simultaneously having the following two properties: \\
  Fix any sufficiently small $\delta >0$, $\epsilon>0$, there has sufficiently large $n'$, 
\begin{enumerate}
\item {\small{$\lim_{n\rightarrow\infty}\Pr\left(D^{n}>\varepsilon(n',\delta)~\Big|\sum_{j=1}^{{n_{v}}}\sum_{i=1}^{n'}\left|\triangle F_{\widetilde{V}^{n}\left|\widetilde{U}^{n}\right.}^{\left(n'\right)}\left(\widetilde{\mathsf{v}}_{j}\left|\widetilde{\mathsf{u}}_{i}\right.\right)-\left[W_{0}^{(n')}\right]_{i,j}\right|>\delta\right)\geq1-\epsilon$}} whenever {\small{$\Pr \Big(\sum_{j=1}^{{n_{v}}}\sum_{i=1}^{n'}\Big|\triangle F_{\widetilde{V}^{n}\left|\widetilde{U}^{n}\right.}^{\left(n'\right)}\left(\widetilde{\mathsf{v}}_{j}\left|\widetilde{\mathsf{u}}_{i}\right.\right)-\left[W_{0}^{(n')}\right]_{i,j}\Big|> \delta \Big) > 0$}}, where $\varepsilon(n',\delta)$ is strictly positive and can be arbitrary small. 
\item {\small{$\lim_{n\rightarrow\infty}\Pr\Big( D^{n}> \mu' (n',\delta) ~\Big| 
  \sum_{j=1}^{{n_{v}}}\sum_{i=1}^{n'}\left|\triangle F_{\widetilde{V}^{n}\left|\widetilde{U}^{n}\right.}^{\left(n'\right)}\left(\widetilde{\mathsf{v}}_{j}\left|\widetilde{\mathsf{u}}_{i}\right.\right)-\left[W_{0}^{(n')}\right]_{i,j}\right|\leq \delta \Big) \leq
  \epsilon$}} whenever {\small{$\Pr \Big( \sum_{j=1}^{{n_{v}}}\sum_{i=1}^{n'}\Big|\triangle F_{\widetilde{V}^{n}\left|\widetilde{U}^{n}\right.}^{\left(n'\right)}\left(\widetilde{\mathsf{v}}_{j}\left|\widetilde{\mathsf{u}}_{i}\right.\right)-\left[W_{0}^{(n')}\right]_{i,j}\Big|<\delta \Big) > 0$,
  where $\mu' (n',\delta) \rightarrow 0$ as $n'\rightarrow \infty$, $\delta \rightarrow 0$}}.
\end{enumerate}
\end{theorem}
The properties 1) and 2) of 
Theorem 1 together imply that we have $D^{n}\rightarrow0$ in probability if the
relay network is safe. To elaborate a little further, 
for a relay network in which the source's observation
channel is non-manipulable, it is theoretically feasible to check
whether the relay have conducted Byzantine attacks. Conversely, if the observation channel is
manipulable, the security of the relay network against Byzantine
attacks may not be guaranteed by only checking the source's
observations.

\emph{Remark 1:} Theorem \ref{thm:main2} gives the detectability of continuous attacks over continuous channels. 
It extends the the detectability of discrete attacks given by \cite{GravesINFOCOM12} into continuous form.
This extension is not trivial based on twofold work of this paper. Firstly, we prove the convergence of $F_{Y^{n}\left|\widetilde{X}^{n}\right.}^{n}\left(t\left|\widetilde{\msf{x}}\right.\right)$ (\ref{c1}) is established for arbitrary sufficient large $n'$ rather than fixed $n'$, where the involved variables $\widetilde{X}$, $\widetilde{U}$, and $\widetilde{V}$ do not follow markov constraint. For completing such proof work,  it is not available to straightforwardly use the methods of  \cite{GravesINFOCOM12}, which processes the discrete variables following markov constraint and having fixed dimensions of their alphabets. 
Secondly, the checking problem for non-manipulability of continuous channel is to check number of solutions to an integral equation. 
The method of \cite{GravesINFOCOM12} for checking non-manipulability of discrete channel is to check number of solutions to an matrix equation. The checking method  given by \cite{GravesINFOCOM12}
cannot be used for integral equation. 

In this paper,  we will show almost all the continuous relay networks are non-manipulable as follows.

\begin{proposition}\label{pro1}\textbf{(Non-manipulability of Networks With Nonzero-Coefficients)}
If and only if $h_1\neq0$, $h_2\neq0$ and $h_3\neq0$, the observation channel $\left(f_{U\left|X\right.}\left(u\left|{x}\right.\right),\; F_{Y\left|V\right.}\left(y\left|{v}\right.\right)\right)$ is non-manipulable. 
\end{proposition}Obviously, the Proposition 1 indicates the non-manipulability is very common for the general AWGN relay networks. 
Jointly considering Theorem 1 and Proposition 1, we attain that in almost all the  
AWGN relay networks, arbitrary attacks are detectable by using physical-layer observation. Furthermore, according to Proposition \ref{pro1}, 
even the coefficient of direct channel, i.e., $h_3$, is very little yet nonzero, the observation channel is still non-manipulable. 
It indicates even the direct channel suffers deep fading,  the signal observed from the direct channel still can be used for attack detection and achieving asymptotically errorless performance. According to Proposition 1, we have the following corollary.

\begin{cor}\textbf{(Manipulability of AWGN Relay Networks Without Direct Channel)}\label{cor1}
If the direct channel does not exist, i.e., $h_3=0$, the observation channel $\left(f_{U\left|X\right.}\left(u\left|{x}\right.\right),\; F_{Y\left|V\right.}\left(y\left|{v}\right.\right)\right)$ is manipulable.
\end{cor}The proof of Corollary \ref{cor1} is completed by setting $\Psi\left(v\left|u\right.\right)=f_{U}\left(v\right)$, then we have {\small{$\int_{-\infty}^{+\infty}\int_{-\infty}^{+\infty}f_{U\left|{X}\right.}\left(u\left|x\right.\right)\Psi\left(v\left|u\right.\right)F_{Y\left|V\right.}\left(y\left|v\right.\right)dudv=\int_{-\infty}^{+\infty}f_{U\left|{X}\right.}\left(u\left|x\right.\right)F_{Y\left|V\right.}\left(y\left|u\right.\right)du$}}. According to Definition 2, the networks without direct channel is manipulable. From Theorem 1, 
the manipulability indicates in such networks, some attacks are not detectable upon observation of the destination.
Recall that the direct channel is specified by $X =h_3S+N_d$. If $h_3=0$, then $X =N_d$ which is always
statistical independent on the observation from relay, $Y^n$. Hence, $X^n$ cannot work as SI to check
$Y^n$ due to its loss of statistic dependence on $Y^n$. It also reflects the vulnerability of lacking SI.
Proposition 1 and Corollary \ref{cor1} jointly illustrate the necessity of utilizing SI. For
detectability of arbitrary attacks, SI is allowed to be little, but cannot be vacancy.

\section{Numerical Examples}
We give three numerical examples in
this section to illustrate the detectability
results of Theorem 1, Proposition 1, and Corollary 1.
In all the examples, we assume that the source
alphabet is binary. We employ decision statistic
{\small{$$D^{n}=\frac{1}{{n_{x}-2}}\frac{1}{{n_{y}-2}}\sum_{k=1}^{{n_{x}-1}}\sum_{m=1}^{{n_{y}-1}}\left|F_{Y^{n}\left|\widetilde{X}^{n}\right.}^{n}\left(t_{m}\left|\widetilde{\mathsf{x}}_{k}\right.\right)-\int_{-\infty}^{+\infty}f_{U\left|\widetilde{X}\right.}\left(u\left|\widetilde{\mathsf{x}}_{k}\right.\right)F_{Y\left|V\right.}\left(t_{m}\left|u\right.\right)du\right|$$}}
for Byzantine attack detection in the examples.

We first consider the setup that $h_1=h_2=h_3=1$. We term the direct channel
as strong in the sense that its coefficient is as large as relaying channel.
According to Proposition 1, the considered channel is non-manipulable.
Then, two different attacks are considered. In the first attack, 
referred to as Attack 1, the relay conducts a non-i.i.d. attack 
by mapping its $i$th input symbol $U_i$ to $V_i$ according 
to $V_i=U_i-1$ when $i$ is odd, and according to $V_i=2U_i-1$
when $i$ is even. In the second attack considered, referred to
as Attack 2, relay makes $V$ follow the distribution
identical to $U$. Meanwhile, $U$ and $V$ are independent 
with each other. Since the observation channel is non-manipulable,
these two attacks are detectable.

We conduct computer simulation to illustrate the detectability of 
both attacks using $D^n$. The empirical CDFs of $D^n$ obtained
from the simulation are plotted in Figs. 4 and 5 for Attack 1 and Attack 2,
respectively. From these two figures, it is observed that there are clear
separations between the empirical CDFs of $D^n$ for the non-malicious
case and the malicious case of Attack 1 and Attack 2 when $n=10^3$. 
This observation verifies sufficiency of non-manipulability for the detectability of
attacks promised by Theorem 1.
\begin{figure}
\begin{centering}
\includegraphics[scale=0.5]{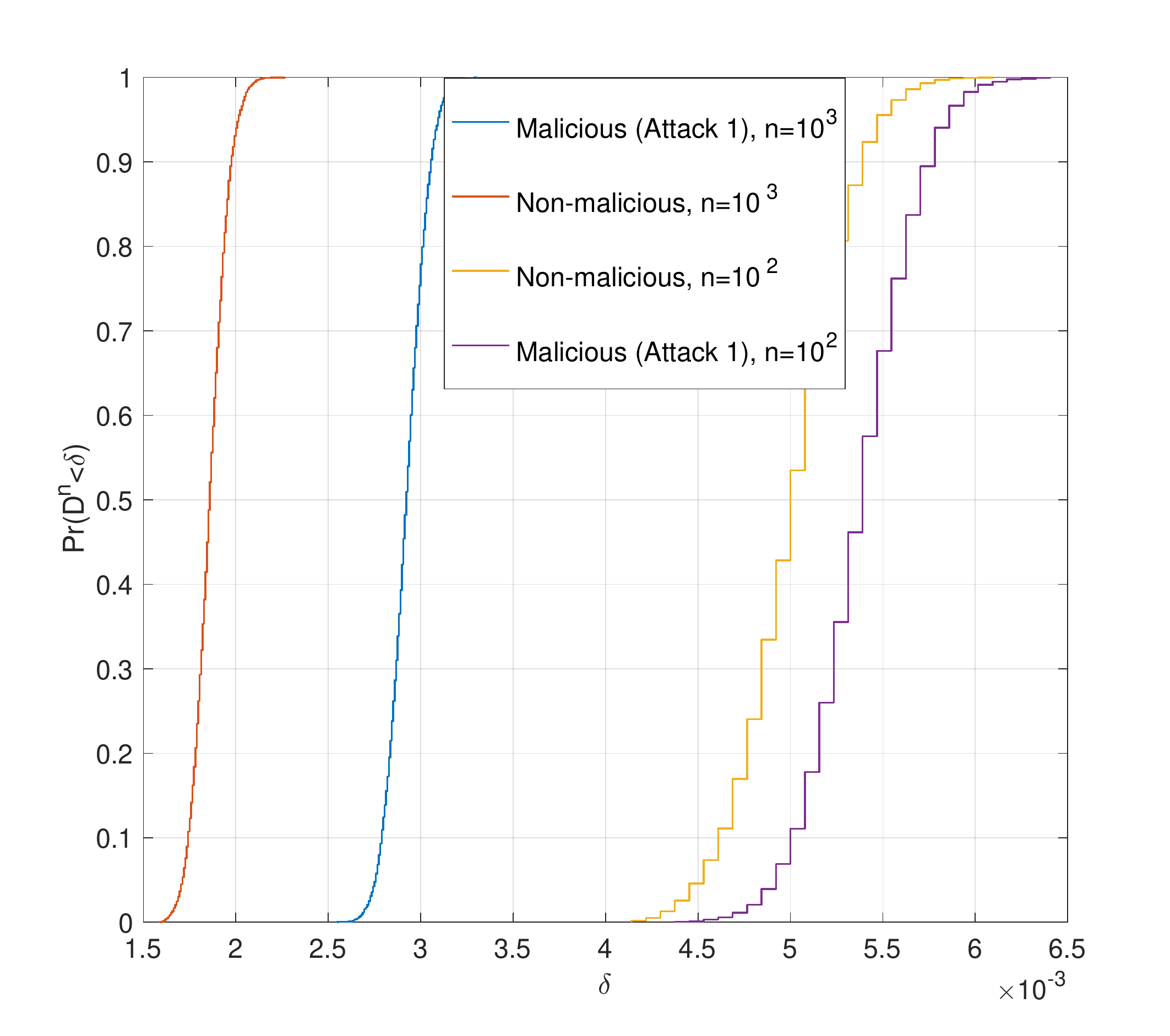}
\par\end{centering}

\caption{Empirical CDFs of $D^n$ for Attack 1 in non-manipulable observation
channel characterized by $h_1=h_2=h_3=1$. }

\end{figure}
\begin{figure}
\begin{centering}
\includegraphics[scale=0.5]{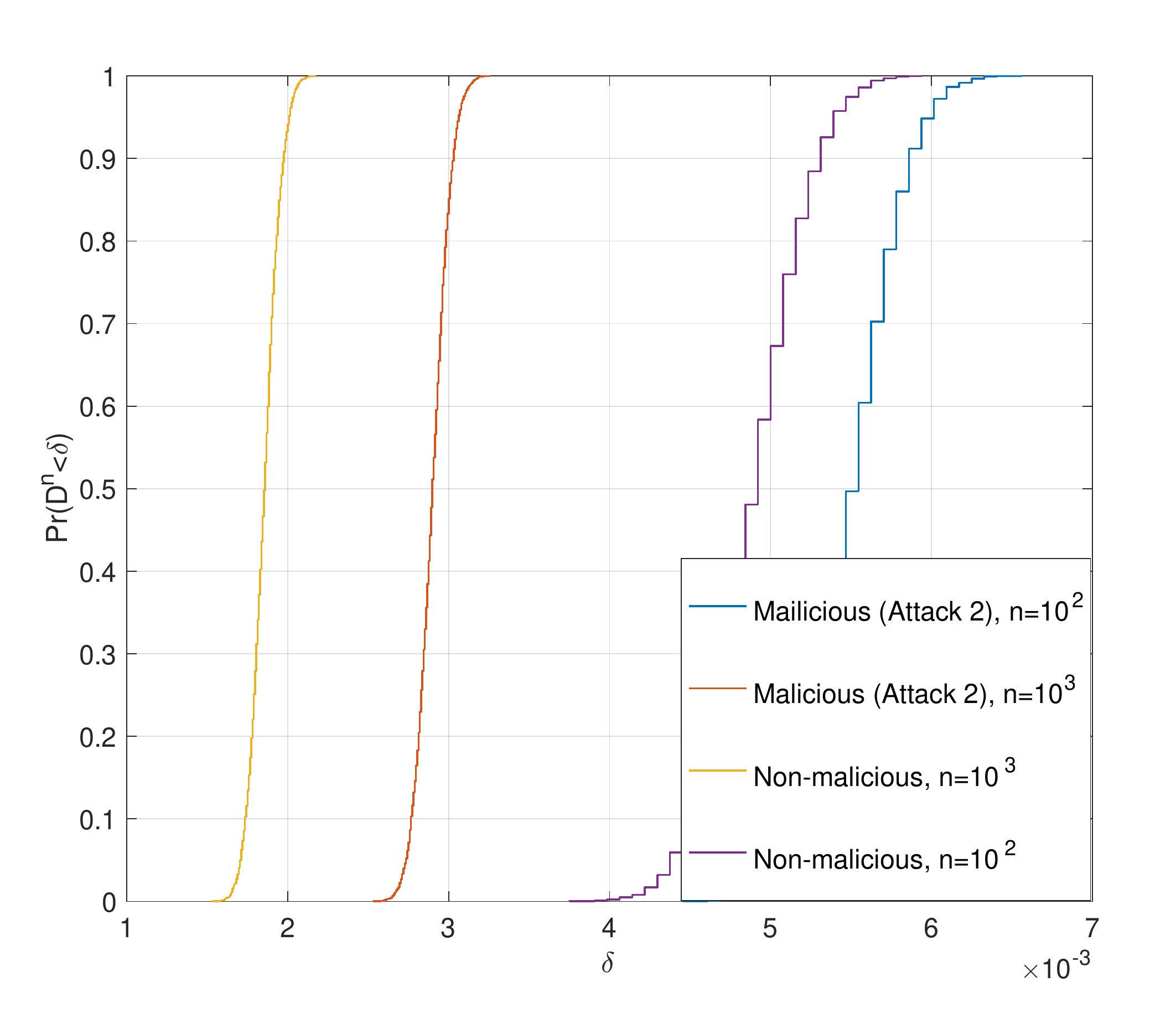}
\par\end{centering}

\caption{Empirical CDFs of $D^n$ for Attack 2 in non-manipulable observation
channel characterized by $h_1=h_2=h_3=1$.}

\end{figure}
Secondly, we consider the setup that $h_1=h_2=1$, $h_3=0.01$. We 
term the direct channel as weak in the sense that its coefficient
is much smaller than relaying channel. According to Proposition 1, 
this network with weak direct channel is still non-manipulable.
Based on Theorem 1, Attack 2 is detectable.
We consider Attack 2 in computer simulation. The empirical CDFs
of $D^n$ are plotted in Fig. 6. Similar to Figs. 4 and 5, it is also 
observed clear separations between the empirical CDF of $D^n$ for the non-malicious
case and the malicious case of Attack 2. Nevertheless, this separation
appears when $n=10^5$. This observation verifies sufficiency of non-manipulability for the detectability of
attacks promised by Theorem 1.
It also indicates the detection with weak direct channel
needs much more observations than it with strong direct channel.
\begin{figure}
\begin{centering}
\includegraphics[scale=0.5]{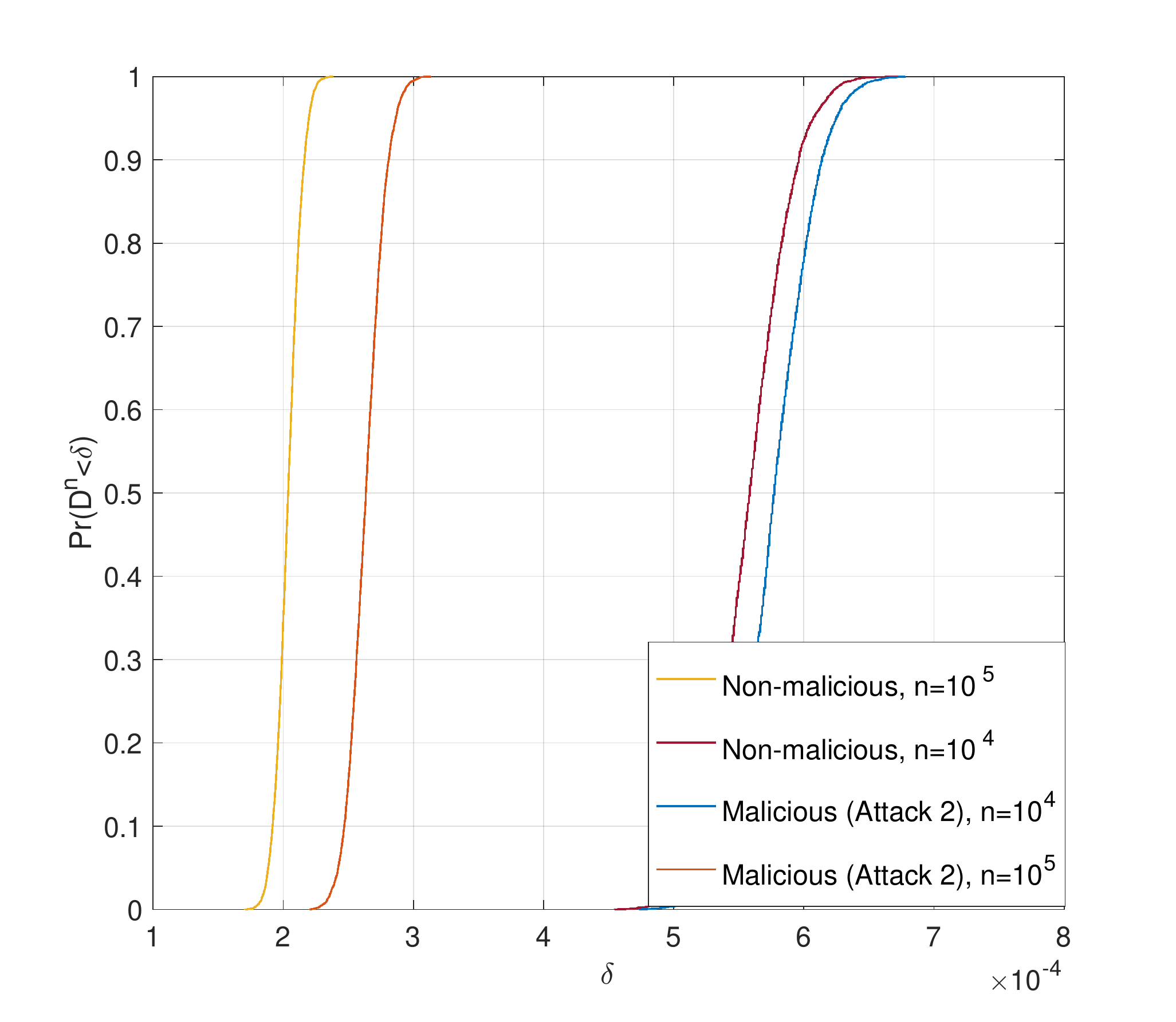}
\par\end{centering}

\caption{Empirical CDFs of $D^n$ for Attack 2 in non-manipulable observation
channel characterized by $h_1=h_2=1, h_3=0.01$.}

\end{figure}

Finally, we consider the setup that $h_1=h_2=1$,  $h_3=0$.
Under this setup, the direct channel is vacancy. According to
Proposition 1 and Corollary 1, this network is manipulable.
Based on Theorem 1, we know that there some attacks are not
attackable in this network. In order to verify this, we simulated 
Attack 2 in such network. The empirical CDFs of $D^n$ obtained
from the simulation is plotted in Fig. 7. It is seen from 
Fig. 7 that the CDFs are indistinguishable, regardless of the value of $n$.
As a result, from $D^n$, Attack 2 is not detectable. This observation 
verifies the necessity of non-manipulability for the detectability of
attacks promised by Theorem 1.

\begin{figure}
\begin{centering}
\includegraphics[scale=0.5]{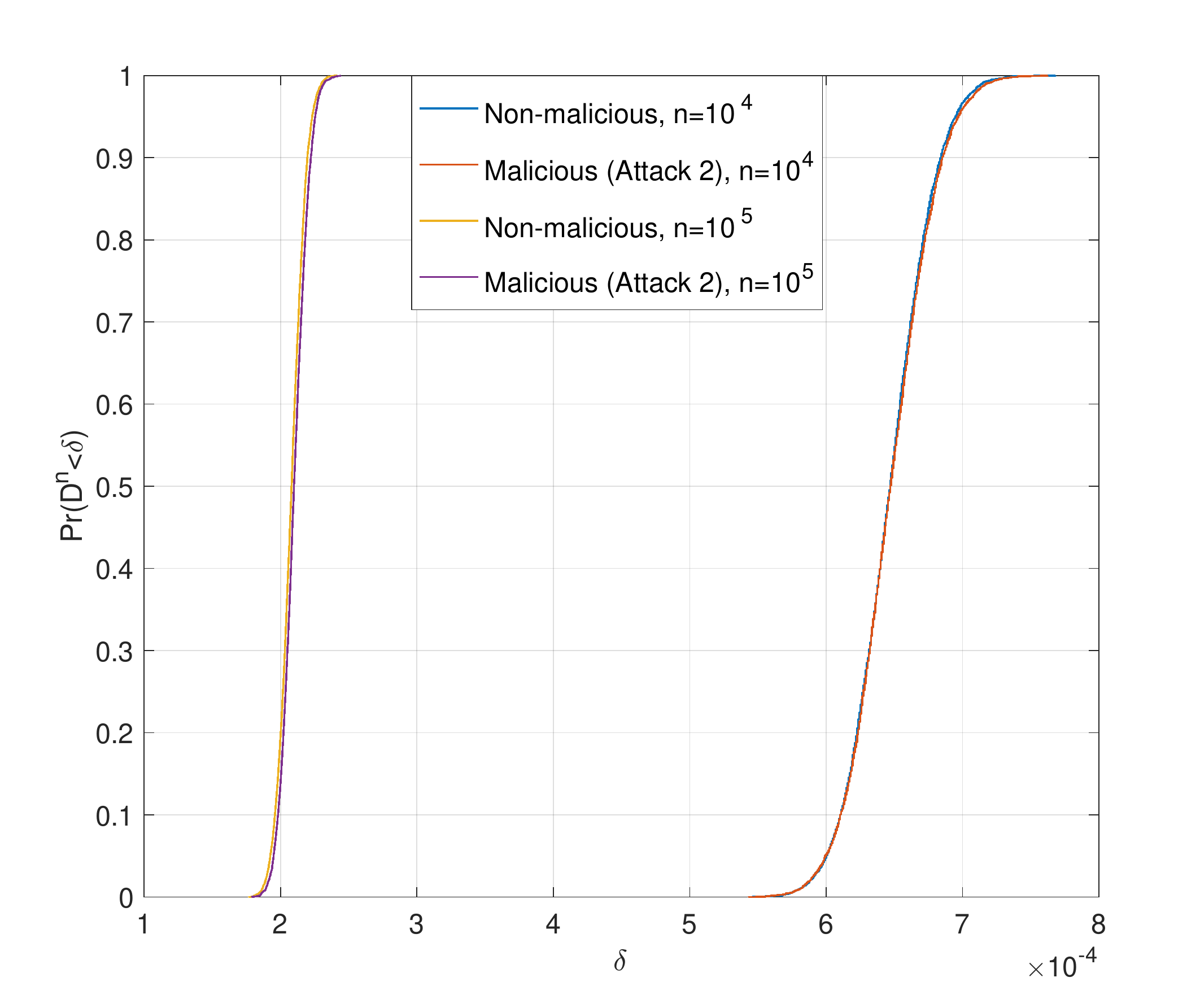}
\par\end{centering}

\caption{Empirical CDFs of $D^n$ for Attack 2 in manipulable observation
channel characterized by $h_1=h_2=1, h_3=0$.}

\end{figure}

\section{conclusions}
This paper considers attack detection problem in wireless relay system. We
prove arbitrary attacks are detectable only using physical-layer observations, as 
long as the system satisfies a non-manipulable condition. Then, we further prove
that if and only if the coefficients of all channels in the relay system are non-zero, the non-manipulability of the wireless system
is guaranteed. It indicates that using physical-layer observations, rather than any other secret-key or AMD code, 
wireless relay systems are generally able to detect arbitrary attacks.
\appendices{}
\section{Proof of Proposition 1}
\begin{lemma}\label{lemp1}
If $h_{1}\neq0$ and $h_{3}\neq0$, there does not exist any i.i.d attack making the statistical distribution of $U$ conditioned on $X$ is equivalent to the statistical distribution of $V$ conditioned on $X$.
\end{lemma}
\begin{IEEEproof}
Let us assume the manipulable wireless channel exists, which indicates there at least one i.i.d attack making the statistical distribution of $U$ conditioned on $X$ is equivalent to the statistical distribution of $V$ conditioned on $X$. Hence, we have $I\left(X;U\right)=I\left(X;V\right)$, where $I\left(\cdot;\cdot\right)$ denotes mutual information between the two input variables. On the other hand, $\left(X,U,V\right)$ forms a Markov chain as $X\rightarrow U\rightarrow V$. From Data-Processing Inequality,  $I\left(X;U\right)=I\left(X;V\right)$ implies the Markov chain $X\rightarrow V\rightarrow U$ is also established. Then, let us use $\mathsf{u}$ and $\mathsf{v}$ to respectively denote arbitrary possible values of $U$ and $V$, we have
\begin{equation}
\label{markov_eq}
\Pr\left(U=\mathsf{u}\left|V=\mathsf{v},X=a\right.\right)=\Pr\left(U=\mathsf{u}\left|V=\mathsf{v},X=b\right.\right)
 \end{equation}where $a$ and $b$ denote arbitrary value of $X$, $a\neq b$.
Furthermore, $\Pr\left(U=\mathsf{u}\left|V=\mathsf{v},X=a\right.\right)$ can be extended as 
{\small{\begin{align}\label{extend_1}
\Pr\left(U=\mathsf{u}\left|V=\mathsf{v},X=a\right.\right)&=\frac{\Pr\left(U=\mathsf{u},V=\mathsf{v},X=a\right)}{\Pr\left(V=\mathsf{v},X=a\right)}\\&\nonumber
=\frac{\Pr\left(X=a\right)\Pr\left(U=\mathsf{u}\left|X=a\right.\right)\Pr\left(V=\mathsf{v}\left|U=\mathsf{u}\right.\right)}{\Pr\left(V=\mathsf{v},X=a\right)}\\&\nonumber
=\frac{f_{U\left|X\right.}\left(\mathsf{u}\left|a\right.\right)\Pr\left(V=\mathsf{v}\left|U=\mathsf{u}\right.\right)}{f_{U\left|X\right.}\left(\mathsf{v}\left|a\right.\right)}\end{align}}}where the second equation follows the fact that $X\rightarrow U\rightarrow V$ also forms a Markov chain. Similarly, $\Pr\left(U=\mathsf{u}\left|V=\mathsf{v},X=b\right.\right)$ can be extended as 
{\small{\begin{align}\label{extend_2}
\Pr\left(U=\mathsf{u}\left|V=\mathsf{v},X=b\right.\right)&=\frac{\Pr\left(U=\mathsf{u},V=\mathsf{v},X=b\right)}{\Pr\left(V=\mathsf{v},{X}=b\right)}\\&\nonumber
=\frac{\Pr\left(X=b\right)\Pr\left(U=\mathsf{u}\left|X=b\right.\right)\Pr\left(V=\mathsf{v}\left|U=\mathsf{u}\right.\right)}{\Pr\left(V=\mathsf{v},X=b\right)}\\&\nonumber
=\frac{f_{U\left|X\right.}\left(\mathsf{u}\left|b\right.\right)\Pr\left(V=\mathsf{v}\left|U=\mathsf{u}\right.\right)}{f_{U\left|X\right.}\left(\mathsf{v}\left|b\right.\right)},\end{align}}}where the second equation is again relying on the fact that
$X\rightarrow U\rightarrow V$ forms a Markov chain. Substituting (\ref{extend_1}) and (\ref{extend_2}) into (\ref{markov_eq}), we have 
{\small{\begin{equation}\label{reshape_1}
\frac{f_{U\left|X\right.}\left(\mathsf{u}\left|a\right.\right)\Pr\left(V=\mathsf{v}\left|U=\mathsf{u}\right.\right)}{f_{U\left|X\right.}\left(\mathsf{v}\left|a\right.\right)}=\frac{f_{U\left|X\right.}\left(\mathsf{u}\left|b\right.\right)\Pr\left(V=\mathsf{v}\left|U=\mathsf{u}\right.\right)}{f_{U\left|X\right.}\left(\mathsf{v}\left|b\right.\right)}.
\end{equation}}}Notice that (\ref{reshape_1}) should be always established for arbitrary value of $a$ and $b$. Thus, we have 
\begin{equation}\label{eq_infinity}
\lim_{a\rightarrow\infty}\frac{f_{U\left|X\right.}\left(\mathsf{u}\left|a\right.\right)\Pr\left(V=\mathsf{v}\left|U=\mathsf{u}\right.\right)}{f_{U\left|X\right.}\left(\mathsf{v}\left|a\right.\right)}=\lim_{b\rightarrow-\infty}\frac{f_{U\left|X\right.}\left(\mathsf{u}\left|b\right.\right)\Pr\left(V=\mathsf{v}\left|U=\mathsf{u}\right.\right)}{f_{U\left|X\right.}\left(\mathsf{v}\left|b\right.\right)}.
\end{equation}which could be further reshaped as 
\begin{equation}\label{eq_infinity}
\Pr\left(V=\mathsf{v}\left|U=\mathsf{u}\right.\right)\lim_{a\rightarrow\infty}\frac{f_{U\left|X\right.}\left(\mathsf{u}\left|a\right.\right)}{f_{U\left|X\right.}\left(\mathsf{v}\left|a\right.\right)}=\Pr\left(V=\mathsf{v}\left|U=\mathsf{u}\right.\right)\lim_{b\rightarrow-\infty}\frac{f_{U\left|X\right.}\left(\mathsf{u}\left|b\right.\right)}{f_{U\left|X\right.}\left(\mathsf{v}\left|b\right.\right)}.
\end{equation}
Without loss of generality, we assume $h_{3}>0$. Then, according to the expression of $f_{U\left|X\right.}\left(u\left|x\right.\right)$, (\ref{eq_infinity}) becomes
{\small{\begin{equation}\label{reshape_2}
\frac{\exp\left(-\frac{\left(\mathsf{u}+h_{1}\right)^{2}}{2}\right)\Pr\left(V=\mathsf{v}\left|U=\mathsf{u}\right.\right)}{\exp\left(-\frac{\left(\mathsf{v}+h_{1}\right)^{2}}{2}\right)}=\frac{\exp\left(-\frac{\left(\mathsf{u}-h_{1}\right)^{2}}{2}\right)\Pr\left(V=\mathsf{v}\left|U=\mathsf{u}\right.\right)}{\exp\left(-\frac{\left(\mathsf{v}-h_{1}\right)^{2}}{2}\right)},\end{equation}}}which indicates 
{\small{\begin{equation}\label{reshape_3}
\frac{\Pr\left(V=\mathsf{v}\left|U=\mathsf{u}\right.\right)}{\exp\left(2h_1\mathsf{u}\right)}=\frac{\Pr\left(V=\mathsf{v}\left|U=\mathsf{u}\right.\right)}{\exp\left(2h_1\mathsf{v}\right)}.
\end{equation}}}Obviously, as $h_{1}\neq0$, if and only if $\Pr\left(V=\mathsf{v}\left|U=\mathsf{u}\right.\right)=0$ for $\mathsf{u}\neq\mathsf{v}$, (\ref{reshape_3}) would be true. Then, for any attacks that $\Pr\left(V=\mathsf{v}\left|U=\mathsf{u}\right.\right)\neq 0$ for $\mathsf{u}\neq\mathsf{v}$, (\ref{reshape_3}) would not be true, which contradicts to the assumption that 
there at least one i.i.d attack making the statistical distribution of $U$ conditioned on $X_1$ is equivalent to the statistical distribution of $V$ conditioned on $X_1$. Hence, the lemma is proved.
\end{IEEEproof}
\begin{lemma}\label{lemp2}
For random variables $Z_1$, $Z_2$, $Z_3$ $Z_4$, $Z_5$, where
$Z_3=Z_1+Z_2$, $Z_5=Z_4+Z_2$, $Z_1$ and $Z_4$ are both stochastic independent 
with $Z_2$, 
if pdf $f_{Z_{3}\left|\widetilde{X}\right.}\left({z}_{3}\left|\widetilde{\msf{x}}\right.\right)=f_{Z_{5}\left|\widetilde{X}\right.}\left({z}_{5}\left|\widetilde{\msf{x}}\right.\right)$, then there must have $f_{Z_{1}\left|\widetilde{X}\right.}\left({z}_{1}\left|\widetilde{\msf{x}}\right.\right)=f_{Z_{4}\left|\widetilde{X}\right.}\left({z}_{4}\left|\widetilde{\msf{x}}\right.\right)$.
\end{lemma}
\begin{IEEEproof}
According to the fact that $Z_3=Z_1+Z_2$,  where $Z_1$ and $Z_2$ are stochastic independent 
with each other, then the characteristic function of $Z_3$ conditioned on $X_1=\widetilde{\msf{x}}$ is expressed by 
{\small{\begin{equation}\label{c1}
\varphi_{Z_{3}\left|\widetilde{X}\right.}\left(t\left|\widetilde{\msf{x}}\right.\right)=\varphi_{Z_{1}\left|\widetilde{X}\right.}\left(t\left|\widetilde{\msf{x}}\right.\right)\varphi_{Z_{2}\left|\widetilde{X}\right.}\left(t\left|\widetilde{\msf{x}}\right.\right),
\end{equation}}}where $\varphi_{Z_{3}\left|\widetilde{X}\right.}\left(t\left|\widetilde{\msf{x}}\right.\right)$, $\varphi_{Z_{1}\left|\widetilde{X}\right.}\left(t\left|\widetilde{\msf{x}}\right.\right)$ and $\varphi_{Z_{2}\left|\widetilde{X}\right.}\left(t\left|\widetilde{\msf{x}}\right.\right)$ denote the characteristic functions of $Z_3$, $Z_2$ and $Z_1$ conditioned on $X_1=\widetilde{\msf{x}}$, respectively.
Similarly, according to the fact that $Z_5=Z_4+Z_2$,  where $Z_4$ and $Z_2$ are stochastic independent 
with each other, then the characteristic function of $Z_5$ conditioned on $X_1=\widetilde{\msf{x}}$ is expressed by
{\small{\begin{equation}\label{c2}
\varphi_{Z_{5}\left|\widetilde{X}\right.}\left(t\left|\widetilde{\msf{x}}\right.\right)=\varphi_{Z_{4}\left|\widetilde{X}\right.}\left(t\left|\widetilde{\msf{x}}\right.\right)\varphi_{Z_{2}\left|\widetilde{X}\right.}\left(t\left|\widetilde{\msf{x}}\right.\right),
\end{equation}}}where $\varphi_{Z_{5}\left|\widetilde{X}\right.}\left(t\left|\widetilde{\msf{x}}\right.\right)$ and $\varphi_{Z_{4}\left|\widetilde{X}\right.}\left(t\left|\widetilde{\msf{x}}\right.\right)$ denote the characteristic functions of $Z_4$ and $Z_2$ conditioned on $X_1=\widetilde{\msf{x}}$, respectively.
Since {\small{$f_{Z_{3}\left|\widetilde{X}\right.}\left(\mathsf{z}_{3}\left|\widetilde{\msf{x}}\right.\right)=f_{Z_{5}\left|\widetilde{X}\right.}\left(\mathsf{z}_{5}\left|\widetilde{\msf{x}}\right.\right)$}}, we have 
{\small{\begin{equation}\label{c3}
\varphi_{Z_{3}\left|\widetilde{X}\right.}\left(t\left|\widetilde{\msf{x}}\right.\right)=\varphi_{Z_{5}\left|\widetilde{X}\right.}\left(t\left|\widetilde{\msf{x}}\right.\right)
\end{equation}}}Substituting (\ref{c1}) and (\ref{c2}) into (\ref{c1}), we get
{\small{\begin{equation}\label{c4}
\varphi_{Z_{4}\left|\widetilde{X}\right.}\left(t\left|\widetilde{\msf{x}}\right.\right)\varphi_{Z_{2}\left|\widetilde{X}\right.}\left(t\left|\widetilde{\msf{x}}\right.\right)=\varphi_{Z_{1}\left|\widetilde{X}\right.}\left(t\left|\widetilde{\msf{x}}\right.\right)\varphi_{Z_{2}\left|\widetilde{X}\right.}\left(t\left|\widetilde{\msf{x}}\right.\right).
\end{equation}}}Since {\small{$\varphi_{Z_{2}\left|\widetilde{X}\right.}\left(t\left|\widetilde{\msf{x}}\right.\right)$}}is characteristic function which always attains strictly non-zero value across $t\in\left(-\infty,+\infty\right)$, then we have 
{\small{\begin{equation}\label{c5}
\varphi_{Z_{4}\left|\widetilde{X}\right.}\left(t\left|\widetilde{\msf{x}}\right.\right)=\varphi_{Z_{1}\left|\widetilde{X}\right.}\left(t\left|\widetilde{\msf{x}}\right.\right).
\end{equation}}}From the knowledge that pdf can be uniquely determined by characteristic function, (\ref{c5}) indicates
{\small{\begin{equation}\label{c6}
f_{Z_{1}\left|\widetilde{X}\right.}\left({z}_{1}\left|\widetilde{\msf{x}}\right.\right)=f_{Z_{4}\left|\widetilde{X}\right.}\left({z}_{4}\left|\widetilde{\msf{x}}\right.\right).
\end{equation}}}
\end{IEEEproof}Let us back to the proof of Proposition 1. Revisiting $Y =h_2V+N_s$, based on Lemma \ref{lemp1},
if and only if the relay is absolutely reliable, i.e., $U=V$, we can get {\small{$f_{U\left|X\right.}\left(u\left|{x}\right.\right)=f_{V\left|X\right.}\left(v\left|{x}\right.\right)$}}. 
Since $h_2$ is a nonzero constant, we can easily get that if and only if $U=V$, {\small{$f_{h_2U\left|X\right.}\left(h_2u\left|\mathsf{x}\right.\right)=f_{h_2V\left|X\right.}\left(h_2v\left|{x}\right.\right)$}} holds true. Jointly consider that Lemma \ref{lemp2} indicates if and only if {\small{$f_{h_2U\left|X\right.}\left(h_2u\left|{x}\right.\right)=f_{h_2V\left|X\right.}\left(h_2v\left|{x}\right.\right)$}}, there exists {\small{$f_{h_2U+N_s\left|X\right.}\left(y\left|{x}\right.\right)=f_{h_2V+N_s\left|X\right.}\left(y\left|{x}\right.\right)$}}. We finally get that if and only if $U=V$, {\small{$f_{h_2U+N_s\left|X\right.}\left(y\left|{x}\right.\right)=f_{h_2V+N_s\left|X\right.}\left(y\left|{x}\right.\right)$}} holds true. Hence, the proposition 1 is proved.

%
%
%

\section{Necessity proof of Theorem 1}
\begin{IEEEproof}
Based on Definition 2, for manipulable observation channel, there exists 
\begin{equation}\label{necc_con}
\int_{-\infty}^{+\infty}\int_{-\infty}^{+\infty}f_{U\left|{X}\right.}\left(u\left|x\right.\right)\Psi\left(v\left|u\right.\right)F_{Y\left|V\right.}\left(y\left|v\right.\right)dudv=\int_{-\infty}^{+\infty}f_{U\left|{X}\right.}\left(u\left|x\right.\right)F_{Y\left|V\right.}\left(y\left|u\right.\right)du
\end{equation}where $\ensuremath{\Psi\left(v\left|u\right.\right)}\neq\Phi\left(v-u\right)$. Then, we consider the following two cases.
\begin{enumerate}
\item In case I, the relay reliably forwards symbols according to $U=V$, the destination 
observes $Y^n$ from the relay, and observes $X^n$ from the source.
\item In case II, the relay conducts i.i.d. attack according to $f_{V\left|U\right.}\left(v\left|u\right.\right)=\Psi\left(v\left|u\right.\right)$. 
For easy description, we use $Y'^n$ and $X'^n$ to denote the destination's observation from the relay and from the source, respectively.
\end{enumerate}
Obviously, in case I, the relay is reliable. Under the condition that case I occurs, decision statistics $D^n$ are 
constructed upon on $\left\{ \ensuremath{Y^{n}},X^{n}\right\}$. 
In case II, the relay is malicious. Under the condition that case II occurs, decision statistics $D^n$ are 
constructed upon on $\left\{ \ensuremath{Y'^{n}},X'^{n}\right\}$. 
According to (\ref{necc_con}), it is not hard to check that  $\left\{ \ensuremath{Y'^{n}},X'^{n}\right\}$ and $\left\{ \ensuremath{Y^{n}},X^{n}\right\} $
follow the same distribution. Then, any decision statistics $D^n$ constructed upon on  $\left\{ \ensuremath{Y'^{n}},X'^{n}\right\}$ and $\left\{ \ensuremath{Y^{n}},X^{n}\right\} $ must follow the same distribution. Hence, if there exists $D^n$ satisfies the second property of Theorem 1, $\lim_{n\rightarrow\infty}\Pr\Big( D^{n}> \mu' (n',\delta) ~\Big| 
  \text{case I} \Big) \leq
  \epsilon$, there must have $\lim_{n\rightarrow\infty}\Pr\Big( D^{n}> \mu' (n',\delta) ~\Big| 
  \text{case II} \Big) \leq
  \epsilon$. It indicates the two properties of Theorem 1 cannot be satisfied simultaneously.
 \end{IEEEproof}

\section{Sufficiency Proof of Theorem~\ref{thm:main2}}
\subsection{Preparations}
For easy description of our proof, we first define some functions and values as follows.
\begin{equation}
\triangle F_{\widetilde{V}^{n}\left|\widetilde{X}^{n}\right.}^{n}\left(v\left|\widetilde{\msf{x}}\right.\right)=\begin{cases}
\begin{array}{cc}
\frac{\sum_{i=1}^{n}1_{i}\left(\widetilde{V}_{i}=\widetilde{\mathsf{v}}_{k}\right)1_{i}\left(\widetilde{X}_{i}=\widetilde{\msf{x}}\right)}{N\left(\widetilde{\msf{x}}\left|\widetilde{X}^{n}\right.\right)}, & N\left(\widetilde{\msf{x}}\left|\widetilde{X}^{n}\right.\right)\neq0,\, v\in\mathcal{B}\left(\widetilde{\mathsf{v}}_{k}\right)\\
0, & otherwise
\end{array}\end{cases}
\end{equation}
\begin{equation}
F_{Y^{n}\left|\widetilde{X}^{n}\right.}^{n}\left(t\left|\widetilde{\msf{x}}\right.\right)=\begin{cases}
\begin{array}{cc}
\frac{\sum_{i=1}^{n}1_{i}\left(Y_{i}<t\right)1_{i}\left(\widetilde{X}_{i}=\widetilde{\msf{x}}\right)}{N\left(\widetilde{\msf{x}}\left|\widetilde{X}^{n}\right.\right)}, &  N\left(\widetilde{\msf{x}}\left|\widetilde{X}^{n}\right.\right)\neq0,\\
0, & otherwise
\end{array}\end{cases}
\end{equation}
\begin{equation}
F_{Y\left|V\right.}\left(t\left|v\right.\right)=\int_{-\infty}^{t}f_{Y\left|V\right.}\left(y\left|v\right.\right)dy
\end{equation}
For $k=1,2,\ldots,n_v$, $\overline{\mathsf{v}}_{k}$ denotes generic value belonging to $\mathcal{B}\left(\widetilde{\mathsf{v}}_{k}\right)$.

In order to prove one convergence property given later, we also
define 
{\small{\begin{align}
&\nonumber\triangle F_{i,i',j,j'}=P_{\widetilde{X}_{i},\widetilde{X}_{i'}\left|\widetilde{V}_{i},\widetilde{V}_{i'},\widetilde{U}_{i},\widetilde{U}_{i'}\right.}\left\{ \widetilde{\mathsf{x}},\,\widetilde{\mathsf{x}}\left|\widetilde{\mathsf{v}}_{k},\,\widetilde{\mathsf{v}}_{k},\,\widetilde{\mathsf{u}}_{j},\,\widetilde{\mathsf{u}}_{j'}\right.\right\} -P_{\widetilde{X}\left|\widetilde{U}\right.}\left(\widetilde{\mathsf{x}}\left|\widetilde{\mathsf{u}}_{j}\right.\right)P_{\widetilde{X}_{i'}\left|\widetilde{V}_{i},\widetilde{V}_{i'},\widetilde{U}_{i},\widetilde{U}_{i'}\right.}\left\{ \widetilde{\mathsf{x}}\left|\widetilde{\mathsf{v}}_{k},\,\widetilde{\mathsf{v}}_{k},\,\widetilde{\mathsf{u}}_{j},\,\widetilde{\mathsf{u}}_{j'}\right.\right\}\\\nonumber
&-P_{\widetilde{X}_{i}\left|\widetilde{V}_{i},\widetilde{V}_{i'},\widetilde{U}_{i},\widetilde{U}_{i'}\right.}\left\{ \widetilde{\mathsf{x}}\left|\widetilde{\mathsf{v}}_{k},\,\widetilde{\mathsf{v}}_{k},\,\widetilde{\mathsf{u}}_{j},\,\widetilde{\mathsf{u}}_{j'}\right.\right\} P_{\widetilde{X}\left|\widetilde{U}\right.}\left(\widetilde{\mathsf{x}}\left|\widetilde{\mathsf{u}}_{j'}\right.\right)+P_{\widetilde{X}\left|\widetilde{U}\right.}\left(\widetilde{\mathsf{x}}\left|\widetilde{\mathsf{u}}_{j}\right.\right)P_{\widetilde{X}\left|\widetilde{U}\right.}\left(\widetilde{\mathsf{x}}\left|\widetilde{\mathsf{u}}_{j'}\right.\right)
\end{align}}}where $i\neq i'$, $j,j'=1,2,\ldots n'$, $i,i'=1,2,\ldots n$. Then, we have the following lemma.
\begin{lemma}\label{lem1}
If we choose $\alpha_{1}=-\beta_{1}$, and  $\beta_{1}=\sqrt{n'}$,
then upon this setup, 
there also exist a upper bound for $F_{i,i'j,j'}$ across $j,j'=2,\ldots n'-1$, $i,i'=1,2,3,\ldots n$, $i\neq i'$. This upper bound only depends on $n'$ rather than $n$. Hence, we denote the upper bound as $\triangle F_{max}\left(n'\right)$. $\triangle F_{max}\left(n'\right)$ has property that  
\begin{equation}
n'^{k}\triangle F_{max}\left(n'\right)\rightarrow0
\end{equation}where $k$ is strictly less than $\frac{1}{2}$, i.e., $k<\frac{1}{2}$.
\end{lemma}
\begin{IEEEproof}
To bound $\triangle F_{i,i'j,j'}$, we have
\begin{equation}
P_{\widetilde{X}\left|\widetilde{U}\right.}\left(\widetilde{\mathsf{x}}\left|\widetilde{\mathsf{u}}_{j}\right.\right)=\frac{\int_{u\in\mathcal{B}\left(\widetilde{\mathsf{u}}_{j}\right)}P_{\widetilde{X}\left|{U}\right.}\left(\widetilde{\mathsf{x}}\left|u\right.\right)f_{U}\left(u\right)du}{\int_{u\in\mathcal{B}\left(\widetilde{\mathsf{u}}_{j}\right)}f_{U}\left(u\right)du}
\end{equation}which indicates
\begin{equation}\label{bound1}
\underset{u\in\mathcal{B}\left(\widetilde{\mathsf{u}}_{j}\right)}{\min}P_{\widetilde{X}\left|U\right.}\left(\widetilde{\mathsf{x}}\left|u\right.\right)\leq P_{\widetilde{X}\left|\widetilde{U}\right.}\left(\widetilde{\mathsf{x}}\left|\widetilde{\mathsf{u}}_{j}\right.\right)\leq\underset{u\in\mathcal{B}\left(\widetilde{\mathsf{u}}_{j}\right)}{\max}P_{\widetilde{X}\left|U\right.}\left(\widetilde{\mathsf{x}}\left|u\right.\right).
\end{equation}On the other hand, since
{\small{\begin{align*}
&P_{\widetilde{X}_{i}\left|\widetilde{V}_{i},\widetilde{V}_{i'},\widetilde{U}_{i},\widetilde{U}_{i'}\right.}\left\{ \widetilde{\mathsf{x}}\left|\widetilde{\mathsf{v}}_{k},\,\widetilde{\mathsf{v}}_{k},\,\widetilde{\mathsf{u}}_{j},\,\widetilde{\mathsf{u}}_{j'}\right.\right\}=\\
&\frac{\int_{v\in\mathcal{B}\left(\widetilde{\mathsf{v}}_{k}\right)}\int_{v'\in\mathcal{B}\left(\widetilde{\mathsf{v}}_{k}\right)}\int_{u\in\mathcal{B}\left(\widetilde{\mathsf{u}}_{j}\right)}\int_{u'\in\mathcal{B}\left(\widetilde{\mathsf{u}}_{j'}\right)}P_{\widetilde{X}\left|U\right.}\left(\widetilde{\mathsf{x}}\left|u\right.\right)f_{U_{i},U_{i'}\left|V_{i},V_{i'}\right.}\left(u,u'\left|v,\, v'\right.\right)f_{V_{i},V_{i'}}\left(v_{i},\, v_{i'}\right)dudu'dvdv'}{\int_{v\in\mathcal{B}\left(\widetilde{\mathsf{v}}_{k}\right)}\int_{v'\in\mathcal{B}\left(\widetilde{\mathsf{v}}_{k'}\right)}\int_{u\in\mathcal{B}\left(\widetilde{\mathsf{u}}_{j}\right)}\int_{u'\in\mathcal{B}\left(\widetilde{\mathsf{u}}_{j'}\right)}f_{U_{i},U_{i'}\left|V_{i},V_{i'}\right.}\left(u,u'\left|v,\, v'\right.\right)f_{V_{i},V_{i'}}\left(v_{i},\, v_{i'}\right)dudu'dvdv'}
\end{align*}}}and 
{\small{\begin{align*}
&P_{\widetilde{X}_{i},\widetilde{X}_{i'}\left|\widetilde{V}_{i},\widetilde{V}_{i'},\widetilde{U}_{i},\widetilde{U}_{i'}\right.}\left\{ \widetilde{\mathsf{x}},\,\widetilde{\mathsf{x}}\left|\widetilde{\mathsf{v}}_{k},\,\widetilde{\mathsf{v}}_{k},\,\widetilde{\mathsf{u}}_{j},\,\widetilde{\mathsf{u}}_{j'}\right.\right\} =\\
&\frac{\int_{v\in\mathcal{B}\left(\widetilde{\mathsf{v}}_{k}\right)}\int_{v'\in\mathcal{B}\left(\widetilde{\mathsf{v}}_{k}\right)}\int_{u\in\mathcal{B}\left(\widetilde{\mathsf{u}}_{j}\right)}\int_{u'\in\mathcal{B}\left(\widetilde{\mathsf{u}}_{j'}\right)}P_{\widetilde{X}\left|U\right.}\left(\widetilde{\mathsf{x}}\left|u\right.\right)P_{\widetilde{X}\left|U\right.}\left(\widetilde{\mathsf{x}}\left|u\right.\right)f_{U_{i},U_{i'}\left|V_{i},V_{i'}\right.}\left(u,u'\left|v,\, v'\right.\right)f_{V_{i},V_{i'}}\left(v_{i},\, v_{i'}\right)dudu'dvdv'}{\int_{v\in\mathcal{B}\left(\widetilde{\mathsf{v}}_{k}\right)}\int_{v'\in\mathcal{B}\left(\widetilde{\mathsf{v}}_{k'}\right)}\int_{u\in\mathcal{B}\left(\widetilde{\mathsf{u}}_{j}\right)}\int_{u'\in\mathcal{B}\left(\widetilde{\mathsf{u}}_{j'}\right)}f_{U_{i},U_{i'}\left|V_{i},V_{i'}\right.}\left(u,u'\left|v,\, v'\right.\right)f_{V_{i},V_{i'}}\left(v_{i},\, v_{i'}\right)dudu'dvdv'},
\end{align*}}}
we have 
\begin{equation}\label{bound21}
\underset{u\in\mathcal{B}\left(\widetilde{\mathsf{u}}_{j}\right)}{\min}P_{\widetilde{X}\left|U\right.}\left(\widetilde{\mathsf{x}}\left|u\right.\right)\leq P_{\widetilde{X}_{i}\left|\widetilde{V}_{i},\widetilde{V}_{i'},\widetilde{U}_{i},\widetilde{U}_{i'}\right.}\left\{ \widetilde{\mathsf{x}}\left|\widetilde{\mathsf{v}}_{k},\,\widetilde{\mathsf{v}}_{k},\,\widetilde{\mathsf{u}}_{j},\,\widetilde{\mathsf{u}}_{j'}\right.\right\} \leq\underset{u\in\mathcal{B}\left(\widetilde{\mathsf{u}}_{j}\right)}{\max}P_{\widetilde{X}\left|U\right.}\left(\widetilde{\mathsf{x}}\left|u\right.\right)
\end{equation}and 
{\small{\begin{align}\label{bound22}
&P_{\widetilde{X}_{i}\left|\widetilde{V}_{i},\widetilde{V}_{i'},\widetilde{U}_{i},\widetilde{U}_{i'}\right.}\left\{ \widetilde{\mathsf{x}}\left|\widetilde{\mathsf{v}}_{k},\,\widetilde{\mathsf{v}}_{k},\,\widetilde{\mathsf{u}}_{j},\,\widetilde{\mathsf{u}}_{j'}\right.\right\} \underset{u\in\mathcal{B}\left(\widetilde{\mathsf{u}}_{j}\right)}{\min}P_{\widetilde{X}\left|U\right.}\left(\widetilde{\mathsf{x}}\left|u\right.\right)\leq P_{\widetilde{X}_{i},\widetilde{X}_{i'}\left|\widetilde{V}_{i},\widetilde{V}_{i'},\widetilde{U}_{i},\widetilde{U}_{i'}\right.}\left\{ \widetilde{\mathsf{x}},\,\widetilde{\mathsf{x}}\left|\widetilde{\mathsf{v}}_{k},\,\widetilde{\mathsf{v}}_{k},\,\widetilde{\mathsf{u}}_{j},\,\widetilde{\mathsf{u}}_{j'}\right.\right\}\\\nonumber
& \leq P_{\widetilde{X}_{i}\left|\widetilde{V}_{i},\widetilde{V}_{i'},\widetilde{U}_{i},\widetilde{U}_{i'}\right.}\left\{ \widetilde{\mathsf{x}}\left|\widetilde{\mathsf{v}}_{k},\,\widetilde{\mathsf{v}}_{k},\,\widetilde{\mathsf{u}}_{j},\,\widetilde{\mathsf{u}}_{j'}\right.\right\} \underset{u\in\mathcal{B}\left(\widetilde{\mathsf{u}}_{j}\right)}{\max}P_{\widetilde{X}\left|U\right.}\left(\widetilde{\mathsf{x}}\left|u\right.\right)
\end{align}}}
Jointly considering (\ref{bound1}), (\ref{bound21}) and (\ref{bound22}), 
$\triangle F_{i,i',j,j'}$ could be bound as
{\small{\begin{equation}
\nonumber\abs{\triangle F_{i,i',j,j'}}\leq\underset{u\in\mathcal{B}\left(\widetilde{\mathsf{u}}_{j}\right)}{\max}2P_{\widetilde{X}\left|U\right.}^{2}\left(\widetilde{\mathsf{x}}\left|u\right.\right)-\underset{u\in\mathcal{B}\left(\widetilde{\mathsf{u}}_{j}\right)}{\min}2P_{\widetilde{X}\left|U\right.}^{2}\left(\widetilde{\mathsf{x}}\left|u\right.\right)\label{reshape}.
\end{equation}}}
Then, we have 
\begin{equation}\label{Fmax}
\underset{j,j'=2,\ldots n'-1,i,i'=1,2,\ldots n,i\neq i'}{\max}\triangle F_{i,i'j,j'}<\underset{j=2,3,\ldots n'-1}{\max}2\left(\underset{u\in\mathcal{B}\left(\widetilde{\mathsf{u}}_{j}\right)}{\max}P_{\widetilde{X}\left|U\right.}^{2}\left(\widetilde{\mathsf{x}}\left|u\right.\right)-\underset{u\in\mathcal{B}\left(\widetilde{\mathsf{u}}_{j}\right)}{\min}P_{\widetilde{X}\left|U\right.}^{2}\left(\widetilde{\mathsf{x}}\left|u\right.\right)\right)\defn\triangle F_{max}\left(n'\right)
\end{equation}Then, we proceed to focus on the property of $F_{max}\left(n'\right)$. 
Revisiting the system model, 
for $j=2,3,\ldots,n'-1$, we have
{\small{\begin{equation}\
\underset{u\in\mathcal{B}\left(\widetilde{\mathsf{u}}_{j}\right)}{\max}P_{\widetilde{X}\left|U\right.}^{2}\left(\widetilde{\mathsf{x}}\left|u\right.\right)-\underset{u\in\mathcal{B}\left(\widetilde{\mathsf{u}}_{j}\right)}{\min}P_{\widetilde{X}\left|U\right.}^{2}\left(\widetilde{\mathsf{x}}\left|u\right.\right)\leq 2P_{\widetilde{X}\left|U\right.}\left(\widetilde{\mathsf{x}}\left|u'_{j}\right.\right){P'}_{\widetilde{X}\left|U\right.}\left(\widetilde{\mathsf{x}}\left|u'_{j}\right.\right)\frac{2\sqrt{n'}}{n'-2}
\end{equation}}}where $u'_{j}\in\mathcal{B}\left(\widetilde{\mathsf{u}}_{j}\right)$, ${P'}_{\widetilde{X}\left|U\right.}\left(\widetilde{\mathsf{x}}\left|u\right.\right)$ is derived function of ${P}_{\widetilde{X}\left|U\right.}\left(\widetilde{\mathsf{x}}\left|u\right.\right)$. The maximum of ${P'}_{\widetilde{X}\left|U\right.}\left(\widetilde{\mathsf{x}}\left|u\right.\right)$ in $\left(-\infty,+\infty\right)$ is bounded. 
\begin{align}\label{limitation1}
&\nonumber\lim_{n'\rightarrow\infty}n'^{k}\left(\underset{u\in\mathcal{B}\left(\widetilde{\mathsf{u}}_{j}\right)}{\max}P_{\widetilde{X}\left|U\right.}^{2}\left(\widetilde{\mathsf{x}}\left|u\right.\right)-\underset{u\in\mathcal{B}\left(\widetilde{\mathsf{u}}_{j}\right)}{\min}P_{\widetilde{X}\left|U\right.}^{2}\left(\widetilde{\mathsf{x}}\left|u\right.\right)\right)\leq\lim_{n'\rightarrow\infty}n'^{k}2P_{\widetilde{X}\left|U\right.}\left(\widetilde{\mathsf{x}}\left|u'_{j}\right.\right){P'}_{\widetilde{X}\left|U\right.}\left(\widetilde{\mathsf{x}}\left|u'_{j}\right.\right)\frac{2\sqrt{n'}}{n'-2}\\&
=0
\end{align}where the last equality follows the fact that the maximum of ${P'}_{\widetilde{X}\left|U\right.}\left(\widetilde{\mathsf{x}}\left|u'_{j}\right.\right)$ and ${P}_{\widetilde{X}\left|U\right.}\left(\widetilde{\mathsf{x}}\left|u'_{j}\right.\right)$ in $\left(-\infty,+\infty\right)$ is bounded, and $k$ is strictly less than 1/2. 
Finally, based on the definition of $F_{max}\left(n'\right)$ in (\ref{Fmax}), the statement of this lemma is immediate. 
\end{IEEEproof}
Similar to the definition of $\triangle F_{i,i',j,j'}$ and Lemma \ref{lem1}, we define 
{\small{\begin{align}
\nonumber G_{i,i',k,k'}&=P_{Y_{i}\left|\widetilde{V}_{i},\widetilde{V}_{i'},\widetilde{X}_{i},\widetilde{X}_{i'}\right.}\left(Y_{i}<t\left|\widetilde{\mathsf{v}}_{k},\widetilde{\mathsf{v}}_{k'},\widetilde{\mathsf{x}},\widetilde{\mathsf{x}}\right.\right)\left(P_{Y_{i'}\left|Y_{i},\widetilde{V}_{i},\widetilde{V}_{i'},\widetilde{X}_{i},\widetilde{X}_{i'}\right.}\left(Y_{i'}<t\left|Y_{i}<t,\widetilde{\mathsf{v}}_{k},\widetilde{\mathsf{v}}_{k'},\widetilde{\mathsf{x}},\widetilde{\mathsf{x}}\right.\right)-F_{Y\left|V\right.}\left(t\left|\overline{\mathsf{v}}_{k'}\right.\right)\right)\\
&-F_{Y\left|V\right.}\left(t\left|\overline{\mathsf{v}}_{k}\right.\right)\left(P_{Y_{i'}\left|\widetilde{V}_{i},\widetilde{V}_{i'},\widetilde{X}_{i},\widetilde{X}_{i'}\right.}\left(Y_{i'}<t\left|\widetilde{\mathsf{v}}_{k},\widetilde{\mathsf{v}}_{k'},\widetilde{\mathsf{x}},\widetilde{\mathsf{x}}\right.\right)-F_{Y\left|V\right.}\left(t\left|\overline{\mathsf{v}}_{k'}\right.\right)\right)
\end{align}}}where $i\neq i'$, $k,k'=2,3,\ldots n_v-1$, $i,i'=1,2,\ldots n$. Then, we have the following lemma.
\begin{lemma}\label{lemA1}
If we choose $\alpha_{2}=-\beta_{2}$, and  $\beta_{2}=\sqrt{n_v}$,
then upon this setup, 
there also exist a upper bound for $G_{i,i',k,k'}$ across $i\neq i'$, $k,k'=2,3,\ldots n_v-1$, $i,i'=1,2,\ldots n$. This upper bound only depends on $n_v$ rather than $n_v$. Hence, we denote the upper bound as $\triangle G_{max}\left(n_v\right)$. $\triangle G_{max}\left(n_v\right)$ has property that  
\begin{equation}
n_{v}^{\tau}\triangle G_{max}\left(n_v\right)\rightarrow0
\end{equation}where $k$ is strictly less than $\frac{1}{2}$, i.e., $\tau<\frac{1}{2}$.
\end{lemma}
\begin{IEEEproof}
Following the proof method employed by Lemma \ref{lem1}, the assertion is direct. More details are omitted 
due to space limitation.
\end{IEEEproof}
\begin{lemma}\label{lem2}
For arbitrary $\widetilde{\mathsf{v}}_{k}$, sufficiently small $\mu$ and $\varepsilon\leq\frac{\mu}{2n'}$, 
{\small{\begin{align}
&\nonumber \Pr\left\{ \left|\triangle F_{\widetilde{V}^{n}\left|\widetilde{X}^{n}\right.}^{n}\left(\widetilde{\mathsf{v}}_{k}\left|\widetilde{\msf{x}}\right.\right)-\int_{-\infty}^{+\infty}f_{U\left|\widetilde{X}\right.}\left(u\left|\widetilde{\msf{x}}\right.\right)\triangle F_{\widetilde{V}^{n}\left|\widetilde{U}^{n}\right.}^{(n')}\left(\widetilde{\mathsf{v}}_{k}\left|u\right.\right)du\right|>\mu\right\}  \\\label{final_upper}
&<\frac{4}{\mu^{2}}\left(\frac{n'^{2}}{n}+\frac{1}{\Pr\left\{ \left(\widetilde{X}^{n},\,\widetilde{U}^{n}\right)\in\mathcal{A}_{\varepsilon}\right\} \left(P_{\widetilde{X}}\left(\widetilde{\mathsf{x}}\right)-\varepsilon\right)^{2}}\triangle F_{max}\left(n'\right)\right)+\Pr\left\{ \left(\widetilde{X}^{n},\,\widetilde{U}^{n}\right)\notin \mathcal{A}_{\varepsilon}\right\}
\end{align}}}
where{\small{\begin{equation}
\mathcal{A}_{\varepsilon}=\left\{ \left({\widetilde{x}}^{n},\widetilde{u}^{n}\right):\:\left|P_{\widetilde{U}\left|\widetilde{X}\right.}\left(\widetilde{\mathsf{u}}\left|\widetilde{\mathsf{x}}\right.\right)-\frac{N\left(\widetilde{\mathsf{u}}\left|\widetilde{U}^{n}\right.\right)P_{\widetilde{X}\left|\widetilde{U}\right.}\left(\widetilde{\mathsf{x}}\left|\widetilde{\mathsf{u}}\right.\right)}{N\left(\widetilde{\mathsf{x}}\left|\widetilde{X}^{n}\right.\right)}\right|<\varepsilon,\;\left|P_{\widetilde{X}}\left(\widetilde{\mathsf{x}}\right)-\frac{N\left(\widetilde{\mathsf{x}}\left|\widetilde{X}^{n}\right.\right)}{n}\right|<\varepsilon\right\}.
\end{equation}}}
\end{lemma}
\begin{IEEEproof}
Notice that
{\small{\begin{align}\label{upper}
&\Pr\left\{ \left|\triangle F_{\widetilde{V}^{n}\left|\widetilde{X}^{n}\right.}^{n}\left(\widetilde{\mathsf{v}}_{k}\left|\widetilde{\msf{x}}\right.\right)-\int_{-\infty}^{+\infty}f_{U\left|\widetilde{X}\right.}\left(u\left|\widetilde{\msf{x}}\right.\right)\triangle F_{\widetilde{V}^{n}\left|\widetilde{U}^{n}\right.}^{(n')}\left(\widetilde{\mathsf{v}}_{k}\left|u\right.\right)du\right|>\mu\right\}  \\\nonumber
&<\Pr\left\{ \left|\triangle F_{\widetilde{V}^{n}\left|\widetilde{X}^{n}\right.}^{n}\left(\widetilde{\mathsf{v}}_{k}\left|\widetilde{\msf{x}}\right.\right)-\int_{-\infty}^{+\infty}f_{U\left|\widetilde{X}\right.}\left(u\left|\widetilde{\msf{x}}\right.\right)\triangle F_{\widetilde{V}^{n}\left|\widetilde{U}^{n}\right.}^{(n')}\left(\widetilde{\mathsf{v}}_{k}\left|u\right.\right)du\right|>\mu\left| \left(\widetilde{X}^{n},\,\widetilde{U}^{n}\right)\in \mathcal{A}_{\varepsilon}\right.\right\} +\Pr\left\{ \left(\widetilde{X}^{n},\,\widetilde{U}^{n}\right)\notin \mathcal{A}_{\varepsilon}\right\} 
\end{align}}}
Firstly notice that after $n'$, $\alpha_{1}$ and $\beta_{1}$ are chosen and fixed properly,
\begin{equation}
\Pr\left\{ \left(\widetilde{X}^{n},\,\widetilde{U}^{n}\right)\notin \mathcal{A}_{\varepsilon}\right\} \rightarrow0
\end{equation}as $n$ approaches to infinity.  Then, focusing on the first item in the right side of (\ref{upper}), $\left(\widetilde{X}^{n},\,\widetilde{U}^{n}\right)\in \mathcal{A}_{\varepsilon}$ indicates $N\left(\widetilde{\mathsf{u}}_{j}\left|\widetilde{U}^{n}\right.\right)>0$ for all $j=1,2,\ldots, n'$. 
Hence, $\frac{\sum_{i=1}^{n}1_{i}\left(\widetilde{V}_{i}=\widetilde{\mathsf{v}}_{k}\right)1_{i}\left(\widetilde{U}_{i}=\widetilde{\mathsf{u}}_{j}\right)}{N\left(\widetilde{\mathsf{u}}_{j}\left|\widetilde{U}^{n}\right.\right)}
 $ is well-defined for all $j=1,2,\ldots, n'$. Under the condition {\small{$ \left(\widetilde{X}^{n},\,\widetilde{U}^{n}\right)\in \mathcal{A}_{\varepsilon}$}}, we have
{\small{\begin{align}\label{upper1}
&\left|\triangle F_{\widetilde{V}^{n}\left|\widetilde{X}^{n}\right.}^{n}\left(\widetilde{\mathsf{v}}_{k}\left|\widetilde{\msf{x}}\right.\right)-\int_{-\infty}^{+\infty}f_{U\left|\widetilde{X}\right.}\left(u\left|\widetilde{\msf{x}}\right.\right)\triangle F_{\widetilde{V}^{n}\left|\widetilde{U}^{n}\right.}^{(n')}\left(\widetilde{\mathsf{v}}_{k}\left|u\right.\right)du\right|\\\nonumber
&=\left|\frac{\sum_{i=1}^{n}1_{i}\left(\widetilde{V}_{i}=\widetilde{\mathsf{v}}_{k}\right)1_{i}\left(\widetilde{U}_{i}=\widetilde{\mathsf{u}}_{j}\right)}{N\left(\widetilde{\mathsf{u}}_{j}\left|\widetilde{U}^{n}\right.\right)}
 -\sum_{j=1}^{n'}\frac{\sum_{i=1}^{n}1_{i}\left(\widetilde{V}_{i}=\widetilde{\mathsf{v}}_{k}\right)1_{i}\left(\widetilde{U}_{i}=\widetilde{\mathsf{u}}_{j}\right)}{N\left(\widetilde{\mathsf{u}}_{j}\left|\widetilde{U}^{n}\right.\right)}\int_{\mathcal{B}\left(\widetilde{\mathsf{u}}_{j}\right)}f_{U\left|\widetilde{X}\right.}\left(u\left|\widetilde{\msf{x}}\right.\right)du\right|\\\nonumber
&\text{=}\left|\frac{\sum_{j=1}^{n'}\sum_{i=1}^{n}1_{i}\left(\widetilde{V}_{i}=\widetilde{\mathsf{v}}_{k}\right)1_{i}\left(\widetilde{X}_{i}=\widetilde{\msf{x}}\right)1_{i}\left(\widetilde{U}_{i}=\widetilde{\mathsf{u}}_{j}\right)}{N\left(\widetilde{\msf{x}}\left|\widetilde{X}^{n}\right.\right)}-\sum_{j=1}^{n'}\frac{\sum_{i=1}^{n}1_{i}\left(\widetilde{V}_{i}=\widetilde{\mathsf{v}}_{k}\right)1_{i}\left(\widetilde{U}_{i}=\widetilde{\mathsf{u}}_{j}\right)}{N\left(\widetilde{\mathsf{u}}_{j}\left|\widetilde{U}^{n}\right.\right)}P_{\widetilde{U}\left|\widetilde{X}\right.}\left(\widetilde{\mathsf{u}}_{j}\left|\widetilde{\msf{x}}\right.\right)\right|
\end{align}}}Substituting (\ref{upper1}) into the first item in the right side of (\ref{upper}), it becomes
{\small{\begin{align}\label{upper2}
&\Pr\left\{ \left|\frac{\sum_{j=1}^{n'}\sum_{i=1}^{n}1_{i}\left(\widetilde{V}_{i}=\widetilde{\mathsf{v}}_{k}\right)1_{i}\left(\widetilde{X}_{i}=\widetilde{\msf{x}}\right)1_{i}\left(\widetilde{U}_{i}=\widetilde{\mathsf{u}}_{j}\right)}{N\left(\widetilde{\msf{x}}\left|\widetilde{X}^{n}\right.\right)}-\sum_{j=1}^{n'}\frac{\sum_{i=1}^{n}1_{i}\left(\widetilde{V}_{i}=\widetilde{\mathsf{v}}_{k}\right)1_{i}\left(\widetilde{U}_{i}=\widetilde{\mathsf{u}}_{j}\right)}{N\left(\widetilde{\mathsf{u}}_{j}\left|\widetilde{U}^{n}\right.\right)}P_{\widetilde{U}\left|\widetilde{X}\right.}\left(\widetilde{\mathsf{u}}_{j}\left|\widetilde{\msf{x}}\right.\right)\right|>\mu\left|\left(\widetilde{X}^{n},\,\widetilde{U}^{n}\right)\in\mathcal{A}_{\varepsilon}\right.\right\} \\\nonumber
&<\Pr\left\{ \left|\sum_{j=1}^{n'}\underbrace{\left(\frac{\sum_{i=1}^{n}1_{i}\left(\widetilde{V}_{i}=\widetilde{\mathsf{v}}_{k}\right)1_{i}\left(\widetilde{X}_{i}=\widetilde{\msf{x}}\right)1_{i}\left(\widetilde{U}_{i}=\widetilde{\mathsf{u}}_{j}\right)}{n\left(P_{\widetilde{X}}\left(\mathsf{x}\right)-\varepsilon\right)}-\frac{\sum_{i=1}^{n}1_{i}\left(\widetilde{V}_{i}=\widetilde{\mathsf{v}}_{k}\right)1_{i}\left(\widetilde{U}_{i}=\widetilde{\mathsf{u}}_{j}\right)P_{\widetilde{X}\left|\widetilde{U}\right.}\left(\widetilde{\msf{x}}\left|\widetilde{\mathsf{u}}_{j}\right.\right)}{n\left(P_{\widetilde{X}}\left(\mathsf{x}\right)-\varepsilon\right)}\right)}_{H_{j}}\right|>\frac{\mu}{2}\left|\left(\widetilde{X}^{n},\,\widetilde{U}^{n}\right)\in\mathcal{A}_{\varepsilon}\right.\right\}  \\\nonumber
&+\sum_{j=1}^{n'}\Pr\left\{ \left|\frac{\sum_{i=1}^{n}1_{i}\left(\widetilde{V}_{i}=\widetilde{\mathsf{v}}_{k}\right)1_{i}\left(\widetilde{U}_{i}=\widetilde{\mathsf{u}}_{j}\right)}{N\left(\widetilde{\mathsf{u}}_{j}\left|\widetilde{U}^{n}\right.\right)}P_{\widetilde{U}\left|\widetilde{X}\right.}\left(\widetilde{\mathsf{u}}_{j}\left|\widetilde{\msf{x}}\right.\right)-\frac{\sum_{i=1}^{n}1_{i}\left(\widetilde{V}_{i}=\widetilde{\mathsf{v}}_{k}\right)1_{i}\left(\widetilde{U}_{i}=\widetilde{\mathsf{u}}_{j}\right)P_{\widetilde{X}\left|\widetilde{U}\right.}\left(\widetilde{\msf{x}}\left|\widetilde{\mathsf{u}}_{j}\right.\right)}{N\left(\widetilde{\msf{x}}\left|\widetilde{X}^{n}\right.\right)}\right|>\frac{\mu}{2n'}\left|\left(\widetilde{X}^{n},\,\widetilde{U}^{n}\right)\in\mathcal{A}_{\varepsilon}\right.\right\} 
\end{align}}}The second item in the right of (\ref{upper2}) can be further bound as 
{\small{\begin{align}\label{upper3}
&\Pr\left\{ \left|\frac{\sum_{i=1}^{n}1_{i}\left(\widetilde{V}_{i}=\widetilde{\mathsf{v}}_{k}\right)1_{i}\left(\widetilde{U}_{i}=\widetilde{\mathsf{u}}_{j}\right)}{N\left(\widetilde{\mathsf{u}}_{j}\left|\widetilde{U}^{n}\right.\right)}P_{\widetilde{U}\left|\widetilde{X}\right.}\left(\widetilde{\mathsf{u}}_{j}\left|\widetilde{\msf{x}}\right.\right)-\frac{\sum_{i=1}^{n}1_{i}\left(\widetilde{V}_{i}=\widetilde{\mathsf{v}}_{k}\right)1_{i}\left(\widetilde{U}_{i}=\widetilde{\mathsf{u}}_{j}\right)P_{\widetilde{X}\left|\widetilde{U}\right.}\left(\widetilde{\msf{x}}\left|\widetilde{\mathsf{u}}_{j}\right.\right)}{N\left(\widetilde{\msf{x}}\left|\widetilde{X}^{n}\right.\right)}\right|>\frac{\mu}{2n'}\left|\left(\widetilde{X}^{n},\,\widetilde{U}^{n}\right)\in \mathcal{A}_{\varepsilon}\right.\right\} \\\nonumber
&\leq\Pr\left\{ \left|P_{\widetilde{U}\left|\widetilde{X}\right.}\left(\widetilde{\mathsf{u}}_{j}\left|\widetilde{\msf{x}}\right.\right)-\frac{N\left(\widetilde{\mathsf{u}}\left|\widetilde{U}^{n}\right.\right)P_{\widetilde{X}\left|\widetilde{U}\right.}\left(\widetilde{\msf{x}}\left|\widetilde{\mathsf{u}}_{j}\right.\right)}{N\left(\widetilde{\msf{x}}\left|\widetilde{X}^{n}\right.\right)}\right|>\frac{\mu}{2n'}\left|\left(\widetilde{X}^{n},\,\widetilde{U}^{n}\right)\in \mathcal{A}_{\varepsilon}\right.\right\} =0
\end{align}}}where the last equality follows the definition of $\mathcal{A}_{\varepsilon}$ and set of $\varepsilon\leq\frac{\mu}{2n'}$. From (\ref{upper3}), the second item in the right of (\ref{upper2}) equals to 0. Then, we proceed to bound the first item in the right of (\ref{upper2}) as
{\small{\begin{equation}\label{upper4}
\Pr\left\{ \left|\sum_{j=1}^{n'}H_{j}\right|>\frac{\mu}{2}\left|\left(\widetilde{X}^{n},\,\widetilde{U}^{n}\right)\in\mathcal{A}_{\varepsilon}\right.\right\} <\frac{4}{\mu^{2}}E_{\mathcal{A}_{\varepsilon}}\left|\sum_{j=1}^{n'}H_{j}\right|^{2} =\frac{4}{\mu^{2}}\sum_{j=1}^{n'}\sum_{j'=1}^{n'}E_{\mathcal{A}_{\varepsilon}}\left(H_{j}H_{j'}\right)
\end{equation}}}which follows the Chebyshev theorem. $E_{\mathcal{A}_{\varepsilon}}\left(\cdot\right)$ indicates the expectation of its input conditioned on $\left(\widetilde{X}^{n},\,\widetilde{U}^{n}\right)\in\mathcal{A}_{\varepsilon}$.
{\small{\begin{align}
&\nonumber E_{\mathcal{A}_{\varepsilon}}\left(H_{j}H_{j'}\right)=\\\nonumber
&\frac{E_{\mathcal{A}_{\varepsilon}}\left(\sum_{i=1}^{n}1_{i}\left(\widetilde{V}_{i}=\widetilde{\mathsf{v}}_{k}\right)1_{i}\left(\widetilde{U}_{i}=\widetilde{\mathsf{u}}_{j}\right)\left(1_{i}\left(\widetilde{X}_{i}=\widetilde{\msf{x}}\right)-P_{\widetilde{X}\left|\widetilde{U}\right.}\left(\widetilde{\msf{x}}\left|\widetilde{\mathsf{u}}_{j}\right.\right)\right)\right)\left(\sum_{i=1}^{n}1_{i}\left(\widetilde{V}_{i}=\widetilde{\mathsf{v}}_{k}\right)1_{i}\left(\widetilde{U}_{i}=\widetilde{\mathsf{u}}_{j'}\right)\left(1_{i}\left(\widetilde{X}_{i}=\widetilde{\msf{x}}\right)-P_{\widetilde{X}\left|\widetilde{U}\right.}\left(\widetilde{\msf{x}}\left|\widetilde{\mathsf{u}}_{j'}\right.\right)\right)\right)}{n^{2}\left(P_{\widetilde{X}}\left(\mathsf{x}\right)-\varepsilon\right)^{2}}\\
&\nonumber \leq\frac{\sum_{i=1}^{n}E_{\mathcal{A}_{\varepsilon}}\left\{ 1_{i}\left(\widetilde{V}_{i}=\widetilde{\mathsf{v}}_{k}\right)1_{i}\left(\widetilde{U}_{i}=\widetilde{\mathsf{u}}_{j}\right)\left(1_{i}\left(\widetilde{X}_{i}=\widetilde{\msf{x}}\right)-P_{\widetilde{X}\left|\widetilde{U}\right.}\left(\widetilde{\msf{x}}\left|\widetilde{\mathsf{u}}_{j}\right.\right)\right)1_{i}\left(\widetilde{U}_{i}=\widetilde{\mathsf{u}}_{j'}\right)\left(1_{i}\left(\widetilde{X}_{i}=\widetilde{\msf{x}}\right)-P_{\widetilde{X}\left|\widetilde{U}\right.}\left(\widetilde{\msf{x}}\left|\widetilde{\mathsf{u}}_{j'}\right.\right)\right)\right\} }{n^{2}\left(P_{\widetilde{X}}\left(\mathsf{x}\right)-\varepsilon\right)^{2}}\\
&\nonumber +\frac{E\sum_{i=1}^{n}\sum_{i=1,i'\neq i}^{n}1_{i}\left(\widetilde{V}_{i}=\widetilde{\mathsf{v}}_{k}\right)1_{i}\left(\widetilde{U}_{i}=\widetilde{\mathsf{u}}_{j}\right)1_{i'}\left(\widetilde{V}_{i'}=\widetilde{\mathsf{v}}_{k}\right)1_{i'}\left(\widetilde{U}_{i'}=\widetilde{\mathsf{u}}_{j'}\right)\left(1_{i}\left(\widetilde{X}_{i}=\widetilde{\msf{x}}\right)-P_{\widetilde{X}\left|\widetilde{U}\right.}\left(\widetilde{\msf{x}}\left|\widetilde{\mathsf{u}}_{j}\right.\right)\right)\left(1_{i'}\left(\widetilde{X}_{i'}=\widetilde{\msf{x}}\right)-P_{\widetilde{X}\left|\widetilde{U}\right.}\left(\widetilde{\msf{x}}\left|\widetilde{\mathsf{u}}_{j'}\right.\right)\right)}{\Pr\left\{ \left(\widetilde{X}^{n},\,\widetilde{U}^{n}\right)\in\mathcal{A}_{\varepsilon}\right\} n^{2}\left(P_{\widetilde{X}}\left(\mathsf{x}\right)-\varepsilon\right)^{2}}\\\nonumber
&\leq\frac{1}{n}+\frac{\sum_{i=1}^{n}\sum_{i=1,i'\neq i}^{n}E\left\{ 1_{i}\left(\widetilde{V}_{i}=\widetilde{\mathsf{v}}_{k}\right)1_{i}\left(\widetilde{U}_{i}=\widetilde{\mathsf{u}}_{j}\right)1_{i'}\left(\widetilde{V}_{i'}=\widetilde{\mathsf{v}}_{k}\right)1_{i'}\left(\widetilde{U}_{i'}=\widetilde{\mathsf{u}}_{j'}\right)1_{i}\left(\widetilde{X}_{i}=\widetilde{\msf{x}}\right)1_{i'}\left(\widetilde{X}_{i'}=\widetilde{\msf{x}}\right)\right\} }{\Pr\left\{ \left(\widetilde{X}^{n},\,\widetilde{U}^{n}\right)\in\mathcal{A}_{\varepsilon}\right\} n^{2}\left(P_{\widetilde{X}}\left(\mathsf{x}\right)-\varepsilon\right)^{2}}\\\nonumber
&-\frac{\sum_{i=1}^{n}\sum_{i=1,i'\neq i}^{n}E\left\{ 1_{i}\left(\widetilde{V}_{i}=\widetilde{\mathsf{v}}_{k}\right)1_{i}\left(\widetilde{U}_{i}=\widetilde{\mathsf{u}}_{j}\right)1_{i'}\left(\widetilde{V}_{i'}=\widetilde{\mathsf{v}}_{k}\right)1_{i'}\left(\widetilde{U}_{i'}=\widetilde{\mathsf{u}}_{j'}\right)1_{i'}\left(\widetilde{X}_{i'}=\widetilde{\msf{x}}\right)\right\} P_{\widetilde{X}\left|\widetilde{U}\right.}\left(\widetilde{\msf{x}}\left|\widetilde{\mathsf{u}}_{j}\right.\right)}{\Pr\left\{ \left(\widetilde{X}^{n},\,\widetilde{U}^{n}\right)\in\mathcal{A}_{\varepsilon}\right\} n^{2}\left(P_{\widetilde{X}}\left(\mathsf{x}\right)-\varepsilon\right)^{2}}\\\nonumber
&-\frac{\sum_{i=1}^{n}\sum_{i=1,i'\neq i}^{n}E\left\{ 1_{i}\left(\widetilde{V}_{i}=\widetilde{\mathsf{v}}_{k}\right)1_{i}\left(\widetilde{U}_{i}=\widetilde{\mathsf{u}}_{j}\right)1_{i'}\left(\widetilde{V}_{i'}=\widetilde{\mathsf{v}}_{k}\right)1_{i'}\left(\widetilde{U}_{i'}=\widetilde{\mathsf{u}}_{j'}\right)1_{i}\left(\widetilde{X}_{i}=\widetilde{\msf{x}}\right)\right\} P_{\widetilde{X}\left|\widetilde{U}\right.}\left(\widetilde{\msf{x}}\left|\widetilde{\mathsf{u}}_{j'}\right.\right)}{\Pr\left\{ \left(\widetilde{X}^{n},\,\widetilde{U}^{n}\right)\in\mathcal{A}_{\varepsilon}\right\} n^{2}\left(P_{\widetilde{X}}\left(\mathsf{x}\right)-\varepsilon\right)^{2}}\\\nonumber
&+\frac{\sum_{i=1}^{n}\sum_{i=1,i'\neq i}^{n}E\left\{ 1_{i}\left(\widetilde{V}_{i}=\widetilde{\mathsf{v}}_{k}\right)1_{i}\left(\widetilde{U}_{i}=\widetilde{\mathsf{u}}_{j}\right)1_{i'}\left(\widetilde{V}_{i'}=\widetilde{\mathsf{v}}_{k}\right)1_{i'}\left(\widetilde{U}_{i'}=\widetilde{\mathsf{u}}_{j'}\right)\right\} P_{\widetilde{X}\left|\widetilde{U}\right.}\left(\widetilde{\msf{x}}\left|\widetilde{\mathsf{u}}_{j}\right.\right)P_{\widetilde{X}\left|\widetilde{U}\right.}\left(\widetilde{\msf{x}}\left|\widetilde{\mathsf{u}}_{j'}\right.\right)}{\Pr\left\{ \left(\widetilde{X}^{n},\,\widetilde{U}^{n}\right)\in\mathcal{A}_{\varepsilon}\right\} n^{2}\left(P_{\widetilde{X}}\left(\mathsf{x}\right)-\varepsilon\right)^{2}}\\\label{upper5}
&=\frac{1}{n}+\frac{\sum_{i=1}^{n}\sum_{i=1,i'\neq i}^{n}P_{\widetilde{V}_{i}, \widetilde{V}_{i'},\widetilde{U}_{i},\widetilde{U}_{i'}}\left\{ \widetilde{\mathsf{v}}_{k},\, \widetilde{\mathsf{v}}_{k},\,\widetilde{\mathsf{u}}_{j},\,\widetilde{\mathsf{u}}_{j}\right\} \triangle F_{i,i'j,j'}}{\Pr\left\{ \left(\widetilde{X}^{n},\,\widetilde{U}^{n}\right)\in\mathcal{A}_{\varepsilon}\right\} n^{2}\left(P_{\widetilde{X}}\left(\mathsf{x}\right)-\varepsilon\right)^{2}}
\end{align}}}Substituting (\ref{upper5}) into (\ref{upper4}), we have
{\small{\begin{align}
&\nonumber\Pr\left\{ \left|\sum_{j=1}^{n'}H_{j}\right|>\frac{\mu}{2}\left|\left(\widetilde{X}^{n},\,\widetilde{U}^{n}\right)\in\mathcal{A}_{\varepsilon}\right.\right\} <\frac{4}{\mu^{2}}\sum_{j=1}^{n'}\sum_{j'=1}^{n'}\left\{ \frac{1}{n}+\frac{\sum_{i=1}^{n}\sum_{i=1,i'\neq i}^{n}P_{V_{i},V_{i'},\widetilde{U}_{i},\widetilde{U}_{i'}}\left\{ \widetilde{\mathsf{v}}_{k},\, \widetilde{\mathsf{v}}_{k},\,\widetilde{\mathsf{u}}_{j},\,\widetilde{\mathsf{u}}_{j}\right\} }{\Pr\left\{ \left(\widetilde{X}^{n},\,\widetilde{U}^{n}\right)\in\mathcal{A}_{\varepsilon}\right\} n^{2}\left(P_{\widetilde{X}}\left(\mathsf{x}\right)-\varepsilon\right)^{2}}\triangle F_{max}\left(n'\right)\right\} \\\nonumber
&\leq\frac{4}{\mu^{2}}\left(\frac{n'^{2}}{n}+\frac{\sum_{i=1}^{n}\sum_{i=1,i'\neq i}^{n}\sum_{j=1}^{n'}\sum_{j'=1}^{n'}P_{\widetilde{U}}\left\{ \widetilde{\mathsf{u}}_{j}\right\} P_{\widetilde{U}}\left\{ \widetilde{\mathsf{u}}_{j'}\right\} }{\Pr\left\{ \left(\widetilde{X}^{n},\,\widetilde{U}^{n}\right)\in\mathcal{A}_{\varepsilon}\right\} n^{2}\left(P_{\widetilde{X}}\left(\mathsf{x}\right)-\varepsilon\right)^{2}}\triangle F_{max}\left(n'\right)\right)\\\label{upper6}
&\leq \frac{4}{\mu^{2}}\left(\frac{n'^{2}}{n}+\frac{1}{\Pr\left\{ \left(\widetilde{X}^{n},\,\widetilde{U}^{n}\right)\in\mathcal{A}_{\varepsilon}\right\} \left(P_{\widetilde{X}}\left(\mathsf{x}\right)-\varepsilon\right)^{2}}\triangle F_{max}\left(n'\right)\right)
\end{align}}}From (\ref{upper6}) (\ref{upper3}) (\ref{upper2}) and (\ref{upper}), we have 
\begin{align}
&\nonumber \Pr\left\{ \left|\triangle F_{V^{n}\left|\widetilde{X}^{n}\right.}^{n}\left(\widetilde{\mathsf{v}}_{k}\left|\widetilde{\msf{x}}\right.\right)-\int_{-\infty}^{+\infty}f_{U\left|\widetilde{X}\right.}\left(u\left|\widetilde{\msf{x}}\right.\right)\triangle F^{(n')}_{V^{n}\left|\widetilde{U}^{n}\right.}\left(\widetilde{\mathsf{v}}_{k}\left|u\right.\right)du\right|>\mu\right\} \\\label{final_upper}
&<\frac{4}{\mu^{2}}\left(\frac{n'^{2}}{n}+\frac{1}{\Pr\left\{ \left(\widetilde{X}^{n},\,\widetilde{U}^{n}\right)\in\mathcal{A}_{\varepsilon}\right\} \left(P_{\widetilde{X}}\left(\mathsf{x}\right)-\varepsilon\right)^{2}}\triangle F_{max}\left(n'\right)\right)+\Pr\left\{ \left(\widetilde{X}^{n},\,\widetilde{U}^{n}\right)\notin \mathcal{A}_{\varepsilon}\right\}
\end{align}The proof is finished.
\end{IEEEproof}
Upon the aforementioned lemmas, the following assertion convergence property can be proved. 
\begin{lemma}\label{Alem2}
For arbitrary $t$, sufficiently small $\mu$ and $\varepsilon\leq\frac{\mu}{4n'n_v}$,
{\small{\begin{align}
\Pr\left\{ \left|F_{Y^{n}\left|\widetilde{X}^{n}\right.}^{n}\left(t\left|\widetilde{\msf{x}}\right.\right)-\sum_{k=1}^{n_{v}}\sum_{j=1}^{n'}P_{\widetilde{U}\left|\widetilde{X}\right.}\left(\widetilde{\mathsf{u}}_{j}\left|\widetilde{\msf{x}}\right.\right)\triangle F_{\widetilde{V}^{n}\left|\widetilde{U}^{n}\right.}^{(n')}\left(\widetilde{\mathsf{v}}_{k}\left|\widetilde{\mathsf{u}}_{j}\right.\right)F_{Y\left|V\right.}\left(t\left|\overline{\mathsf{v}}_{k}\right.\right)\right|>\mu\right\}  
\end{align}}}
\end{lemma}
\begin{IEEEproof}
Notice that 
{\small{\begin{align}\label{fupper1}
&\Pr\left\{ \left|F_{Y^{n}\left|\widetilde{X}^{n}\right.}^{n}\left(t\left|\widetilde{\msf{x}}\right.\right)-\sum_{k=1}^{n_{v}}\sum_{j=1}^{n'}P_{\widetilde{U}\left|\widetilde{X}\right.}\left(\widetilde{\mathsf{u}}_{j}\left|\widetilde{\msf{x}}\right.\right)\triangle F_{\widetilde{V}^{n}\left|\widetilde{U}^{n}\right.}^{(n')}\left(\widetilde{\mathsf{v}}_{k}\left|\widetilde{\mathsf{u}}_{j}\right.\right)F_{Y\left|V\right.}\left(t\left|\overline{\mathsf{v}}_{k}\right.\right)\right|>\mu\right\}< \\\nonumber &\hspace{-20pt}
\Pr\left\{ \left|F_{Y^{n}\left|\widetilde{X}^{n}\right.}^{n}\left(t\left|\widetilde{\msf{x}}\right.\right)-\sum_{k=1}^{n_{v}}\triangle F_{\widetilde{V}^{n}\left|\widetilde{X}^{n}\right.}^{n}\left(\widetilde{\mathsf{v}}_{k}\left|\widetilde{\msf{x}}\right.\right)F_{Y\left|V\right.}\left(t\left|\overline{\mathsf{v}}_{k}\right.\right)\right|>\frac{\mu}{2}\right\} +\Pr\left\{ \sum_{k=1}^{n_{v}}\left|\sum_{j=1}^{n'}P_{\widetilde{U}\left|\widetilde{X}\right.}\left(\widetilde{\mathsf{u}}_{j}\left|\widetilde{\msf{x}}\right.\right)\triangle F_{\widetilde{V}^{n}\left|\widetilde{U}^{n}\right.}^{(n')}\left(\widetilde{\mathsf{v}}_{k}\left|\widetilde{\mathsf{u}}_{j}\right.\right)-\triangle F_{\widetilde{V}^{n}\left|\widetilde{X}^{n}\right.}^{n}\left(\widetilde{\mathsf{v}}_{k}\left|\widetilde{\msf{x}}\right.\right)\right|>\frac{\mu}{2}\right\}. 
\end{align}}}
According to Lemma \ref{lem2}, the second item in the right side of (\ref{fupper1}) can be bound as 
{\small{\begin{align}\label{fupper2}
\nonumber&\Pr\left\{ \sum_{k=1}^{n_{v}}\left|\sum_{j=1}^{n'}P_{\widetilde{U}\left|\widetilde{X}\right.}\left(\widetilde{\mathsf{u}}_{j}\left|\widetilde{\msf{x}}\right.\right)\triangle F_{\widetilde{V}^{n}\left|\widetilde{U}^{n}\right.}^{(n')}\left(\widetilde{\mathsf{v}}_{k}\left|\widetilde{\mathsf{u}}_{j}\right.\right)-\triangle F_{\widetilde{V}^{n}\left|\widetilde{X}^{n}\right.}^{n}\left(\widetilde{\mathsf{v}}_{k}\left|\widetilde{\msf{x}}\right.\right)\right|>\frac{\mu}{2}\right\}< \\
&\frac{16}{\mu^{2}}\left(\frac{n'^{2}n_{v}^{2}}{n}+\frac{n_{v}^{2}}{\Pr\left\{ \left(\widetilde{X}^{n},\,\widetilde{U}^{n}\right)\in\mathcal{A}_{\varepsilon}\right\} \left(P_{\widetilde{X}}\left(\mathsf{x}\right)-\varepsilon\right)^{2}}\triangle F_{max}\left(n'\right)\right)+\Pr\left\{ \left(\widetilde{X}^{n},\,\widetilde{U}^{n}\right)\notin T_{\left[\widetilde{X},\widetilde{U}\right]_{\varepsilon}}^{n}\right\}. 
\end{align}}}Then, we just need to focus on the first item in the right side of (\ref{fupper1}). 
Following the similar steps employed in the proof of (\ref{lem2}), we have 
{\small{\begin{align}\label{fupper3}
&\nonumber \Pr\left\{ \left|F_{Y^{n}\left|\widetilde{X}^{n}\right.}^{n}\left(t\left|\widetilde{\msf{x}}\right.\right)-\sum_{k=1}^{n_{v}}\triangle F_{\widetilde{V}^{n}\left|\widetilde{X}^{n}\right.}^{n}\left(\widetilde{\mathsf{v}}_{k}\left|\widetilde{\msf{x}}\right.\right)F_{Y\left|V\right.}\left(t\left|\overline{\mathsf{v}}_{k}\right.\right)\right|>\frac{\mu}{2}\right\}<\\
&\hspace{-20pt}\Pr\left\{ \left|\sum_{k=1}^{n_{v}}{\left(\frac{\sum_{i=1}^{n}1_{i}\left(\widetilde{V}_{i}=\widetilde{\mathsf{v}}_{k}\right)1_{i}\left(\widetilde{X}_{i}=\widetilde{\msf{x}}\right)1_{i}\left(Y_{i}<t\right)}{N\left(\widetilde{\msf{x}}\left|X^{n}\right.\right)}-\frac{\sum_{i=1}^{n}1_{i}\left(\widetilde{V}_{i}=\widetilde{\mathsf{v}}_{k}\right)1_{i}\left(\widetilde{X}_{i}=\widetilde{\msf{x}}\right)}{N\left(\widetilde{\msf{x}}\left|X^{n}\right.\right)}F_{Y\left|V\right.}\left(t\left|\overline{\mathsf{v}}_{k}\right.\right)\right)}\right|>\frac{\mu}{2}\left|\widetilde{X}^{n}\in T_{\left[\widetilde{X}\right]_{\varepsilon}}^{n}\right.\right\} +\Pr\left\{ \widetilde{X}^{n}\notin T_{\left[\widetilde{X}\right]_{\varepsilon}}^{n}\right\}, 
\end{align}}}where {\small{$\Pr\left\{ \widetilde{X}^{n}\notin T_{\left[\widetilde{X}\right]_{\varepsilon}}^{n}\right\} \rightarrow0$}} as $n$ approaches to infinity. Then, we focus on bounding the first item in the right side of (\ref{fupper3}). 
Following the Chebyshev theorem, we attain 
{\small{\begin{align}\label{fupper4}
&\nonumber \Pr\left\{ \left|\sum_{k=1}^{n_{v}}\frac{\sum_{i=1}^{n}1_{i}\left(\widetilde{V}_{i}=\widetilde{\mathsf{v}}_{k}\right)1_{i}\left(\widetilde{X}_{i}=\widetilde{\msf{x}}\right)1_{i}\left(Y_{i}<t\right)}{N\left(\widetilde{\msf{x}}\left|X^{n}\right.\right)}-\frac{\sum_{i=1}^{n}1_{i}\left(\widetilde{V}_{i}=\widetilde{\mathsf{v}}_{k}\right)1_{i}\left(\widetilde{X}_{i}=\widetilde{\msf{x}}\right)}{N\left(\widetilde{\msf{x}}\left|X^{n}\right.\right)}F_{Y\left|V\right.}\left(t\left|\overline{\mathsf{v}}_{k}\right.\right)\right|>\frac{\mu}{2}\left|\widetilde{X}^{n}\in T_{\left[\widetilde{X}\right]_{\varepsilon}}^{n}\right.\right\}< \\\nonumber
&\frac{1}{\Pr\left\{ \widetilde{X}^{n}\in T_{\left[\widetilde{X}\right]_{\varepsilon}}^{n}\right\} }\Pr\left\{ \left|\sum_{k=1}^{n_{v}}\underbrace{\left(\frac{\sum_{i=1}^{n}1_{i}\left(\widetilde{V}_{i}=\widetilde{\mathsf{v}}_{k}\right)1_{i}\left(\widetilde{X}_{i}=\widetilde{\msf{x}}\right)1_{i}\left(Y_{i}<t\right)}{n\left(P_{\widetilde{X}}\left(\widetilde{\msf{x}}\right)-\varepsilon\right)}-\frac{\sum_{i=1}^{n}1_{i}\left(\widetilde{V}_{i}=\widetilde{\mathsf{v}}_{k}\right)1_{i}\left(\widetilde{X}_{i}=\widetilde{\msf{x}}\right)}{n\left(P_{\widetilde{X}}\left(\widetilde{\msf{x}}\right)-\varepsilon\right)}F_{Y\left|V\right.}\left(t\left|\overline{\mathsf{v}}_{k}\right.\right)\right)}_{H_{k}}\right|>\frac{\mu}{2}\right\} \\
&\leq\frac{4}{\mu^{2}}\frac{1}{\Pr\left\{ \widetilde{X}^{n}\in T_{\left[\widetilde{X}\right]_{\varepsilon}}^{n}\right\} }\sum_{k=1}^{n_{v}}\sum_{k'=1}^{n_{v}}E\left(H_{k}H_{k'}\right). 
\end{align}}}
{\small{\begin{align}\label{fupper5}
&\nonumber E\left(H_{k}H_{k'}\right)=\frac{E\left(\sum_{i=1}^{n}1_{i}\left(\widetilde{V}_{i}=\widetilde{\mathsf{v}}_{k}\right)1_{i}\left(\widetilde{X}_{i}=\widetilde{\msf{x}}\right)\left(1_{i}\left(Y_{i}<t\right)-F_{Y\left|V\right.}\left(t\left|\overline{\mathsf{v}}_{k}\right.\right)\right)\right)\left(\sum_{i=1}^{n}1_{i}\left(\widetilde{V}_{i}=\widetilde{\mathsf{v}}_{k}\right)1_{i}\left(\widetilde{X}_{i}=\widetilde{\msf{x}}\right)\left(1_{i}\left(Y_{i}<t\right)-F_{Y\left|V\right.}\left(t\left|\overline{\mathsf{v}}_{k'}\right.\right)\right)\right)}{n^{2}\left(P_{\widetilde{X}}\left(\widetilde{\msf{x}}\right)-\varepsilon\right)^{2}}\\&\nonumber 
<\frac{1}{n\left(P_{\widetilde{X}}\left(\widetilde{\msf{x}}\right)-\varepsilon\right)^{2}}+\\&\nonumber 
\hspace{30pt}\frac{E\left(\sum_{i=1}^{n}\sum_{i\text{'}=1,i\neq i}^{n}1_{i}\left(\widetilde{V}_{i}=\widetilde{\mathsf{v}}_{k}\right)1_{i}\left(\widetilde{X}_{i}=\widetilde{\msf{x}}\right)\left(1_{i}\left(Y_{i}<t\right)-F_{Y\left|V\right.}\left(t\left|\overline{\mathsf{v}}_{k}\right.\right)\right)1_{i'}\left(\widetilde{V}_{i'}=\widetilde{\mathsf{v}}_{k'}\right)1_{i}\left(\widetilde{X}_{i'}=\widetilde{\msf{x}}\right)\left(1_{i'}\left(Y_{i'}<t\right)-F_{Y\left|V\right.}\left(t\left|\overline{\mathsf{v}}_{k}\right.\right)\right)\right)}{n^{2}\left(P_{\widetilde{X}}\left(\widetilde{\msf{x}}\right)-\varepsilon\right)^{2}}\\&\nonumber 
<\frac{5}{n\left(P_{\widetilde{X}}\left(\widetilde{\msf{x}}\right)-\varepsilon\right)^{2}}+\frac{\sum_{i=2}^{n-1}\sum_{i\text{'}=2,i\neq i}^{n-1}E\left(1_{i}\left(\widetilde{V}_{i}=\widetilde{\mathsf{v}}_{k}\right)1_{i}\left(\widetilde{X}_{i}=\widetilde{\msf{x}}\right)1_{i'}\left(\widetilde{V}_{i'}=\widetilde{\mathsf{v}}_{k'}\right)1_{i}\left(\widetilde{X}_{i'}=\widetilde{\msf{x}}\right)1_{i}\left(Y_{i}<t\right)1_{i'}\left(Y_{i'}<t\right)\right)}{n^{2}\left(P_{\widetilde{X}}\left(\widetilde{\msf{x}}\right)-\varepsilon\right)^{2}}\\&\nonumber
-\frac{\sum_{i=2}^{n-1}\sum_{i\text{'}=2,i\neq i}^{n-1}E\left(1_{i}\left(\widetilde{V}_{i}=\widetilde{\mathsf{v}}_{k}\right)1_{i}\left(\widetilde{X}_{i}=\widetilde{\msf{x}}\right)1_{i'}\left(\widetilde{V}_{i'}=\widetilde{\mathsf{v}}_{k'}\right)1_{i}\left(\widetilde{X}_{i'}=\widetilde{\msf{x}}\right)1_{i}\left(Y_{i}<t\right)F_{Y\left|V\right.}\left(t\left|\overline{\mathsf{v}}_{k'}\right.\right)\right)}{n^{2}\left(P_{\widetilde{X}}\left(\widetilde{\msf{x}}\right)-\varepsilon\right)^{2}}\\&\nonumber
-\frac{\sum_{i=2}^{n-1}\sum_{i\text{'}=2,i\neq i}^{n-1}E\left(1_{i}\left(\widetilde{V}_{i}=\widetilde{\mathsf{v}}_{k}\right)1_{i}\left(\widetilde{X}_{i}=\widetilde{\msf{x}}\right)1_{i'}\left(\widetilde{V}_{i'}=\widetilde{\mathsf{v}}_{k'}\right)1_{i}\left(\widetilde{X}_{i'}=\widetilde{\msf{x}}\right)F_{Y\left|V\right.}\left(t\left|\overline{\mathsf{v}}_{k}\right.\right)1_{i'}\left(Y_{i'}<t\right)\right)}{n^{2}\left(P_{\widetilde{X}}\left(\widetilde{\msf{x}}\right)-\varepsilon\right)^{2}}\\&\nonumber
+\frac{\sum_{i=2}^{n-1}\sum_{i\text{'}=2,i\neq i}^{n-1}E\left(1_{i}\left(\widetilde{V}_{i}=\widetilde{\mathsf{v}}_{k}\right)1_{i}\left(\widetilde{X}_{i}=\widetilde{\msf{x}}\right)1_{i'}\left(\widetilde{V}_{i'}=\widetilde{\mathsf{v}}_{k'}\right)1_{i}\left(\widetilde{X}_{i'}=\widetilde{\msf{x}}\right)F_{Y\left|V\right.}\left(t\left|\overline{\mathsf{v}}_{k}\right.\right)F_{Y\left|V\right.}\left(t\left|\overline{\mathsf{v}}_{k'}\right.\right)\right)}{n^{2}\left(P_{\widetilde{X}}\left(\widetilde{\msf{x}}\right)-\varepsilon\right)^{2}}\\&\nonumber
<\frac{5}{n\left(P_{\widetilde{X}}\left(\widetilde{\msf{x}}\right)-\varepsilon\right)^{2}}+\frac{\sum_{i=2}^{n-1}\sum_{i\text{'}=2,i\neq i}^{n-1}P_{\widetilde{V}_{i},\widetilde{V}_{i'},X_{1,i},X_{1,i'}}\left(\widetilde{\mathsf{v}}_{k},\widetilde{\mathsf{v}}_{k'},\widetilde{\msf{x}},\widetilde{\msf{x}}\right)\triangle G_{max}}{n^{2}\left(P_{\widetilde{X}}\left(\widetilde{\msf{x}}\right)-\varepsilon\right)^{2}}\\& 
<\frac{5}{n\left(P_{\widetilde{X}}\left(\widetilde{\msf{x}}\right)-\varepsilon\right)^{2}}+\frac{P_{\widetilde{V}_{i},\widetilde{V}_{i'},X_{1,i},X_{1,i'}}\left(\widetilde{\mathsf{v}}_{k},\widetilde{\mathsf{v}}_{k'},\widetilde{\msf{x}},\widetilde{\msf{x}}\right)\triangle G_{max}}{\left(P_{\widetilde{X}}\left(\widetilde{\msf{x}}\right)-\varepsilon\right)^{2}}
\end{align}}}Substituting (\ref{fupper5}) into (\ref{fupper4}), we get 
{\small{\begin{align}\label{fupper6}
&\nonumber \Pr\left\{ \left|\sum_{k=1}^{n_{v}}\frac{\sum_{i=1}^{n}1_{i}\left(\widetilde{V}_{i}=\widetilde{\mathsf{v}}_{k}\right)1_{i}\left(\widetilde{X}_{i}=\widetilde{\msf{x}}\right)1_{i}\left(Y_{i}<t\right)}{N\left(\widetilde{\msf{x}}\left|X^{n}\right.\right)}-\frac{\sum_{i=1}^{n}1_{i}\left(\widetilde{V}_{i}=\widetilde{\mathsf{v}}_{k}\right)1_{i}\left(\widetilde{X}_{i}=\widetilde{\msf{x}}\right)}{N\left(\widetilde{\msf{x}}\left|X^{n}\right.\right)}F_{Y\left|V\right.}\left(t\left|\overline{\mathsf{v}}_{k}\right.\right)\right|>\frac{\mu}{2}\left|\widetilde{X}^{n}\in T_{\left[\widetilde{X}\right]_{\varepsilon}}^{n}\right.\right\} \\
&<\frac{4}{\mu^{2}}\frac{1}{\Pr\left\{ \widetilde{X}^{n}\in T_{\left[\widetilde{X}\right]_{\varepsilon}}^{n}\right\} }\left(\frac{5n_{v}^{2}}{n\left(P_{\widetilde{X}}\left(\widetilde{\msf{x}}\right)-\varepsilon\right)^{2}}+\frac{\triangle G_{max}}{\left(P_{\widetilde{X}}\left(\widetilde{\msf{x}}\right)-\varepsilon\right)^{2}}\right)
\end{align}}}From (\ref{fupper6}), (\ref{fupper1}), (\ref{fupper2}), and (\ref{fupper3}), we finally obtain
{\small{\begin{align*}
&\Pr\left\{ \left|F_{Y^{n}\left|\widetilde{X}^{n}\right.}^{n}\left(t\left|\widetilde{\msf{x}}\right.\right)-\sum_{k=1}^{n_{v}}\sum_{j=1}^{n'}P_{\widetilde{U}\left|\widetilde{X}\right.}\left(\widetilde{\mathsf{u}}_{j}\left|\widetilde{\msf{x}}\right.\right)\triangle F_{\widetilde{V}^{n}\left|\widetilde{U}^{n}\right.}^{(n')}\left(\widetilde{\mathsf{v}}_{k}\left|\widetilde{\mathsf{u}}_{j}\right.\right)F_{Y\left|V\right.}\left(t\left|\overline{\mathsf{v}}_{k}\right.\right)\right|>\mu\right\}\\&
<\frac{4}{\mu^{2}}\frac{1}{\Pr\left\{ \widetilde{X}^{n}\in T_{\left[\widetilde{X}\right]_{\varepsilon}}^{n}\right\} }\left(\frac{5n_{v}^{2}}{n\left(P_{\widetilde{X}}\left(\widetilde{\msf{x}}\right)-\varepsilon\right)^{2}}+\frac{\triangle G_{max}}{\left(P_{\widetilde{X}}\left(\widetilde{\msf{x}}\right)-\varepsilon\right)^{2}}\right)+\Pr\left\{ \widetilde{X}^{n}\notin T_{\left[\widetilde{X}\right]_{\varepsilon}}^{n}\right\}\\&
+\frac{16}{\mu^{2}}\left(\frac{n'^{2}n_{v}^{2}}{n}+\frac{n_{v}^{2}}{\Pr\left\{ \left(\widetilde{X}^{n},\,\widetilde{U}^{n}\right)\in\mathcal{A}_{\varepsilon}\right\} \left(P_{\widetilde{X}}\left(\mathsf{x}\right)-\varepsilon\right)^{2}}\triangle F_{max}\left(n'\right)\right)+\Pr\left\{ \left(\widetilde{X}^{n},\,\widetilde{U}^{n}\right)\notin T_{\left[\widetilde{X},\widetilde{U}\right]_{\varepsilon}}^{n}\right\}
\end{align*}}}
The proof is finished.
\end{IEEEproof}
Upon the convergence property, the following lemma is immediate. 
\begin{lemma}\label{lem3}
For sequence $t_{1}, t_{2}, \ldots, t_{n_y-1}$, upon setup that $\beta_{3}=-\alpha_{3}=\sqrt{n_{y}}$, $n_{y}^{2}=\sqrt{n_{v}}$, and $n_{v}^{2}=\sqrt{n'}$, we have
\begin{enumerate}
\item Fix $\mu$ to arbitrary small value, there has
{\small{\begin{equation}
\lim_{n\rightarrow\infty,n'\rightarrow\infty}\Pr\left\{ \frac{\beta_{3}-\alpha_{3}}{n_{y}-2}\frac{\beta_{4}-\alpha_{4}}{n_{x}-2}\sum_{i=1}^{n_{x}-1}\sum_{m=1}^{n_{y}-1}\left|F_{Y^{n}\left|\widetilde{X}^{n}\right.}^{n}\left(t_{m}\left|\widetilde{\msf{x}}_i\right.\right)-\sum_{k=1}^{n_{v}}\sum_{j=1}^{n'}P_{\widetilde{U}\left|\widetilde{X}\right.}\left(\widetilde{\mathsf{u}}_{j}\left|\widetilde{\msf{x}}_i\right.\right)\triangle F_{\widetilde{V}^{n}\left|\widetilde{U}^{n}\right.}^{(n')}\left(\widetilde{\mathsf{v}}_{k}\left|\widetilde{\mathsf{u}}_{j}\right.\right)F_{Y\left|V\right.}\left(t_m\left|\overline{\mathsf{v}}_{k}\right.\right)\right|>\mu\right\} =0.
\end{equation}}}
\item Fix $n'$ to arbitrary large value, and $\epsilon$ to arbitrary small value, there has
{\small{\begin{equation}
\Pr\left\{ \frac{\beta_{3}-\alpha_{3}}{n_{y}-2}\frac{\beta_{4}-\alpha_{4}}{n_{x}-2}\sum_{i=1}^{n_{x}-1}\sum_{m=1}^{n_{y}-1}\left|F_{Y^{n}\left|\widetilde{X}^{n}\right.}^{n}\left(t_m\left|\widetilde{\msf{x}}_{i}\right.\right)-\sum_{k=1}^{n_{v}}\sum_{j=1}^{n'}P_{\widetilde{U}\left|\widetilde{X}\right.}\left(\widetilde{\mathsf{u}}_{j}\left|\widetilde{\msf{x}}_i\right.\right)\triangle F_{\widetilde{V}^{n}\left|\widetilde{U}^{n}\right.}^{(n')}\left(\widetilde{\mathsf{v}}_{k}\left|\widetilde{\mathsf{u}}_{j}\right.\right)F_{Y\left|V\right.}\left(t_m\left|\overline{\mathsf{v}}_{k}\right.\right)\right|>\mu_{n'}\right\} \leq\epsilon
\end{equation}}}where $n$ approaches to infinity, $\lim_{n'\rightarrow\infty}\mu_{n'}=0$.
 \end{enumerate}
\end{lemma}
\begin{IEEEproof}
For arbitrary small $\mu$, we have
{\small{\begin{align}\label{noname}
&\Pr\left\{ \frac{\beta_{3}-\alpha_{3}}{n_{y}-2}\frac{\beta_{4}-\alpha_{4}}{n_{x}-2}\sum_{i=1}^{n_{x}-1}\sum_{m=1}^{n_{y}-1}\left|F_{Y^{n}\left|\widetilde{X}^{n}\right.}^{n}\left(t_{m}\left|\widetilde{\msf{x}}_i\right.\right)-\sum_{k=1}^{n_{v}}\sum_{j=1}^{n'}P_{\widetilde{U}\left|\widetilde{X}\right.}\left(\widetilde{\mathsf{u}}_{j}\left|\widetilde{\msf{x}}_i\right.\right)\triangle F_{\widetilde{V}^{n}\left|\widetilde{U}^{n}\right.}^{(n')}\left(\widetilde{\mathsf{v}}_{k}\left|\widetilde{\mathsf{u}}_{j}\right.\right)F_{Y\left|V\right.}\left(t_m\left|\overline{\mathsf{v}}_{k}\right.\right)\right|>\mu\right\} \\
&\nonumber\leq\sum_{i=1}^{n_{x}-1}\sum_{m=1}^{n_y-1}\Pr\left\{ \left|F_{Y^{n}\left|\widetilde{X}^{n}\right.}^{n}\left(t_{m}\left|\widetilde{\msf{x}}_{i}\right.\right)-\sum_{k=1}^{n_{v}}\sum_{j=1}^{n'}P_{\widetilde{U}\left|\widetilde{X}\right.}\left(\widetilde{\mathsf{u}}_{j}\left|\widetilde{\msf{x}}_{i}\right.\right)\triangle F_{\widetilde{V}^{n}\left|\widetilde{U}^{n}\right.}^{(n')}\left(\widetilde{\mathsf{v}}_{k}\left|\widetilde{\mathsf{u}}_{j}\right.\right)F_{Y\left|V\right.}\left(t\left|\overline{\mathsf{v}}_{k}\right.\right)\right|>\frac{\mu(n_y-2)(n_x-2)}{(\beta_{3}-\alpha_{3})(\beta_{4}-\alpha_{4})(n_y-1)(n_x-1)}\right\} \\
&\nonumber< \frac{4(n_{y}-1)^{2}(n_{x}-1)^{2}}{\mu^{2}(n_{y}-2)^{2}(n_{x}-2)^{2}}\frac{1}{\Pr\left\{ \widetilde{X}^{n}\in T_{\left[\widetilde{X}\right]_{\varepsilon}}^{n}\right\} }\left(\frac{5n_{v}^{2}(\beta_{3}-\alpha_{3})^{2}(\beta_{4}-\alpha_{4})^{2}n_{y}n_{x}}{n\left(P_{\widetilde{X}}\left(\widetilde{\msf{x}}\right)-\varepsilon\right)^{2}}+\frac{n_{y}n_{x}(\beta_{3}-\alpha_{3})^{2}(\beta_{4}-\alpha_{4})^{2}\triangle G_{max}}{\left(P_{\widetilde{X}}\left(\widetilde{\msf{x}}\right)-\varepsilon\right)^{2}}\right)+n_{x}n_{y}\Pr\left\{ \widetilde{X}^{n}\notin T_{\left[\widetilde{X}\right]_{\varepsilon}}^{n}\right\} \\
&\hspace{-40pt}\nonumber +\frac{16(n_{y}-1)^{2}(n_{x}-1)^{2}}{\mu^{2}(n_{y}-2)^{2}(n_{x}-2)^{2}}\left(\frac{n'^{2}n_{v}^{2}(\beta_{3}-\alpha_{3})^{2}(\beta_{4}-\alpha_{4})^{2}n_{x}n_{y}}{n}+\frac{n_{v}^{2}(\beta_{3}-\alpha_{3})^{2}(\beta_{4}-\alpha_{4})^{2}n_{x}n_{y}}{\Pr\left\{ \left(\widetilde{X}^{n},\,\widetilde{U}^{n}\right)\in\mathcal{A}_{\varepsilon}\right\} \left(P_{\widetilde{X}}\left(\mathsf{x}\right)-\varepsilon\right)^{2}}\triangle F_{max}\left(n'\right)\right)+n_{x}n_{y}\Pr\left\{ \left(\widetilde{X}^{n},\,\widetilde{U}^{n}\right)\notin T_{\left[\widetilde{X},\widetilde{U}\right]_{\varepsilon}}^{n}\right\} 
\end{align}}}where the last inequality follows Lemma \ref{Alem2}. 
Upon the setup that $\beta_{3}=-\alpha_{3}=\sqrt{n_{y}}$, $n_{y}^{2}=\sqrt{n_{v}}$, and $n_{v}^{2}=\sqrt{n'}$,
we have 
$$n_{v}^{2}(\beta_{3}-\alpha_{3})^{2}(\beta_{4}-\alpha_{4})^{2}n_{x}n_{y}<{n'}^{1/2},$$
Then, $n_{v}^{2}(\beta_{3}-\alpha_{3})^{2}(\beta_{4}-\alpha_{4})^{2}n_{x}n_{y}\triangle F_{max}\left(n'\right)\rightarrow0$ as $n'\rightarrow\infty$
according to Lemma \ref{lem1}, and 
$$n_{y}n_{x}(\beta_{3}-\alpha_{3})^{2}(\beta_{4}-\alpha_{4})^{2}<{n_v}^{1/2}$$
$n_{y}n_{x}(\beta_{3}-\alpha_{3})^{2}(\beta_{4}-\alpha_{4})^{2}\triangle G_{max}\rightarrow0$ as $n'\rightarrow\infty$ according to Lemma \ref{lemA1}. With these results, 
the upper bound of (\ref{noname}) approaches to 0, as $n'\rightarrow\infty$, $n\rightarrow\infty$. Hence, the first
assertion is proved.
Furthermore, again relying on (\ref{noname}), let us set 
{\small{\begin{align*}
\hspace{-100pt}&\epsilon= \frac{4(n_{y}-1)^{2}(n_{x}-1)^{2}}{\mu_{n'}^{2}(n_{y}-2)^{2}(n_{x}-2)^{2}}\frac{1}{\Pr\left\{ \widetilde{X}^{n}\in T_{\left[\widetilde{X}\right]_{\varepsilon}}^{n}\right\} }\left(\frac{5n_{v}^{2}(\beta_{3}-\alpha_{3})^{2}(\beta_{4}-\alpha_{4})^{2}n_{y}n_{x}}{n\left(P_{\widetilde{X}}\left(\widetilde{\msf{x}}\right)-\varepsilon\right)^{2}}+\frac{n_{y}n_{x}(\beta_{3}-\alpha_{3})^{2}(\beta_{4}-\alpha_{4})^{2}\triangle G_{max}}{\left(P_{\widetilde{X}}\left(\widetilde{\msf{x}}\right)-\varepsilon\right)^{2}}\right)+n_{x}n_{y}\Pr\left\{ \widetilde{X}^{n}\notin T_{\left[\widetilde{X}\right]_{\varepsilon}}^{n}\right\}  \\
&+\frac{16(n_{y}-1)^{2}(n_{x}-1)^{2}}{\mu_{n'}^{2}(n_{y}-2)^{2}(n_{x}-2)^{2}}\left(\frac{n'^{2}n_{v}^{2}(\beta_{3}-\alpha_{3})^{2}(\beta_{4}-\alpha_{4})^{2}n_{x}n_{y}}{n}+\frac{n_{v}^{2}(\beta_{3}-\alpha_{3})^{2}(\beta_{4}-\alpha_{4})^{2}n_{x}n_{y}}{\Pr\left\{ \left(\widetilde{X}^{n},\,\widetilde{U}^{n}\right)\in\mathcal{A}_{\varepsilon}\right\} \left(P_{\widetilde{X}}\left(\mathsf{x}\right)-\varepsilon\right)^{2}}\triangle F_{max}\left(n'\right)\right)+n_{x}n_{y}\Pr\left\{ \left(\widetilde{X}^{n},\,\widetilde{U}^{n}\right)\notin T_{\left[\widetilde{X},\widetilde{U}\right]_{\varepsilon}}^{n}\right\} 
\end{align*}}}As $\epsilon$ is fixed, and $n$ approaches to infinity, it is readily available to get $\lim_{n'\rightarrow\infty}\mu_{n'}=0$. The second statement can be proved.
 
\end{IEEEproof}

\subsection{Sufficiency Proof}
Let us go back to the proof of theorem 1. With the aforementioned lemmas, we will show the decision statistic {\small{$D^{n}=\frac{1}{{n_{x}-2}}\frac{1}{{n_{y}-2}}\sum_{k=1}^{{n_{x}-1}}\sum_{m=1}^{{n_{y}-1}}\left|F_{Y^{n}\left|\widetilde{X}^{n}\right.}^{n}\left(t_{m}\left|\widetilde{\mathsf{x}}_{k}\right.\right)-\int_{-\infty}^{+\infty}f_{U\left|\widetilde{X}\right.}\left(u\left|\widetilde{\mathsf{x}}_{k}\right.\right)F_{Y\left|V\right.}\left(t_{m}\left|u\right.\right)du\right|$}} simultaneously satisfies the properties stated by Theorem 1.

Upon $n'$, we define
function 
{\small{\begin{align}
&\nonumber M^{\left(n'\right)}\left(W^{\left(n'\right)}\right)=\frac{\beta_{4}-\alpha_{4}}{{n_{x}-2}}\frac{\beta_{3}-\alpha_{3}}{{n_{y}-2}}\\&\nonumber\sum_{k=1}^{{n_{x}-1}}\sum_{m=1}^{{n_{y}-1}}\left|\int_{-\infty}^{+\infty}f_{U\left|\widetilde{X}\right.}\left(u\left|\widetilde{\mathsf{x}}_k\right.\right)F_{Y\left|V\right.}\left(t_{m}\left|u\right.\right)du-\sum_{j=1}^{n_{v}}\sum_{i=1}^{n'}P_{\widetilde{U}\left|\widetilde{X}\right.}\left(\widetilde{\mathsf{u}}_{i}\right)w_{i,j,k}F_{Y\left|V\right.}\left(t_{m}\left|\overline{\mathsf{v}}_{m,j,k}\right.\right)\right|^{2}
\end{align}}}where $W^{\left(n'\right)}$ is a matrix variable $\left[W^{\left(n'\right)}\right]_{i,j,k}=w_{i,j,k}$, $i=1,2,\ldots,n'$, $j=1,2,\ldots,n_v$, 
 $k=1,2,\ldots,n_x$
and 
\begin{equation}
F_{Y\left|V\right.}\left(t_{m}\left|\overline{\mathsf{v}}_{m,j,k}\right.\right)=\frac{\int_{u\in\mathcal{B}\left(\widetilde{\mathsf{v}}_{j}\right)}f_{U\left|\widetilde{X}\right.}\left(u\left|\widetilde{\mathsf{x}}_k\right.\right)F_{Y\left|V\right.}\left(t_{m}\left|u\right.\right)du}{\int_{u\in\mathcal{B}\left(\widetilde{\mathsf{v}}_{j}\right)}f_{U\left|\widetilde{X}\right.}\left(u\left|\widetilde{\mathsf{x}}_k\right.\right)du}.
\end{equation}
According to the definition of $P_{\widetilde{U}\left|\widetilde{X}\right.}\left(\widetilde{\mathsf{u}}_{i}\right)$
and $F_{Y\left|V\right.}\left(t_{m}\left|\overline{\mathsf{v}}_{m,j,k}\right.\right)=\frac{\int_{u\in\mathcal{B}\left(\widetilde{\mathsf{v}}_{j}\right)}f_{U\left|\widetilde{X}\right.}\left(u\left|\widetilde{\mathsf{x}}_k\right.\right)F_{Y\left|V\right.}\left(t_{m}\left|u\right.\right)du}{\int_{u\in\mathcal{B}\left(\widetilde{\mathsf{v}}_{j}\right)}f_{U\left|\widetilde{X}\right.}\left(u\left|\widetilde{\mathsf{x}}_k\right.\right)du}$, 
$M^{\left(n'\right)}\left(W^{\left(n'\right)}\right)=0$ has one solution
in the point that $W_{0}^{\left(n'\right)}$ defined as {\small{$$\left[W_{0}^{\left(n'\right)}\right]_{i,j,k}=\begin{cases}
\begin{array}{cc}
1, & \mathcal{B}\left(\widetilde{\mathsf{u}}_{i}\right)\subseteq\mathcal{B}\left(\widetilde{\mathsf{v}}_{j}\right)\\
0, & otherwise
\end{array}\end{cases}
 $$}}$i=1,\ldots n', j=1,\ldots n'-1, k=1,2,\ldots,n_x$ in the domain $\mathcal{D}^{(n')}=\big\{ W^{(n')}:\,0\leq w_{i,j,k}\leq1,\sum_{j=1}^{n_{v}}w_{i,j,k}=1,j=1,2,\ldots n_{v},\, i=1,\ldots n',  k=1,2,\ldots,n_x\big\}$. We also define $\mathcal{D}_{s}^{(n')}$ as {\small{$\mathcal{D}_{s}^{(n')}=\big\{ W^{(n')}:\,\left|W^{(n')}-W_{0}^{(n')}\right|\geq\delta,\, W^{(n')}\in\mathcal{D}^{(n')}\big\}$}},
and the infimum of $M^{\left(n'\right)}\left(W^{\left(n'\right)}\right)$ over ${D}_{s}^{(n')}$ is denoted as $\lambda^{(n')}\left(\delta\right)$.
\begin{lemma}\label{lem5}
If $\lambda^{(n')}\left(\delta\right)$ is strictly positive, then $M^{\left(n'\right)}\left(W^{\left(n'\right)}\right)=0$ has single solution
in the point $W_{0}^{\left(n'\right)}$. Moreover, $\lambda^{(n')}\left(\delta\right)\rightarrow0,\delta\rightarrow0$.
\end{lemma}
\begin{IEEEproof}
Using the observation that $M^{\left(n'\right)}\left(W^{\left(n'\right)}\right)$ is a convex function, the proof of this lemma follows our previous work.
\end{IEEEproof}
\begin{lemma}\label{lem6}
If the wireless channel is non-manipulable, then for arbitrary small $\delta$, there exist sufficient large $n_0$, such that for any $n'>n_{0}$, $\lambda^{(n')}\left(\delta\right)\geq\mu_{n'}$.
\end{lemma}
\begin{IEEEproof}
Notice that for arbitrary $W_{f}^{(n')}\in\mathcal{D}^{(n')}$, 
there exist a pdf function $f\left(t\left|u\right.\right)$ which satisfies
{\small{\begin{equation}
\left[W_{f}^{(n')}\right]_{i,j,k}=\frac{\int_{u\in\mathcal{B}\left(\widetilde{\mathsf{u}}_{i}\right)}\int_{v\in\mathcal{B}\left(\widetilde{\mathsf{v}}_{j}\right)}f\left(v\left|u\right.\right)f_{U\left|\widetilde{X}\right.}\left(u\left|\widetilde{\mathsf{x}}_k\right.\right)dudv}{\int_{u\in\mathcal{B}\left(\widetilde{\mathsf{u}}_{i}\right)}f_{U\left|\widetilde{X}\right.}\left(u\left|\widetilde{\mathsf{x}}_k\right.\right)du}.
\end{equation}}}Then, according to the condition that 
$\frac{\triangle_{n'_{k}}}{\triangle_{n'_{k-1}}}=k$ and 
$\frac{\triangle_{n'_{k}}}{\triangle_{n'_{1}}}=s_{k}$, fixing $f\left(t\left|u\right.\right)$,
$\left|W_{f}^{(n')}-W_{0}^{(n')}\right|\geq\delta$ implies $\left|W_{f}^{(n'_{k})}-W_{0}^{(n'_{k})}\right|\geq\delta$
for $n'_{k}>n'$.
For the sake of proof, we define a function set 
{\small{\begin{equation}
\mathcal{F}=\left\{ f\left(v\left|u\right.\right):\lim_{n'\rightarrow\infty}\sum_{k=1}^{n_{x}}\sum_{j=1}^{n_{v}}\sum_{i=1}^{{n'}}\left|\frac{\int_{u\in\mathcal{B}\left(\widetilde{\mathsf{u}}_{i}\right)}\int_{v\in\mathcal{B}\left(\widetilde{\mathsf{v}}_{j}\right)}f\left(v\left|u\right.\right)f_{U\left|\widetilde{X}\right.}\left(u\left|\widetilde{\mathsf{x}}_k\right.\right)dudv}{\int_{u\in\mathcal{B}\left(\widetilde{\mathsf{u}}_{i}\right)}f_{U\left|\widetilde{X}\right.}\left(u\left|\widetilde{\mathsf{x}}_k\right.\right)du}-\left[W_{0}^{\left(n'\right)}\right]_{i.j,k}\right|\geq\delta\right\}.
\end{equation}}}Then, we define $\mathcal{\widetilde{D}}_{s}^{(n')}$ as
{\small{\begin{equation}
\mathcal{\widetilde{D}}_{s}^{(n')}=\left\{ W^{(n')}:\left[W^{(n')}\right]_{i,j,k}=\frac{\int_{u\in\mathcal{B}\left(\widetilde{\mathsf{u}}_{i}\right)}\int_{v\in\mathcal{B}\left(\widetilde{\mathsf{v}}_{j}\right)}f\left(v\left|u\right.\right)f_{U\left|\widetilde{X}\right.}\left(u\left|\widetilde{\mathsf{x}}_k\right.\right)dudv}{\int_{u\in\mathcal{B}\left(\widetilde{\mathsf{u}}_{i}\right)}f_{U\left|\widetilde{X}\right.}\left(u\left|\widetilde{\mathsf{x}}_k\right.\right)du},\, f\left(v\left|u\right.\right)\in\mathcal{F}\right\} 
\end{equation}}}Obviously, $\mathcal{D}_{s}^{(n')}\subseteq\mathcal{\widetilde{D}}_{s}^{(n')}$, hence, we get $\lambda^{(n')}\left(\delta\right)\geq\widetilde{\lambda}^{(n')}\left(\delta\right)$ where $\widetilde{\lambda}^{(n')}\left(\delta\right)$ is infimum of $M^{\left(n'\right)}\left(W^{\left(n'\right)}\right)$ across $\mathcal{\widetilde{D}}_{s}^{(n')}$. 

For arbitrary $f\left(v\left|u\right.\right)\in\mathcal{F}$, we define $\mathsf{\widehat{v}}_{m,j,k}^{(n')}$ which satisfies
{\small{\begin{equation}
F_{Y\left|V\right.}\left(t_{m}\left|\mathsf{\widehat{v}}_{m,j,k}^{(n')}\right.\right)=\frac{\int_{u\in\mathcal{B}\left(\widetilde{\mathsf{u}}_{i}\right)}\int_{v\in\mathcal{B}\left(\widetilde{\mathsf{v}}_{j}\right)}f\left(v\left|u\right.\right)f_{U\left|\widetilde{X}\right.}\left(u\left|\widetilde{\mathsf{x}}_k\right.\right)F_{Y\left|V\right.}\left(t_{m}\left|v\right.\right)dudv}{\int_{u\in\mathcal{B}\left(\widetilde{\mathsf{u}}_{i}\right)}\int_{v\in\mathcal{B}\left(\widetilde{\mathsf{v}}_{j}\right)}f\left(v\left|u\right.\right)f_{U\left|\widetilde{X}\right.}\left(u\left|\widetilde{\mathsf{x}}_k\right.\right)dudv}.
\end{equation}}}Then, we have
{\small{\begin{align}
&\nonumber -\left|\int_{-\infty}^{+\infty}f_{U\left|\widetilde{X}\right.}\left(u\left|\widetilde{\mathsf{x}}_k\right.\right)F_{Y\left|V\right.}\left(t_{m}\left|u\right.\right)du-\sum_{j=1}^{n_{v}}\sum_{i=1}^{n'}P_{\widetilde{U}\left|\widetilde{X}\right.}\left(\widetilde{\mathsf{u}}_{i}\right)w_{i,j,k}F_{Y\left|V\right.}\left(t_{m}\left|\overline{\mathsf{v}}_{m,j,k}\right.\right)\right|^{2}\\
&\nonumber +\left|\int_{-\infty}^{+\infty}f_{U\left|\widetilde{X}\right.}\left(u\left|\widetilde{\mathsf{x}}_k\right.\right)F_{Y\left|V\right.}\left(t_{m}\left|u\right.\right)du-\sum_{j=1}^{n_{v}}\sum_{i=1}^{n'}P_{\widetilde{U}\left|\widetilde{X}\right.}\left(\widetilde{\mathsf{u}}_{i}\right)w_{i,j,k}F_{Y\left|V\right.}\left(t_{m}\left|\mathsf{\widehat{v}}_{m,j,k}^{(n')}\right.\right)\right|^{2}=\\\nonumber
&\left(2\int_{-\infty}^{+\infty}f_{U\left|\widetilde{X}\right.}\left(u\left|\widetilde{\mathsf{x}}_k\right.\right)F_{Y\left|V\right.}\left(t_{m}\left|u\right.\right)du-\sum_{j=1}^{n_{v}}\sum_{i=1}^{n'}P_{\widetilde{U}\left|\widetilde{X}\right.}\left(\widetilde{\mathsf{u}}_{i}\right)w_{i,j,k}F_{Y\left|V\right.}\left(t_{m}\left|\overline{\mathsf{v}}_{m,j,k}\right.\right)-\sum_{j=1}^{n_{v}}\sum_{i=1}^{n'}P_{\widetilde{U}\left|\widetilde{X}\right.}\left(\widetilde{\mathsf{u}}_{i}\right)w_{i,j,k}F_{Y\left|V\right.}\left(t_{m}\left|\mathsf{\widehat{v}}_{m,j,k}^{(n')}\right.\right)\right)\\&\nonumber
\left(\sum_{j=1}^{n_{v}}\sum_{i=1}^{n'}P_{\widetilde{U}\left|\widetilde{X}\right.}\left(\widetilde{\mathsf{u}}_{i}\right)w_{i,j,k}F_{Y\left|V\right.}\left(t_{m}\left|\overline{\mathsf{v}}_{m,j,k}\right.\right)-\sum_{j=1}^{n_{v}}\sum_{i=1}^{n'}P_{\widetilde{U}\left|\widetilde{X}\right.}\left(\widetilde{\mathsf{u}}_{i}\right)w_{i,j,k}F_{Y\left|V\right.}\left(t_{m}\left|\mathsf{\widehat{v}}_{m,j,k}^{(n')}\right.\right)\right)\\&\nonumber
<2\max_{j=1,2,\ldots,n_v, m=1,\ldots,n_{y}}\left|\max_{v\in\mathcal{B}\left(\widetilde{\mathsf{v}}_{j}\right)}F_{Y\left|V\right.}\left(t_{m}\left|v\right.\right)-\min_{v\in\mathcal{B}\left(\widetilde{\mathsf{v}}_{j}\right)}F_{Y\left|V\right.}\left(t_{m}\left|v\right.\right)\right|
\end{align}}}Then, we get 
{\small{\begin{align}\label{temp}
&\nonumber\left|\int_{-\infty}^{+\infty}f_{U\left|\widetilde{X}\right.}\left(u\left|\widetilde{\mathsf{x}}_k\right.\right)F_{Y\left|V\right.}\left(t_{m}\left|u\right.\right)du-\sum_{j=1}^{n_{v}}\sum_{i=1}^{n'}P_{\widetilde{U}\left|\widetilde{X}\right.}\left(\widetilde{\mathsf{u}}_{i}\right)w_{i,j,k}F_{Y\left|V\right.}\left(t_{m}\left|\mathsf{\widehat{v}}_{m,j,k}^{(n')}\right.\right)\right|^{2}\\\nonumber & -2\max_{j=1,2,\ldots,n_v,  m=1,\ldots,n_{y}}\left|\max_{v\in\mathcal{B}\left(\widetilde{\mathsf{v}}_{j}\right)}F_{Y\left|V\right.}\left(t_{m}\left|v\right.\right)-\min_{v\in\mathcal{B}\left(\widetilde{\mathsf{v}}_{j}\right)}F_{Y\left|V\right.}\left(t_{m}\left|v\right.\right)\right|
\\& <\left|\int_{-\infty}^{+\infty}f_{U\left|\widetilde{X}\right.}\left(u\left|\widetilde{\mathsf{x}}_k\right.\right)F_{Y\left|V\right.}\left(t_{m}\left|v\right.\right)du-\sum_{j=1}^{n_{v}}\sum_{i=1}^{n'}P_{\widetilde{U}\left|\widetilde{X}\right.}\left(\widetilde{\mathsf{u}}_{i}\right)w_{i,j,k}F_{Y\left|V\right.}\left(t_{m}\left|\overline{\mathsf{v}}_{m,j,k}\right.\right)\right|^{2}
\end{align}}}From (\ref{temp}), we have 
{\small{\begin{equation}\label{temp1}
\widehat{M}^{\left(n'\right)}\left(W_f^{\left(n'\right)}\right)\leq M^{\left(n'\right)}\left(W_f^{\left(n'\right)}\right)+\gamma_{n'}
\end{equation}}}where 
{\small{\begin{align}
&\nonumber M^{\left(n'\right)}\left(W^{\left(n'\right)}\right)=\frac{\beta_{4}-\alpha_{4}}{{n_{x}-2}}\frac{\beta_{3}-\alpha_{3}}{{n_{y}-2}}\\&\nonumber\sum_{k=1}^{{n_{x}-1}}\sum_{m=1}^{{n_{y}-1}}\left|\int_{-\infty}^{+\infty}f_{U\left|\widetilde{X}\right.}\left(u\left|\widetilde{\mathsf{x}}_k\right.\right)F_{Y\left|V\right.}\left(t_{m}\left|u\right.\right)du-\sum_{j=1}^{n_{v}}\sum_{i=1}^{n'}P_{\widetilde{U}\left|\widetilde{X}\right.}\left(\widetilde{\mathsf{u}}_{i}\right)w_{i,j,k}F_{Y\left|V\right.}\left(t_{m}\left|\widehat{\mathsf{v}}_{m,j,k}^{(n')}\right.\right)\right|^{2}
\\
&\gamma_{n'}=\max_{j=1,\ldots,n_{v},m=1,\ldots,n_{y}}4\sqrt{n_{y}}\sqrt{n_{x}}\left|\max_{v\in\mathcal{B}\left(\widetilde{\mathsf{v}}_{j}\right)}F_{Y\left|V\right.}\left(t_{m}\left|v\right.\right)-\min_{v\in\mathcal{B}\left(\widetilde{\mathsf{v}}_{m,j}\right)}F_{Y\left|V\right.}\left(t_{m}\left|v\right.\right)\right|
\end{align}}}According to the stochastic property of channel model, we have $\lim_{n'\rightarrow\infty}\gamma_{n'}=0$.
In order to prove $\widehat{M}^{\left(n'\right)}\left(W_f^{\left(n'\right)}\right)-\gamma_{n'}>\mu_{n'}$, for arbitrary $f\left(v\left|u\right.\right)\in\mathcal{F}$, we assume for arbitrary large $n'_0$, there exist $n'>n'_0$, such that 
$\widehat{\lambda}^{(n')}\left(\delta\right)-\gamma_{n'}\leq\mu_{n'}$. In other words, 
there exist a sequence denoted as $\widehat{n}_1<
\widehat{n}_2<\ldots, \infty$ by which 
{\small{\begin{equation}
\widehat{M}^{\left(\widehat{n}_{k}\right)}\left(W_{f}^{\left(\widehat{n}_{k}\right)}\right)-\gamma_{\widehat{n}_{k}}\leq\mu_{\widehat{n}_{k}}.
\end{equation}}}
Then, we have
\begin{equation}
\lim_{k\rightarrow\infty}\widehat{M}^{\left(\widehat{n}_{k}\right)}\left(W_{f}^{\left(\widehat{n}_{k}\right)}\right)\leq\lim_{k\rightarrow\infty}(\mu_{\widehat{n}_{k}}+\gamma_{\widehat{n}_{k}})=0
\end{equation}From the expressions of $W_{f}^{(n')}$ and $\widehat{M}^{\left(n'\right)}\left(W^{\left(n'\right)}\right)$, we get there is a division manner for $t\in\left(-\infty,+\infty\right)$ characterized by $\widehat{n}_1<
\widehat{n}_2<\ldots, \infty$ such that
{\small{\begin{align}\label{key1}
\nonumber&\lim_{\widehat{n}_{k}\rightarrow\infty}\frac{\beta_{4}-\alpha_{4}}{{n_{x}-2}}\frac{\beta_{3}-\alpha_{3}}{{n_{y}-2}}\\&\sum_{k=1}^{{n_{x}-1}}\sum_{m=1}^{{n_{y}-2}}\left|\int_{-\infty}^{+\infty}f_{U\left|\widetilde{X}\right.}\left(u\left|\widetilde{\mathsf{x}}_k\right.\right)F_{Y\left|V\right.}\left(t_{m}^{(\widehat{n}_{k})}\left|u\right.\right)du-\int_{-\infty}^{+\infty}\int_{-\infty}^{+\infty}f_{U\left|\widetilde{X}\right.}\left(u\left|\widetilde{\mathsf{x}}_k\right.\right)f\left(v\left|u\right.\right)F_{Y\left|V\right.}\left(t_{m}^{(\widehat{n}_{k})}\left|v\right.\right)dudv\right|^{2}=0\end{align}}}
On the other hand, from the definition of $\mathcal{F}$ and the condition that the wireless channel is non-manipulable, we get
\begin{equation}
\int_{-\infty}^{+\infty}f_{U\left|{X}\right.}\left(u\left|x\right.\right)F_{Y\left|V\right.}\left(t\left|u\right.\right)du\neq\int_{-\infty}^{+\infty}\int_{-\infty}^{+\infty}f_{U\left|{X}\right.}\left(u\left|x\right.\right)f\left(v\left|u\right.\right)F_{Y\left|V\right.}\left(t\left|u\right.\right)dvdu\end{equation}Hence, if $\int_{-\infty}^{+\infty}\int_{-\infty}^{+\infty}\left|\int_{-\infty}^{+\infty}f_{U\left|{X}\right.}\left(u\left|x\right.\right)F_{Y\left|V\right.}\left(t\left|u\right.\right)du-\int_{-\infty}^{+\infty}\int_{-\infty}^{+\infty}f_{U\left|{X}\right.}\left(u\left|x\right.\right)f\left(v\left|u\right.\right)F_{Y\left|V\right.}\left(t\left|u\right.\right)dvdu\right|^{2}dxdt$ can be integrated, we must have
\begin{equation}
\int_{-\infty}^{+\infty}\int_{-\infty}^{+\infty}\left|\int_{-\infty}^{+\infty}f_{U\left|{X}\right.}\left(u\left|x\right.\right)F_{Y\left|V\right.}\left(t\left|u\right.\right)du-\int_{-\infty}^{+\infty}\int_{-\infty}^{+\infty}f_{U\left|{X}\right.}\left(u\left|x\right.\right)f\left(v\left|u\right.\right)F_{Y\left|V\right.}\left(t\left|u\right.\right)dvdu\right|^{2}dxdt>0,
\end{equation}which indicates there is no division manner for $t\in\left(-\infty,+\infty\right)$ making (\ref{key1}) be true. It contradicts with the meaning
of (\ref{key1}).
We proceed to examine another case that if {\small{$\int_{-\infty}^{+\infty}\int_{-\infty}^{+\infty}\left|\int_{-\infty}^{+\infty}f_{U\left|{X}\right.}\left(u\left|x\right.\right)F_{Y\left|V\right.}\left(t\left|u\right.\right)du-\int_{-\infty}^{+\infty}\int_{-\infty}^{+\infty}f_{U\left|{X}\right.}\left(u\left|x\right.\right)f\left(v\left|u\right.\right)F_{Y\left|V\right.}\left(t\left|u\right.\right)dvdu\right|^{2}dxdt$}} cannot be integrated, since {\small{$\left|\int_{-\infty}^{+\infty}f_{U\left|{X}\right.}\left(u\left|x\right.\right)F_{Y\left|V\right.}\left(t\left|u\right.\right)du-\int_{-\infty}^{+\infty}\int_{-\infty}^{+\infty}f_{U\left|{X}\right.}\left(u\left|x\right.\right)f\left(v\left|u\right.\right)F_{Y\left|V\right.}\left(t\left|u\right.\right)dvdu\right|^{2}>0$}}, we have
{\small{\begin{equation}
\lim_{\widetilde{\beta}\rightarrow\infty,\widetilde{\alpha}\rightarrow-\infty}\lim_{\beta\rightarrow\infty,\alpha\rightarrow-\infty}\int_{\widetilde{\alpha}}^{\widetilde{\beta}}\int_{\alpha}^{\beta}\left|\int_{-\infty}^{+\infty}f_{U\left|{X}\right.}\left(u\left|x\right.\right)F_{Y\left|V\right.}\left(t\left|u\right.\right)du-\int_{-\infty}^{+\infty}\int_{-\infty}^{+\infty}f_{U\left|{X}\right.}\left(u\left|x\right.\right)f\left(v\left|u\right.\right)F_{Y\left|V\right.}\left(t\left|u\right.\right)dvdu\right|^{2}dtdx=\infty
\end{equation}}}Hence, there has $\alpha'$, $\beta'$, $\widetilde{\alpha}'$ and $\widetilde{\beta}'$ by which 
\begin{equation}\label{key2}
\int_{\widetilde{\alpha}'}^{\widetilde{\beta}'}\int_{\alpha'}^{\beta'}\left|\int_{-\infty}^{+\infty}f_{U\left|{X}\right.}\left(u\left|x\right.\right)F_{Y\left|V\right.}\left(t\left|u\right.\right)du-\int_{-\infty}^{+\infty}\int_{-\infty}^{+\infty}f_{U\left|{X}\right.}\left(u\left|x\right.\right)f\left(v\left|u\right.\right)F_{Y\left|V\right.}\left(t\left|u\right.\right)dvdu\right|^{2}dtdx>0\end{equation} 
Meanwhile, (\ref{key1}) indicates
\begin{align}\label{key3}
\nonumber&\lim_{\widehat{n}_{k}\rightarrow\infty}\frac{\beta_{4}-\alpha_{4}}{{n_{x}-2}}\frac{\beta_{3}-\alpha_{3}}{{n_{y}-2}}\sum_{\widetilde{\mathsf{x}}_k\in\left[\widetilde{\alpha}',\widetilde{\beta}'\right]}\sum_{t_{m}\in\left[\alpha',\beta'\right]}\\&\left|\int_{-\infty}^{+\infty}f_{U\left|\widetilde{X}\right.}\left(u\left|\widetilde{\mathsf{x}}_k\right.\right)F_{Y\left|V\right.}\left(t_{m}^{(\widehat{n}_{k})}\left|u\right.\right)du-\int_{-\infty}^{+\infty}\int_{-\infty}^{+\infty}f_{U\left|\widetilde{X}\right.}\left(u\left|\widetilde{\mathsf{x}}_k\right.\right)f\left(v\left|u\right.\right)F_{Y\left|V\right.}\left(t_{m}^{(\widehat{n}_{k})}\left|v\right.\right)dudv\right|^{2}=0.
\end{align}However, (\ref{key2}) indicates there is no division manner for $t\in\left(\alpha',\beta'\right)$ and $x_2\in\left(\widehat{\alpha}',\widehat{\beta}'\right)$ making (\ref{key3}) be true. 
Hence, the contradiction happens. Due to these contradictions, we attain the assumption that for arbitrary large $n'_0$, there exist $n'>n'_0$, such that 
$\widehat{M}^{\left(n'\right)}\left(W_{f}^{\left(n'\right)}\right)-\gamma_{n'}\leq\mu_{n'}$ is not right. Therefore, we have there exist $n'_0$, for any $n'>n'_0$, there has $\widehat{M}^{\left(n'\right)}\left(W_{f}^{\left(n'\right)}\right)-\gamma_{n'}>\mu_{n'}$. 
Applying the aforementioned derivation to each function belonging 
to $\mathcal{F}$, we get there exist $n_0$, for any $n'>n_0$, $\widehat{M}^{\left(n'\right)}\left(W_{f}^{\left(n'\right)}\right)-\gamma_{n'}>\mu_{n'}$ is available 
for all possible functions of $\mathcal{F}$. According to (\ref{temp1}), we have 
\begin{equation}
M^{\left(n'\right)}\left(W_f^{\left(n'\right)}\right)<\mu_{n'}
\end{equation}for each function belonging 
to $\mathcal{F}$. 
Since $\widetilde{\lambda}^{(n')}\left(\delta\right)$ is infimum of ${M}^{\left(n'\right)}\left(W^{\left(n'\right)}\right)$ across $\mathcal{\widetilde{D}}_{s}^{(n')}$, we thus have 
\begin{equation}
\widetilde{\lambda}^{(n')}\left(\delta\right)>\mu_{n'}
\end{equation}Revisiting $\lambda^{(n')}\left(\delta\right)\geq\widetilde{\lambda}^{(n')}\left(\delta\right)$, we get
\begin{equation}
{\lambda}^{(n')}\left(\delta\right)>\mu_{n'}
\end{equation}
Finally, the proof is completed. 
\end{IEEEproof}
\begin{lemma}\label{p2}
Fixing arbitrary small $\epsilon$ and $\delta$, if there exist $n'_0$ such that {\small{$$\Pr\{\underbrace{\sum_{j=1}^{{n_{v}}}\sum_{i=1}^{n'_{0}}\left|\triangle F^{(n'_0)}_{\widetilde{V}^{n}\left|\widetilde{U}^{n}\right.}\left(\widetilde{\mathsf{v}}_{j}\left|\widetilde{\mathsf{u}}_{i}\right.\right)-W_{0}^{(n'_{0})}\right|}_{R\left(U^{n},V^{n},n'_{0}\right)}>\delta\}>0$$}}then, we have for $n'>n'_0$, {\small{$\Pr\left\{ D^{n}<\varepsilon\left(n', \delta\right)\left|{R\left(U^{n},V^{n},n'\right)}>\delta\right.\right\}$}} is well-defined and 
$$\Pr\left\{ D^{n}<\varepsilon\left(n', \delta\right)\left|{R\left(U^{n},V^{n},n'\right)}>\delta\right.\right\} \leq\epsilon$$ where $n\rightarrow\infty$, 
$n'$ is sufficient large so as to satisfy the properties given by lemma 4 and lemma 6. $\varepsilon\left(n', \delta\right)$ is strictly positive and can be arbitrary small value.
\end{lemma}

\begin{IEEEproof}
According to lemma 7, there exist $\mu_{n'}$ such that
{\small{\begin{align}\label{p2_1}
\nonumber&\Pr\left\{ \frac{\beta_{4}-\alpha_{4}}{{n_{x}-2}}\frac{\beta_{3}-\alpha_{3}}{{n_{y}-2}}\sum_{k=1}^{{n_{x}-1}}\sum_{m=1}^{n_{y}-1}\left|F_{Y^{n}\left|\widetilde{X}^{n}\right.}^{n}\left(t\left|\widetilde{\mathsf{x}}_k\right.\right)-\sum_{j=1}^{n_{v}}\sum_{i=1}^{n'}P_{\widetilde{U}\left|\widetilde{X}\right.}\left(\widetilde{\mathsf{u}}_{i}\left|\widetilde{\msf{x}}\right.\right)\triangle F_{\widetilde{V}^{n}\left|\widetilde{U}^{n}\right.}^{(n')}\left(\widetilde{\mathsf{v}}_{j}\left|\widetilde{\mathsf{u}}_{i}\right.\right)F_{Y\left|V\right.}\left(t_{m}\left|\overline{\mathsf{v}}_{m,j,k}\right.\right)\right|>\mu_{n'}\right\}  \\& \leq\epsilon\Pr\left\{ R\left(U^{n},V^{n},n'_{0}\right)>\delta\right\}\leq\epsilon\Pr\left\{ R\left(U^{n},V^{n},n'\right)>\delta\right\},
\end{align}}}where $\mu_{n'}\rightarrow0$ as $n'\rightarrow\infty, n\rightarrow\infty$. The last inequality follows the fact that $R\left(U^{n},V^{n},n'_{0}\right)>\delta$ implies $R\left(U^{n},V^{n},n'\right)>\delta$ according to the definition of $\triangle F_{\widetilde{V}^{n}\left|\widetilde{U}^{n}\right.}^{(n')}\left(\widetilde{\mathsf{v}}_{j}\left|\widetilde{\mathsf{u}}_{i}\right.\right)$. 

Then, if {\small{$$\underbrace{\frac{\beta_{4}^{(n')}-\alpha_{4}^{(n')}}{{n_{x}-2}}\frac{\beta_{3}^{(n')}-\alpha_{3}^{(n')}}{{n_{y}-2}}\sum_{k=1}^{{n_{x}-1}}\sum_{m=1}^{n_{y}-1}\left|F_{Y^{n}\left|\widetilde{X}^{n}\right.}^{n}\left(t\left|\widetilde{\mathsf{x}}_k\right.\right)-\sum_{j=1}^{n_{v}}\sum_{i=1}^{n'}P_{\widetilde{U}\left|\widetilde{X}\right.}\left(\widetilde{\mathsf{u}}_{i}\left|\widetilde{\msf{x}}\right.\right)\triangle F_{\widetilde{V}^{n}\left|\widetilde{U}^{n}\right.}^{(n')}\left(\widetilde{\mathsf{v}}_{j}\left|\widetilde{\mathsf{u}}_{i}\right.\right)F_{Y\left|V\right.}\left(t_{m}\left|\overline{\mathsf{v}}_{m,j,k}\right.\right)\right|}_{G\left(V^{n},U^{n},\widetilde{X}^{n}\right)}<\mu_{n'},$$}}we must have
{\small{\begin{align}\label{p2_2}
\nonumber&\left(\beta_{3}^{(n')}-\alpha_{3}^{(n')}\right)\left(\beta_{4}^{(n')}-\alpha_{4}^{(n')}\right)D^{n}\\&\geq{\frac{\beta_{4}^{(\widehat{n}_{k})}-\alpha_{4}^{(\widehat{n}_{k})}}{{n_{x}-2}}\frac{\beta_{3}^{(\widehat{n}_{k})}-\alpha_{3}^{(\widehat{n}_{k})}}{{n_{y}-2}}\sum_{k=1}^{{n_{x}-1}}\sum_{m=1}^{n_{y}-1}\left|\int_{-\infty}^{+\infty}f_{U\left|\widetilde{X}\right.}\left(u\left|\widetilde{\mathsf{x}}_k\right.\right)F_{Y\left|V\right.}\left(t_{m}\left|u\right.\right)du-\sum_{j=1}^{n_{v}}\sum_{i=1}^{n'}P_{\widetilde{U}\left|\widetilde{X}\right.}\left(\widetilde{\mathsf{u}}_{i}\left|\widetilde{\msf{x}}\right.\right)\triangle F_{\widetilde{V}^{n}\left|\widetilde{U}^{n}\right.}^{(n')}\left(\widetilde{\mathsf{v}}_{j}\left|\widetilde{\mathsf{u}}_{i}\right.\right)F_{Y\left|V\right.}\left(t_{m}\left|\overline{\mathsf{v}}_{m,j,k}\right.\right)\right|}-\mu_{n'}\\\nonumber
&\geq{\frac{\beta_{4}^{(n')}-\alpha_{4}^{(n')}}{{n_{x}-2}}\frac{\beta_{3}^{(n')}-\alpha_{3}^{(n')}}{{n_{y}-2}}\sum_{k=1}^{{n_{x}-1}}\sum_{m=1}^{n_{y}-1}\left|\int_{-\infty}^{+\infty}f_{U\left|\widetilde{X}\right.}\left(u\left|\widetilde{\mathsf{x}}_k\right.\right)F_{Y\left|V\right.}\left(t_{m}\left|u\right.\right)du-\sum_{j=1}^{n_{v}}\sum_{i=1}^{n'}P_{\widetilde{U}\left|\widetilde{X}\right.}\left(\widetilde{\mathsf{u}}_{i}\left|\widetilde{\msf{x}}\right.\right)\triangle F_{\widetilde{V}^{n}\left|\widetilde{U}^{n}\right.}^{(n')}\left(\widetilde{\mathsf{v}}_{j}\left|\widetilde{\mathsf{u}}_{i}\right.\right)F_{Y\left|V\right.}\left(t_{m}\left|\overline{\mathsf{v}}_{m,j,k}\right.\right)\right|^2}-\mu_{n'}
\end{align}}}On the other hand, if {\small{$\underbrace{\sum_{j=1}^{{n_{v}}}\sum_{i=1}^{n'}\left|\triangle F_{\widetilde{V}^{n}\left|\widetilde{U}^{n}\right.}^{\left(n'\right)}\left(\widetilde{\mathsf{v}}_{j}\left|\widetilde{\mathsf{u}}_{i}\right.\right)-W_{0}^{(n')}\right|}_{R\left(U^{n},V^{n},n'\right)}>\delta,$}} according to lemma \ref{lem5}, the right side of (\ref{p2_2}) becomes
{\small{\begin{equation}\label{p2_3}
\left(\beta_{3}^{(n')}-\alpha_{3}^{(n')}\right)\left(\beta_{4}^{(n')}-\alpha_{4}^{(n')}\right)D^{n}\geq\lambda^{(n')}\left(\delta\right)-\mu_{n'}
\end{equation}}}which can be reshaped as 
{\small{\begin{equation}\label{p2_4}
D^{n}\geq\frac{\lambda^{(n')}\left(\delta\right)-\mu_{n'}}{\left(\beta_{3}^{(n')}-\alpha_{3}^{(n')}\right)\left(\beta_{4}^{(n')}-\alpha_{4}^{(n')}\right)}
\end{equation}}}Define $\varepsilon\left(n',\delta\right)=\frac{\lambda^{(n')}\left(\delta\right)-\mu_{n'}}{\left(\beta_{3}^{(n')}-\alpha_{3}^{(n')}\right)\left(\beta_{4}^{(n')}-\alpha_{4}^{(n')}\right)}$,
according to lemma \ref{lem6}, $\varepsilon\left(n',\delta\right)>0$ as $n'$ is sufficient large.
From the properties of $\mu_{n'}$ and $\lambda^{(n')}\left(\delta\right)$, 
$\varepsilon\left(n',\delta\right)$ can be arbitrarily small.  

Upon (\ref{p2_2}), (\ref{p2_3}) and (\ref{p2_4}), we have
{\small{\begin{align*}
&\Pr\left\{ D^{n}\geq\varepsilon\left(n',\delta\right),R\left(U^{n},V^{n},n'\right)\geq\delta\right\} \\
&\geq\Pr\left\{ D^{n}\geq\varepsilon\left(n',\delta\right),R\left(U^{n},V^{n},n'\right)\geq\delta,G\left(V^{n},U^{n},\widetilde{X}^{n}\right)\leq\mu'\right\}\\
&=\Pr\left\{ R\left(U^{n},V^{n},n'\right)\geq\delta,G\left(V^{n},U^{n},\widetilde{X}^{n}\right)\leq\mu_{n'}\right\}\\ 
&\geq\Pr\left\{ R\left(U^{n},V^{n},n'\right)\geq\delta\right\} -\Pr\left\{ G\left(V^{n},U^{n},\widetilde{X}^{n}\right)\geq\mu_{n'}\right\} 
\end{align*}}}where the equation follows the logic from (\ref{p2_2}), (\ref{p2_3}) to (\ref{p2_4}). Then, we have
{\small{\begin{equation}
\Pr\left\{ D^{n}\geq\varepsilon\left(n',\delta\right)\left|R\left(U^{n},V^{n},n'\right)\geq\delta\right.\right\} =\frac{\Pr\left\{ D^{n}\geq\varepsilon\left(n',\delta\right),R\left(U^{n},V^{n},n'\right)\geq\delta\right\} }{\Pr\left\{ R\left(U^{n},V^{n},n'\right)\geq\delta\right\} }>1-\frac{\Pr\left\{ G\left(V^{n},U^{n},\widetilde{X}^{n}\right)\geq\mu_{n'}\right\} }{\Pr\left\{ R\left(U^{n},V^{n},n'\right)\geq\delta\right\} }\geq1-\epsilon
\end{equation}}}where the last inequality follows (\ref{p2_1}).
The proof is finished.
\end{IEEEproof}
The first property of theorem 1 is direct result from lemma {\ref{p2}}. 


We proceed to prove the second property of theorem 1. For arbitrary small $\delta$, $\mu$ and $\mu'(n',\delta)=\mu+\frac{n_{x}-1}{n_{x}-2}\frac{n_{y}-1}{n_{y}-2}\delta$, we have
{\small{\begin{align}
&\nonumber \Pr\left\{ D^{n}\leq\mu'(n',\delta)\bigcap\sum_{j=1}^{{n_{v}}}\sum_{i=1}^{n'}\left|\triangle F_{\widetilde{V}^{n}\left|\widetilde{U}^{n}\right.}^{\left(n'\right)}\left(\widetilde{\mathsf{v}}_{j}\left|\widetilde{\mathsf{u}}_{i}\right.\right)-W_{0}^{(n')}\right|\leq\delta\right\} \geq\\
&\nonumber \Pr\big\{ D^{n}\leq\mu'(n',\delta)\bigcap\frac{1}{{n_{x}-2}}\frac{1}{{n_{y}-2}}\sum_{k=1}^{{n_{x}-1}}\sum_{m=1}^{n_{y}-1}\left|F_{Y^{n}\left|\widetilde{X}^{n}\right.}^{n}\left(t\left|\widetilde{\mathsf{x}}_k\right.\right)-\sum_{j=1}^{n_{v}}\sum_{i=1}^{n'}P_{\widetilde{U}\left|\widetilde{X}\right.}\left(\widetilde{\mathsf{u}}_{i}\left|\widetilde{\msf{x}}\right.\right)\triangle F_{\widetilde{V}^{n}\left|\widetilde{U}^{n}\right.}^{(n')}\left(\widetilde{\mathsf{v}}_{j}\left|\widetilde{\mathsf{u}}_{i}\right.\right)F_{Y\left|V\right.}\left(t_{m}\left|\overline{\mathsf{v}}_{m,j,k}\right.\right)\right|\leq\mu\\&\nonumber \bigcap\sum_{j=1}^{{n_{v}}}\sum_{i=1}^{n'}\left|\triangle F_{\widetilde{V}^{n}\left|\widetilde{U}^{n}\right.}^{\left(n'\right)}\left(\widetilde{\mathsf{v}}_{j}\left|\widetilde{\mathsf{u}}_{i}\right.\right)-W_{0}^{(n')}\right|\leq\delta\big\} \\
&\nonumber =\Pr\big\{ \frac{1}{{n_{x}-2}}\frac{1}{{n_{y}-2}}\sum_{k=1}^{{n_{x}-1}}\sum_{m=1}^{n_{y}-1}\left|F_{Y^{n}\left|\widetilde{X}^{n}\right.}^{n}\left(t\left|\widetilde{\mathsf{x}}_k\right.\right)-\sum_{j=1}^{n_{v}}\sum_{i=1}^{n'}P_{\widetilde{U}\left|\widetilde{X}\right.}\left(\widetilde{\mathsf{u}}_{i}\left|\widetilde{\msf{x}}\right.\right)\triangle F_{\widetilde{V}^{n}\left|\widetilde{U}^{n}\right.}^{(n')}\left(\widetilde{\mathsf{v}}_{j}\left|\widetilde{\mathsf{u}}_{i}\right.\right)F_{Y\left|V\right.}\left(t_{m}\left|\overline{\mathsf{v}}_{m,j,k}\right.\right)\right|\leq\mu\\&\nonumber \bigcap\sum_{j=1}^{{n_{v}}}\sum_{i=1}^{n'}\left|\triangle F_{\widetilde{V}^{n}\left|\widetilde{U}^{n}\right.}^{\left(n'\right)}\left(\widetilde{\mathsf{v}}_{j}\left|\widetilde{\mathsf{u}}_{i}\right.\right)-W_{0}^{(n')}\right|\leq\delta\big\}  \\
&\nonumber \geq\Pr\left\{\sum_{j=1}^{{n_{v}}}\sum_{i=1}^{n'}\left|\triangle F_{\widetilde{V}^{n}\left|\widetilde{U}^{n}\right.}^{\left(n'\right)}\left(\widetilde{\mathsf{v}}_{j}\left|\widetilde{\mathsf{u}}_{i}\right.\right)-W_{0}^{(n')}\right|\leq\delta\right\} -\\&\nonumber\Pr\left\{\frac{1}{{n_{x}-2}}\frac{1}{{n_{y}-2}}\sum_{k=1}^{{n_{x}-1}}\sum_{m=1}^{n_{y}-1}\left|F_{Y^{n}\left|\widetilde{X}^{n}\right.}^{n}\left(t\left|\widetilde{\mathsf{x}}_k\right.\right)-\sum_{j=1}^{n_{v}}\sum_{i=1}^{n'}P_{\widetilde{U}\left|\widetilde{X}\right.}\left(\widetilde{\mathsf{u}}_{i}\left|\widetilde{\msf{x}}\right.\right)\triangle F_{\widetilde{V}^{n}\left|\widetilde{U}^{n}\right.}^{(n')}\left(\widetilde{\mathsf{v}}_{j}\left|\widetilde{\mathsf{u}}_{i}\right.\right)F_{Y\left|V\right.}\left(t_{m}\left|\overline{\mathsf{v}}_{m,j,k}\right.\right)\right|>\mu\right\}\label{p1}
\end{align}}}where the equality firstly follows the fact that 
\begin{align}
&\nonumber D^{n}\leq\frac{1}{{n_{x}-2}}\frac{1}{{n_{y}-2}}\sum_{k=1}^{{n_{x}-1}}\sum_{m=1}^{n_{y}-1}\left|F_{Y^{n}\left|\widetilde{X}^{n}\right.}^{n}\left(t\left|\widetilde{\mathsf{x}}_k\right.\right)-\sum_{j=1}^{n_{v}}\sum_{i=1}^{n'}P_{\widetilde{U}\left|\widetilde{X}\right.}\left(\widetilde{\mathsf{u}}_{i}\left|\widetilde{\msf{x}}\right.\right)\triangle F_{\widetilde{V}^{n}\left|\widetilde{U}^{n}\right.}^{(n')}\left(\widetilde{\mathsf{v}}_{j}\left|\widetilde{\mathsf{u}}_{i}\right.\right)F_{Y\left|V\right.}\left(t_{m}\left|\overline{\mathsf{v}}_{m,j,k}\right.\right)\right|\\
&\nonumber+\frac{1}{{n_{x}-2}}\frac{1}{{n_{y}-2}}\sum_{k=1}^{{n_{x}-1}}\sum_{m=1}^{n_{y}-1}\left|\sum_{j=1}^{n_{v}}\sum_{i=1}^{n'}P_{\widetilde{U}\left|\widetilde{X}\right.}\left(\widetilde{\mathsf{u}}_{i}\left|\widetilde{\mathsf{x}}_k\right.\right)\left(\triangle F_{\widetilde{V}^{n}\left|\widetilde{U}^{n}\right.}^{(n')}\left(\widetilde{\mathsf{v}}_{j}\left|\widetilde{\mathsf{u}}_{i}\right.\right)-\left[W_{0}^{(n')}\right]_{i,j}\right)F_{Y\left|V\right.}\left(t_{m}\left|\overline{\mathsf{v}}_{m,j,k}\right.\right)\right|\\
&<\mu+\frac{n_{x}-1}{n_{x}-2}\frac{n_{y}-1}{n_{y}-2}\delta=\mu'(n', \delta).
\end{align}Hence, the equality in (\ref{p1}) is established. Upon (\ref{p1}) and lemma 3, the property 2 in Theorem 1 is direct.

\bibliographystyle{IEEEtran}

\begin{thebibliography}{10}


\bibitem{Buttyan2006Security}
L.~Buttyan and J-P.~Hubaux, \emph{ Security and Cooperation in Wireless
  Networks}, Cambridge University Press, 2007.

\bibitem{bloch2011physical}
M.~Bloch and J.~Barros,\emph{ Physical-Layer Security: From Information
  Theory to Security Engineering}, Cambridge University Press, 2011.


\bibitem{papadimitratos2006secure}
P.~Papadimitratos and Z. J.~Haas,
  ``Secure data communication in mobile ad hoc networks," \emph{
    IEEE J. Sel. Areas Commun.}, vol.~24, no.~2, pp. 343--356,
  Feb. 2006.


\bibitem{hu2005ariadne} Y. C.~Hu, A.~Perrig, and D. B.~Johnson,
  ``Ariadne: A secure on-demand routing protocol for ad hoc networks,"
 \emph{ Wirel.  Netw.}, vol.~11, no. 1-2, pp. 21--38,  Jan. 2005.
 

 \bibitem{ho2008byzantine} T.~Ho, B.~Leong, R.~Koetter,
  M.~M{\'e}dard, M. ~Effros, and D. R.~Karger, ``Byzantine
  modification detection in multicast networks with random network
  coding,"\emph{ IEEE Trans. Inf. Theory}, vol.~54, no.~6,
  pp. 2798--2803,  Jun. 2008.

\bibitem{kosut2009nonlinear}
O.~Kosut, L.~Tong, and D.~Tse, ``Nonlinear network coding is necessary
to combat general Byzantine attacks," in \emph{Proc. 47th Annu. Allerton Conf. on Commun., Control, and Compu.}, Monticello, IL, Oct. 2009, pp. 593--599.
 

 \bibitem{mao2007tracing}
Y.~Mao and M.~Wu, ``Tracing malicious relays in cooperative wireless
communications," \emph{ IEEE Trans. Inf. Forens. Security}, vol.~2,
no.~2, pp.198--212,  Jun. 2007.

\bibitem{Tradeoff}
T. Khalaf, S. Kim, and A. Abdel-Hakim, ``Tradeoff between reliability and security in multiple access relay networks under falsified data injection attack," \emph{ IEEE Trans. Inf. Forens. Security}, vol.~9, no.~3, pp. 386--396, Mar. 2014.

\bibitem{noncoherent}
L.-C.~Lo, Z.-J.~Wang, and W. J.~Huang, ``Noncoherent misbehavior detection in space-time coded cooperative networks,'' in \emph{Proc. IEEE Intl. Conf. Acoustics, Speech and Signal Processing (ICASSP)}, Kyoto, Mar. 2012, pp. 3061--3064.

\bibitem{nonherentsCL}
L.-C. ~Lo, W.~J.~Huang, R.~Y.~Chang, and W.-H.~Chung, ``Noncoherent detection of misbehaving relays in decode-and-forward cooperative networks," \emph{IEEE Comm. Letters}, vol.~19, no.~9, pp.1536--1539, Sept. 2015.

\bibitem{KimTWC}
S. W. Kim, ``Physical integrity check in cooperative relay communications," \emph{IEEE Trans. on Wirel. commun.}, vol. 14, no.11, pp. 6401--6413,
Nov. 2015.

\bibitem{KimCL}
S. W. Kim, T. Khalaf and S. Kim, ``MAP Detection of Misbehaving Relay in Wireless Multiple Access Relay Networks," \emph{IEEE Comm. Letters}, vol. 15, no. 3, pp. 340-342, March 2011.



\bibitem{OFDM}
W.~Hou, X.~Wang, and A.~Refaey, ``Misbehavior detection in amplify-and-forward cooperative OFDM systems," in \emph{Proc. IEEE Intl. Conf. Commun. (ICC)}, Budapest, 2013, Jun. 2013, pp. 5345--5349.

\bibitem{ARQ}
S. Dehnie and S. Tomasin, "Detection of Selfish Nodes in Networks Using CoopMAC Protocol with ARQ,"  \emph{IEEE Trans. on Wirel. commun.}, vol. 9, no. 7, pp. 2328-2337, July 2010.



\bibitem{he2013strong}
X.~He, and A.~Yener, ``Strong secrecy and reliable Byzantine detection
in the presence of an untrusted relay," \emph{ IEEE
  Trans. Inf. Theory}, vol.~59, no.~1, pp. 177--192,  Jan. 2013.


\bibitem{GravesINFOCOM12}
E.~Graves and T. F.~Wong, ``Detection of channel degradation attack by intermediary node in linear networks,"  in \emph{ Proc. IEEE Intl. Conf. Comput. Commun. (INFOCOM)}, Orlando, FL, Mar. 2012, pp. 747--755.
%

\bibitem{GravesIT12}
E.~Graves and T. F.~Wong,  ``Detectability of symbol manipulation by an amplify-and-forward relay,"  in \emph{ arXiv preprint
arXiv:1205.2681}, 2012.

\bibitem{GravesISIT13}
E.~Graves and T. F.~Wong,  ``A coding approach to guarantee information
integrity against a Byzantine relay,"  in \emph{ Proc. IEEE
  Intl. Symp. Inf. Theory (ISIT)}, Istanbul, Jul. 2013, pp. 2780--2784.
 \bibitem{CaoTIFS} 
  R. Cao, T. F. Wong, T. Lv, H. Gao and S. Yang, ``Detecting Byzantine Attacks Without Clean Reference," \emph{ IEEE Trans. Inf. Forens. Security}, vol. 11, no. 12, pp. 2717-2731, Dec. 2016.

 \end{thebibliography}

\end{document}